\documentclass[11pt,a4paper]{article}
\usepackage{jheppub}
\usepackage{amsmath}
\usepackage{amssymb}
\usepackage{amsthm}
\usepackage{mathrsfs}
\usepackage{graphicx}
\usepackage{fancyhdr}
\usepackage{array}
\usepackage{latexsym}
\usepackage[all]{xy}
\usepackage{eufrak}
\usepackage{euscript}
\usepackage{enumerate}
\usepackage{dsfont}
\usepackage{slashed}
\usepackage{hyperref}

\def\[{\left\lbrack}
\def\]{\right\rbrack}

\def\({\left(}
\def\){\right)}

\newcommand{\be}{\begin{equation}}
\newcommand{\ee}{\end{equation}}
\newcommand{\ea}{\end{eqnarray}}
\newcommand{\ba}{\begin{eqnarray}}

\newcommand{\Dslash}{D\!\!\!\!/}

\title{Confinement, brane symmetry and the Julia-Toulouse Approach for condensation of defects}

\author[a,b]{L. S. Grigorio,}
\author[c]{M. S. Guimaraes}%
\author[a]{R. Rougemont}
\author[a]{and C. Wotzasek}
\affiliation[a]{Instituto de F\'\i sica, Universidade Federal do Rio de Janeiro,\\21941-972, Rio de Janeiro, Brazil}
\affiliation[b]{Centro Federal de Educa\c{c}\~ao Tecnol\'ogica Celso Suckow da Fonseca,\\28635-000, Nova Friburgo, Brazil}
\affiliation[c]{Departamento de F\'\i sica Te\'orica, Instituto de F\'\i sica, UERJ - Universidade do
Estado do Rio de Janeiro,\\Rua S\~ ao Francisco Xavier 524, 20550-013 Maracan\~ a, Rio de Janeiro,
Brazil}
\emailAdd{leogrigorio@if.ufrj.br}
\emailAdd{msguimaraes@uerj.br}
\emailAdd{romulo@if.ufrj.br}
\emailAdd{clovis@if.ufrj.br}

\abstract{In this work the phenomenon of charge confinement is approached in various contexts. An universal criterion for the identification of this phenomenon in Abelian gauge theories is suggested: the so-called spontaneous breaking of the brane symmetry. This local symmetry has its most common manifestation in the Dirac string ambiguity present in the electromagnetic theory with monopoles. The spontaneous breaking of the brane symmetry means that the Dirac string becomes part of a brane invariant observable which hides the realization of such a symmetry and develops energy content in the confinement regime. The establishment of this regime can be reached through the condensation of topological defects. The effective theory of the confinement regime can be obtained with the Julia-Toulouse prescription which (originally introduced as the dual mechanism to the Abelian Higgs Mechanism) is generalized in this paper in order to become fully compatible with Elitzur's theorem and describe more general condensates which may break Lorentz and discrete spacetime symmetries. This generalized approach for the condensation of defects is presented here through a series of different applications.}
\arxivnumber{1102.3933}
\begin{document}
\maketitle

%\begin{abstract}
%In this work the phenomenon of charge confinement is approached in various contexts. An universal criterion for the identification of this phenomenon in Abelian gauge theories is suggested: the so-called spontaneous breaking of the brane symmetry. This local symmetry has its most common manifestation in the Dirac string ambiguity present in the electromagnetic theory with monopoles. The spontaneous breaking of the brane symmetry means that the Dirac string becomes part of a brane invariant observable which hides the realization of such a symmetry and develops energy content in the confinement regime. The establishment of this regime can be reached through the condensation of topological defects. The effective theory of the confinement regime can be obtained with the Julia-Toulouse prescription which (originally introduced as the dual mechanism to the Abelian Higgs Mechanism) is generalized in this paper in order to become fully compatible with Elitzur's theorem and describe more general condensates which may break Lorentz and discrete spacetime symmetries. This generalized approach for the condensation of defects is presented here through a series of different applications.
%\end{abstract}
%
%%\pacs{Valid PACS appear here}
%% PACS, the Physics and Astronomy Classification Scheme.
%% Valid PACS numbers may be entered using the \verb+\pacs{#1} command.
%
%%\keywords{Topological defects, brane symmetry, confinement.}
%% Use showkeys class option if keyword display desired
%
%%%%%%%%%%%%%%%%%%%%%%%%%%%%%%%%%%%%%%%%%%%%%%%%%%%%%%%%%%%%%%
%
%\maketitle

\tableofcontents

\section{Introduction}
\label{sec:intro}

The main purpose of this work is to discuss the formulation of Abelian gauge theories in the presence of defects. We develop a general procedure to address the condensation of these structures. This procedure is built on the work of Julia and Toulouse \cite{jt} (latter extended by Quevedo and Trugenberger \cite{qt}) and also on the work of Banks, Myerson and Kogut \cite{Banks:1977cc}, which was further developed by Kleinert \cite{mvf}. We here generalize and unify their results and apply them to a variety of systems with the purpose of describing the physical consequences of taking into account the collective behavior of such defects. In this section we provide an overview of the main concepts that will be discussed in this paper.

Defects are represented by singular configurations of the gauge field. They can be viewed as a low energy manifestation of topological structures of the underlying theory.  The most convenient way to operationally define them is through duality. Defects are the dual representation of a classical charge which couples minimally with the gauge field. Since duality modifies the nature of the coupling we say that defects couple non-minimally.

Duality plays a fundamental role in our formulation. In its most basic definition, we say that a physical system has a dual description if it admits an equivalent mathematical representation. ``Equivalent'' here means that every observable of the original formulation of the theory can be mapped into observables of the dual description.

The simplest system that allows such a construction is the electromagnetic theory in vacuum. The dual map in this case amounts to simply exchange the electric field by the magnetic field and vice-versa. This is in fact a special case in which the system is self-dual. Electromagnetic duality survives in the presence of electric and magnetic charges if they also participate in the map. An extremely important property arises in this case by requiring that the system makes sense as a quantum system: electric charge must be quantized in integer multiples of the inverse of the magnetic charge \cite{dirac}. This suggests that, in a quantum field theory formulation of this system, a regime of strong electric coupling will be mapped by duality in a magnetic formulation of the theory at weak coupling. This is very difficult to prove in general but there are strong hints of this behavior coming from supersymmetric models \cite{Osborn:1979tq, Vafa:1994tf, Seiberg:1994rs}.

Electric and magnetic charges manifest themselves in different ways in a given regime of the theory. In the weak electric coupling regime, the electric charges can be identified with the excitations of a quantum field and constitute the appropriate degrees of freedom. Magnetic charges, on the other hand, manifest themselves as nonpertubative structures and the knowledge about the quantum formulation of such structures is very limited.

A productive way to deal with charges and defects is to look how the gauge field responds to their presence. We can consider the dynamical reasons responsible for the establishment of a given configuration of these objects as external information. The electric or magnetic character manifests in the way these objects couple to the fields of the theory. An electric charge couples minimally with the electromagnetic field defining the holonomy of this field along the charge's trajectory. A magnetic charge induces singularities in the gauge field whose characteristic manifestation is the presence of nontrivial fluxes in the theory due to the violation of Bianchi's identity. Mathematically, these structures will be represented in this work by introducing $p$-currents in the theory. This allows the development of a formalism capable of describing the consequences of the process of condensation of charges and defects in the system (the dynamical reasons responsible for such a condensation, however, are beyond the scope of this formalism and constitute a separate issue). A version of this formalism was put forward originally by Julia and Toulouse \cite{jt} and later applied to general theories involving $p$-forms by Quevedo and Trugenberger \cite{qt} , who formulated it as the dual of the Abelian Higgs Mechanism. It is called the \emph{Julia-Toulouse Approach (JTA) for defects condensation}.

The main objective of this work is to generalize this formalism. Such generalization, which we shall continue to call simply as the JTA, involves two fundamental aspects: it fully takes into account the fact that local symmetries can never be broken, according to Elitzur's theorem \cite{elitzur}, and it deals with general sorts of condensates of defects. Along the way we expect to reveal the important universal aspects of the formalism and it will become clear that it is best understood as a very general prescription to find the effective theory describing the system in a state in which defects are condensed.

The phenomenon of condensation of defects is here understood in a generalized sense. It corresponds to the proliferation of these objects in the system until their collective behavior becomes describable by a field theory. This makes it possible to incorporate in the formalism a variety of systems sharing the very same fundamental aspects. The present work is comprised of many examples of sort. In fact, the JTA will be presented here through the detailed discussion of such examples. The generalization has as one of its main ingredients a formal description of the system embedded in an ensemble of defects, which extends the treatment developed in \cite{Banks:1977cc}. This description allows us to develop a more accurate definition of the Julia-Toulouse prescription by providing an effective parametrization of the process of condensation and/or dilution of topological defects in the system (the Generalized Poisson's Identity (GPI) introduced in \cite{dafdc}, which establishes an order-disorder map, plays an important role here). This has lead us to the natural interpretation of the JTA as a generator of effective field theories.

Since the prescription falls within the general principles of effective theories \cite{efftheory}, symmetries play a pivotal role in the whole formulation. When we define a theory describing gauge fields and defects, there is an extra redundancy in the formulation besides the usual gauge redundancy itself. This extra redundancy has its most famous manifestation  in the theory of the magnetic monopole as the ambiguity associated with the Dirac string configuration \cite{dirac} and is commonly regarded as a special kind of gauge symmetry. In a lattice formulation this redundancy manifests itself as large gauge transformations due to the periodic character of the gauge field, which is an angular variable in this environment. Kleinert \cite{Kleinert:1992eb, mvf} made the important observation that this redundancy is much better treated if considered as an independent ambiguity of the system, one that only exists due to the presence of defects. This is not just a matter of semantics and, as we shall see, it is very productive to understand things this way. Since we will encounter this extra local symmetry in many examples, always associated with the redundant configuration of localized structures (hypersurfaces which we shall generically call as branes) in spacetime, we found useful to devise a name for it: we call it \emph{brane symmetry}.

Being a redundancy (a local symmetry) of the theory, like the gauge symmetry, the brane symmetry can have different realizations. In particular, it can be ``spontaneously broken", which means that its realization can be hidden into brane invariants in the so-called ``broken regime". As it will be discussed in the coming sections, the condensation of defects may induce a phase of the system displaying ``brane symmetry breaking". This will be identified with the main signature of confinement in all models exhibiting this phenomenon discussed in this work. The main property of this ``breaking" is the fact that the unphysical brane becomes part of a brane invariant observable which develops energy content. The observability of the brane in a confining regime is not new in the literature \cite{Pisarski:1986gr, Diamantini:1993iu, mvf, oxman}. In this work we offer a unifying and precise description that allows for a more systematic treatment of this phenomenon. In particular, the careful treatment of the brane symmetry given here, with the use of the concept of ``brane invariant'' maintaining the symmetry in any regime we consider (being its realization explicit or hidden), is the point that makes the JTA fully compatible with Elitzur's theorem \cite{jt-cho}.

We may thus summarize the two main contributions presented in this work:
\begin{itemize}
\item The realization that the spontaneous breaking of the brane symmetry is a signature of confinement in theories involving condensation of topological defects;

\item The development of a ensemble formulation of Abelian theories in the presence of topological defects, which provides an effective parametrization of the condensation-dilution process through the order-disorder map (or flux-current map) resulting in a consistent and precise implementation of the original Julia-Toulouse prescription.
\end{itemize}
After reviewing in section \ref{sec:dual} the concepts of duality, $p$-currents, Poincar\`e's Duality and brane symmetry and its connection with charge quantization and the Poisson identity (generalized to deal with $p$-currents \cite{dafdc}) , we develop the two main points summarized above by carefully working out a variety of applications:
\begin{itemize}
\item In section \ref{sec:1} we present the Julia-Toulouse approach in its generalized form using the relativistic superconductor, as described by the Abelian Higgs model. This is a well known example of a confining theory and, via the ensemble formulation, we can construct the concept of spontaneous breaking of the brane symmetry by calculating in a detailed way the confining potential. The topological defects represented by $p$-currents in the effective theory can be traced back to the vortex solutions of their ultraviolet completion, the Abelian Higgs model, by its nontrivial fluxes;

\item In section \ref{sec:2} we study another well known example of a confining theory: compact QED in $3D$ as introduced by Polyakov \cite{Polyakov:1975rs,Polyakov:1996nc}. We show that this system can be naturally described within the Julia-Toulouse formalism. In this example, the sum over brane configurations in the ensemble formulation can be explicitly computed yielding the well known Sine-Gordon model as an effective theory for the instanton gas. These branes are naturally identified with the infrared manifestation of the topological solutions of the $SO(3)$ Georgi-Glashow model. As a confining theory, it displays the spontaneous breaking of the brane symmetry and mass generation, characterized by the rank jump phenomenon (the rank jump phenomenon is a signature of the JTA and will be discussed in section III, but we can briefly say that a $(p + 1)$-form replaces the original $p$-form when condensation occurs);

\item In section \ref{sec:3} we review a previous result derived by some of us \cite{Gamboa:2008ne}, this time within the ensemble formulation at the level of the partition function, which shows that quantum fermionic fluctuations can be conveniently interpreted as a condensate. The JTA provides a dual picture for the radiative corrections responsible for the induction of the Chern-Simons term in the low energy effective action of quantum electrodynamics in $3D$ (QED$_3$). The Maxwell-Chern-Simons (MCS) theory in $(2+1)D$ is then interpreted as an electric condensate that breaks the $P$ and $T$ symmetries;

\item In section \ref{sec:4} we take a detour to illustrate the generality of the formalism by studying a system which exhibits no confinement: the fractional quantum Hall system. The hierarchy structure \cite{Haldane:1983xm,Wen:1995qn} obeyed by the states of this system is based on a sequence of condensations which are shown to be very well captured by the Julia-Toulouse approach. Since there is no confinement, we show that there is no brane symmetry breaking in this case;

\item In section \ref{sec:5} we study the Maxwell-Chern-Simons theory in $(2+1)D$ in the presence of external instantons \cite{Grigorio:2008gd}. This is a controversial issue and we use the effective representation of the system as a condensate and the ensemble formulation that parametrizes the condensation of defects to offer a proper interpretation of this system. The confinement of external instantons in this system \cite{Diamantini:1993iu,Pisarski:1986gr} is obtained as a consequence of the spontaneous breaking of the brane symmetry and the mass quantization is also discussed;

\item In section \ref{sec:6}, the JTA is used to discuss the inclusion of monopoles in the Carroll-Field-Jackiw model \cite{Carroll:1989vb}, which constitutes an emblematic theory that violates Lorentz symmetry and is expected to play a role as an effective description of the chiral magnetic effect in hot QCD matter \cite{Kharzeev:2009fn}.
\end{itemize}
We close the paper in section \ref{sec:conc} with an overview of the main points discussed throughout this work.

For the most part of this paper we shall work with differential forms. For a detailed presentation of the subject suitable for physicists, see for example \cite{nakahara, nash}.

In this paper we shall use $\hbar=c=1$.

\section{Duality, $p$-currents, brane symmetry and charge quantization}
\label{sec:dual}

We begin this section reviewing the formulation of electromagnetism in vacuum. We consider it as defined on a four dimensional manifold ${\cal M}_4$. We shall work with Minkowski spacetime $\mathrm{I\!R}^{1,3}$ with metric signature $(-, +, +, +)$, but the results here discussed hold in more general backgrounds if the spacetime under consideration does not have a boundary or if the fields vanish on the boundary. The extension of the considerations made here to an arbitrary number of dimensions is straightforward and will be given later in this section.

The Maxwell equations defining the spacetime evolution of the electromagnetic fields in the vacuum are:
\begin{align}
\label{c1s1e01}
dF &= 0,\\
\label{c1s1e02}
d{ }^\ast F &= 0,
\end{align}
where ${}^\ast$ denotes the dual Hodge star operator and $F$ is the 2-form electromagnetic field strength tensor,
\begin{align}
\label{c1s1e03}
F = \frac 12 F_{\mu\nu}dx^{\mu}\wedge dx^{\nu}.
\end{align}

Due to the Poincar\`e's Lemma, the equation $dF=0$ allows us to locally define a 1-form electromagnetic potential $A$ such that,
\begin{align}
\label{c1s1e04}
F=dA,
\end{align}
which leads us to recognize the equation (\ref{c1s1e01}) as the \textit{Bianchi's identity}. The definition of the 1-form $A$ introduces an ambiguity into the theory since the equations of motion are invariant under the transformation:
\begin{align}
\label{c1s1eB}
A \rightarrow A + d\lambda,
\end{align}
where $\lambda$ is a 0-form. The invariance of the theory under the transformation (\ref{c1s1eB}) is called \textit{gauge symmetry}, although this invariance does not constitute a physical symmetry, being actually a redundancy in the variables describing the system.

Equation (\ref{c1s1e02}) describes the dynamics of the electromagnetic fields in vacuum and is obtained as the stationary point of the Maxwell action with respect to the variations of the gauge potencial $A$:
\begin{align}
\label{c1s1e05}
S_{EM}=\int_{{\cal M}_4} -\frac{1}{2e^2} dA \wedge {}^\ast dA \equiv -\frac{1}{2e^2} (dA, dA).
\end{align}

At the level of the action we can obtain a dual formulation to (\ref{c1s1e05}) as follows: notice that an action physically equivalent to (\ref{c1s1e05}) is given by:
\begin{align}
\label{c1s1e06}
S_{M}=\int_{{\cal M}_4}  G \wedge {}^\ast dA + \frac{e^2}{2} G \wedge {}^\ast G,
\end{align}
where $G$ is an auxiliary non-dynamical 2-form. Hence, the so-called \textit{master action} $S_{M}$ constitutes a trivial enlargement of the configuration space of the original theory. The auxiliary character of $G$ means that its equation of motion suffices to specify it as a function only of $A$:
\begin{align}
\label{c1s1e07}
\delta_{G} S_{M}= (\delta G, dA + e^2 G) = 0 \Rightarrow G = -\frac{1}{e^2}dA,
\end{align}
where $\delta_G$ denotes the variation of the action $S_M$ with respect to the variation $\delta G$ of $G$. This equation means that all the information about the dynamics is contained in $A$ and thus we can simply substitute this value of $G$ into the action, which leads us back to $S_{EM}$, given by eq. (\ref{c1s1e05}).

On the other hand, we can integrate by parts  the first term of the master action:
\begin{align}
\label{c1s1e08}
S_{M}= (G, dA) + \frac{e^2}{2} (G, G) = (d^{\dagger} G, A) + \frac{e^2}{2} (G, G),
\end{align}
where $d^{\dagger}$ is the exterior coderivative, the adjoint of the exterior derivative ($d^{\dagger} = {}^\ast d {}^\ast$ in $\mathrm{I\!R}^{1,3}$). The gauge potential $A$ becomes a Lagrange multiplier imposing the constraint:
\begin{align}
\label{c1s1e09}
d^{\dagger} G = 0 \Rightarrow d {}^\ast G = 0,
\end{align}
which is solved noticing that locally, using Poincar\`e's Lemma,
\begin{align}
\label{c1s1e10}
{}^\ast G = dB,
\end{align}
where $B$ is a 1-form. In $\mathrm{I\!R}^{1,3}$ we have then,
\begin{align}
\label{c1s1e11}
S_{M} \rightarrow \frac{e^2}{2} (G, G) = - \frac{e^2}{2} ( {}^\ast G, {}^\ast G) = - \frac{e^2}{2} ( dB, dB) = \tilde{S}_{EM},
\end{align}
which is the well known result revealing  the self-duality of Maxwell's theory in vacuum. Notice that the coupling constant is inverted due to the duality. This is a remarkable property of duality. Although trivially realized here, this property becomes highly non-trivial in an interacting theory. One example is the 2-dimensional Ising model where, as showed by Kramers and Wannier \cite{Kramers:1941kn}, there is a dual map connecting the physics at high tempetatures with the physics at low temperatures. The fixed point of this map corresponds to the critical temperature of the phase transition. Montonen and Olive \cite{Montonen:1977sn} proposed that this electromagnetic self-duality could be extended to non-Abelian theories. This conjecture, however, is only feasible in supersymmetric theories: ${\cal N} = 4$ \cite{Osborn:1979tq, Vafa:1994tf} and also ${\cal N} = 2$ \cite{Seiberg:1994rs} super Yang-Mills. In this context, this duality is called $S$-duality, inverting the regimes of weak and strong coupling (see \cite{Harvey:1996ur, AlvarezGaume:1996mv, DiVecchia:1998ky} for reviews on the subject). More recently, these concepts were incorpored in the study of the so-called Langlands program in number theory \cite{Kapustin:2006pk}. There are also many similarities with $T$-duality in string theories, which relates theories with compactification radius $R$ with theories with compactification radius $\frac{\ell^2_s}{R}$, where $\ell_s$ is the string length scale (see \cite{Alvarez:1994dn} for a review). In general, the inversion of parameters is a defining characteristic of the duality process we are interested in.

The very same reasoning used above can be applied to show, for example, the duality between the Proca theory and the massive Kalb-Ramond model in $\mathrm{I\!R}^{1,3}$ or the duality between the self-dual action and the Maxwell-Chern-Simons theory in $\mathrm{I\!R}^{3}$.

These considerations about duality are naturally extended to general $p$-forms in arbitrary dimensions $D$. In fact, it is easy to show that for massless $p$-forms described by Maxwell-like theories, the dual pair $A_{p}$ and $\tilde{A}_{q}$ obeys the relation $p+q+2=D$, where $p$ e $q$ are the ranks of the forms. For massive $p$-forms the relation is $p+q+1=D$. Detailed investigations about duality transformations and the corresponding structure of the duality groups were worked out in \cite{Deser:1997mz,Wotzasek:1998rj,Noronha:2003vp}.

In order to have a non-trivial electromagnetic theory the presence of sources is required. We now recall how localized sources (also called \textit{classical currents}) can be introduced into the theory.

Electric sources are naturally incorpored into  equation (\ref{c1s1e02}):
\begin{align}
\label{c1s2e01}
d{}^\ast F = e{}^\ast J_e,
\end{align}
where $J_e = J_{e\mu}dx^{\mu}$ is the electric current 1-form. This implies the following modification in the action (\ref{c1s1e05}):
\begin{align}
\label{c1s2e02}
S_{EM}=\int_{{\cal M}_4} -\frac 12 dA \wedge {}^\ast dA + eA\wedge {}^\ast J_e.
\end{align}

The introduction of magnetic charges, however, is a non-trivial task since the presence of magnetic monopoles in the electromagnetic theory violates Bianchi's identity and this affects the definition of the gauge potential $A$, which is a fundamental quantity in a quantum formulation of the theory. There are at least two equivalent ways of introducing magnetic charges. One possibility, related to the ideas of Wu and Yang \cite{Wu:1975es}, consists in recognizing that the space where the theory is defined is non-trivial in the presence of monopoles. The insertion of monopoles is characterized by the presence of ``holes" in the space as seen by the electromagnetic field. The manifold is no longer topologically equivalent to $\mathrm{I\!R}^{1,3}$ (or $\mathrm{I\!R}^{4}$ in the Euclidean case) and there are surfaces that do not contract to a point, introducing non-trivial homology and homotopy groups associated with the manifold. Specifically, in the vicinity of a monopole the magnetic flux through the 2-sphere $S^2$ is non-trivial and this, together with the fact that $dF=0$, results in a non-trivial cohomology class defined by $F$ which is recognized as the \textit{first Chern class} of the $U(1)$ complex vector bundle. This means that it is not possible to write $F=dA$ with $A$ globally defined. The procedure is then to define different potential $A$ in different regions of the manifold. Since $F$ is well defined in the whole spacetime, the potentials are related by a gauge transformation in the intersections of the regions where they are defined. A very rich topological structure emerges from these observations although we also see that we end with an operational asymmetry between the definition of electric and magnetic charges.

Another equivalent possibility was originally introduced by Dirac \cite{dirac} and it is this approach that we shall adopt in this paper. It consists in studying the consequences of an explicit violation of Bianchi's identity,
\begin{align}
\label{c1s2e03}
dF = g {}^\ast J_g,
\end{align}
where $J_g = J_{g\mu}dx^{\mu}$ is the magnetic current 1-form. It is still possible to introduce a gauge potential $A$  if we use the concept of $p$-currents, which we discuss in the sequel.

A localized particle in a spatial manifold $\mathcal{M}_3$ is a geometric point in this space and it is called a 0-brane. Its world-line in a spacetime manifold $\mathcal{M}_4$ is a 1-brane. Similarly, a string is a $1$-brane in $\mathcal{M}_3$ which traces out a surface, a $2$-brane, in its evolution in spacetime $\mathcal{M}_4$. These concepts are naturally generalized to other objects of higher dimensionality called $p$-branes. A $p$-brane in a spatial manifold $\mathcal{M}_D$, creates a $(p+1)$-brane through its evolution in a spacetime manifold $\mathcal{M}_{D+1}$, which is its world (hyper)surface. This geometric view can be formalized introducing the concept of \emph{Poincar\`e's Duality} (discussions about these concepts can be seen in \cite{Bekaert:2002cz}).

If ${\cal N}_p$ is a $p$-surface ($dim\;{\cal N}_p=p$) contained in ${\cal M}_D$ ($dim\;{\cal M}_D=D$) and defined by the equations $X^{\mu} = X^{\mu}(y^{a})$, where $y^{a}$, $a=0,1,\ldots,p-1$, parametrize the surface ${\cal N}_p$, then the $p$-form:
\begin{align}
\label{c1s2e04}
J_p = J_{p\;\mu_1,\ldots,\mu_p} dx^{\mu_1} \wedge \ldots \wedge dx^{\mu_p},
\end{align}
with:
\begin{align}
\label{c1s2e05}
J^{\mu_1,\ldots,\mu_p}_p (x) = \int_{{\cal N}_p} \delta^{D}(x-X(y^{a})) \frac{\partial X^{\mu_1}}{\partial y^{a_1}}\ldots \frac{\partial X^{\mu_p}}{\partial y^{a_p}}  dy^{a_1} \wedge \ldots \wedge dy^{a_p},
\end{align}
defines the $\emph{p}$-\emph{current} Poincar\`e -dual to the surface ${\cal N}_p$, a fact we denote as:
\begin{align}
\label{c1s2e06}
{\bf P}({\cal N}_p) = {}^\ast J_p.
\end{align}
Notice that a $p$-current is a $p$-form whose components are distributions. This definition constitutes an isomorphism between the $p$-branes and the corresponding $p$-currents. Thus, in this work we sometimes refer to the $p$-currents as $p$-branes and vice-versa. Some properties follow:
\begin{enumerate}
 \item For any $p$-form $A_p$:
 \begin{align}
\label{c1s2e07}
\int_{{\cal M}_D} A_p \wedge {}^\ast J_p =  \int_{{\cal N}_p} A_p.
\end{align}
This property is often taken as the definition of $p$-currents.

 \item If $\partial$ is the boundary operator such that $\partial {\cal A}_p = {\cal B}_{p-1}$ denotes the $(p-1)$-surface which is the boundary of ${\cal A}_p$, then for any $p$-surface ${\cal N}_p$:
  \begin{align}
\label{c1s2e08}
{\bf P}({\partial\cal N}_p) = (-1)^p d({\bf P}({\cal N}_p)).
\end{align}
This property is easily verified. For an arbitrary $(p-1)$-form $A_{p-1}$ we have:
\begin{align}
\label{c1s2eA}
\int_{{\cal M}_D} A_{p-1} \wedge {\bf P}({\partial\cal N}_p) &=  \int_{{\partial\cal N}_p} A_{p-1} = \int_{{\cal N}_p} d A_{p-1}\nonumber\\
&= \int_{{\cal M}_D} d A_{p-1} \wedge {\bf P}({\cal N}_p)\nonumber\\
&= (-1)^p \int_{{\cal M}_D} A_{p-1} \wedge d{\bf P}({\cal N}_p),
\end{align}
where we have used Stokes theorem in the first line.

\item \textit{Generalized Poisson's Identity (GPI)}. If $A_p$ is an arbitrary $p$-form then:
\begin{align}
\label{c1s2eB}
\sum_{\{{\cal N}_p\}} \delta({}^\ast A_p - {\bf P}({\cal N}_p)) = \sum_{\{{\cal Q}_{(D-p)}\}} e^{2\pi i\int_{{\cal M}_D} {\bf P}({\cal Q}_{(D-p)}) \wedge {}^\ast A_p},
\end{align}
where the formal sum is taken over all configurations of the indicated branes. The GPI (\ref{c1s2eB}) is a generalization of the Poisson's identity \cite{Banks:1977cc,mvf} for an arbitrary number of dimensions and for general $p$-forms. It is discussed in details in Appendix A of \cite{dafdc}. We say that the surfaces $\mathcal{N}_p$ and $\mathcal{Q}_{(D-p)}$ (or their respective Poincar\`e -dual currents) are Poisson-dual to each other.

 \item If two surfaces contained in ${\cal M}_D$ have complementary dimensions (\emph{i.e.}, if the sum of their dimensions is equal to $dim\;{\cal M}_D=D$) then, in general, they will intercept in a determined number of points and it is possible to assign an integer to these intersections which is called \textit{intersection number}, defined by:
  \begin{align}
\label{c1s2e09}
I ({\cal A}_p,  {\cal B}_{D-p}) = \int_{{\cal M}_D} {\bf P}({\cal A}_p) \wedge {\bf P}({\cal B}_{D-p}) = n \in \mathds{Z}.
\end{align}
      The associated sign to the number is a consequence of the orientation defined by the spacetime ${\cal M}_D$.
 \end{enumerate}

Related to this last property there is the definition of \textit{linking number}: if ${\cal A}_p$ and ${\cal B}_{D-p-1}$ are surfaces contained in ${\cal M}_D$ such that $dim\;{\cal A}_p + dim\;{\cal B}_{D-p-1} + 1 = dim\;{\cal M}_D$ and if $\partial {\cal C}_{p+1} = {\cal A}_p$, then the linking number between ${\cal A}_p$ and ${\cal B}_{D-p-1}$ is defined by:
\begin{align}
\label{c1s2e10}
L({\cal A}_p,  {\cal B}_{D-p-1}) = I ({\cal C}_{p+1},  {\cal B}_{D-p-1}).
\end{align}

After this brief \emph{intermezzo} we can look back at the equations (\ref{c1s2e06}) and (\ref{c1s2e08}) and notice that they consolidate the geometric idea we have about current conservation: we say that a current is conserved when its world surface does not have a boundary, that is,
\begin{align}
\label{c1s2e11}
 d({\bf P}({\cal N}_p)) = d {}^\ast J_p = 0.
\end{align}

In the electromagnetic case, the electric (\ref{c1s2e01}) and the magnetic (\ref{c1s2e03}) currents are conserved since $d^2 =0$. These observations allow us to introduce again the potential $A$, since generalizing the Poincar\`e's Lemma for $p$-currents we can write:
\begin{align}
\label{c1s2e12}
d {}^\ast J_p = 0 \Rightarrow {}^\ast J_p = d {}^\ast \Sigma_{p+1},
\end{align}
where $\Sigma_{p+1}$ is a $(p+1)$-current. The $(p+1)$-brane Poincar\`e -dual to $\Sigma_{p+1}$ has as its boundary the brane Poincar\`e -dual to $J_p$. Hence, (\ref{c1s2e03}) leads us to:
\begin{align}
\label{c1s2e13}
dF = g{}^\ast J_g = g d {}^\ast \Sigma_g  \Rightarrow d(F - g{}^\ast \Sigma_g) = 0 \Rightarrow F= dA + g{}^\ast \Sigma_g.
\end{align}
The surface Poincar\`e -dual to $\Sigma_g$ has as its boundary the surface Poincar\`e -dual to the magnetic current and, thus, it is recognized as the surface traced by the Dirac string in spacetime \cite{dirac}. We shall call these branes or their corresponding Poincar\`e -dual currents as \textit{Dirac branes}.

Notice that $F$, being an observable field, must be well defined. Hence, $A$ must be singular where $\Sigma_g$ is nonzero in order to keep $F$ regular.

The definition of the Dirac brane introduces another ambiguity into the theory beyond the gauge ambiguity. We see that the observable $J_g$ remains invariant under the transformation:
\begin{align}
\label{c1s2e14}
{}^\ast \Sigma_g  \rightarrow {}^\ast \Sigma_g  + d{}^\ast \Lambda_g,
\end{align}
where $\Lambda_g$ is a 3-current. In order to keep the consistency of the theory with $F$ being an observable, we see that under (\ref{c1s2e14}), $A$ transforms as:
\begin{align}
\label{c1s2e15}
A \rightarrow  A - g{}^\ast \Lambda_g.
\end{align}
We call this kind of transformation as \textit{brane transformation}. The importance of this symmetry and its independence regarding the gauge symmetry was emphasized by Kleinert in \cite{Kleinert:1992eb, mvf}.

The action (\ref{c1s2e02}) must be generalized in order to obtain a brane invariant action. We define:
\begin{align}
\label{c1s2e16}
S_{EM}=\int_{{\cal M}_4} -\frac 12 (dA + g{}^\ast \Sigma_g) \wedge {}^\ast (dA + g{}^\ast \Sigma_g) + eA\wedge {}^\ast J_e,
\end{align}
where $\Sigma_g$ is the 2-current such that ${}^\ast J_g = d{}^\ast\Sigma_g$ is the magnetic current. Indeed, extremizing the action with respect to variations of $A$ we obtain the equation of motion:
\begin{align}
\label{c1s2e17}
d{}^\ast F = e{}^\ast J_e,
\end{align}
where now $F = dA + g{}^\ast \Sigma_g$ is the physical electromagnetic field. From this definition of $F$ it also follows the identity:
\begin{align}
\label{c1s2e18}
d F = g{}^\ast J_g.
\end{align}
The action does not seem to be invariant under brane transformations due to the last term. Indeed, the variation $\delta_B$ of the action due to the transformation (\ref{c1s2e14}) and (\ref{c1s2e15}) is:
\begin{align}
\label{c1s2e19}
\delta_B S_{EM} &= - eg \int_{{\cal M}_4}   {}^\ast \Lambda_g \wedge {}^\ast J_e = - eg \int_{{\cal M}_4}   {\bf P}({\cal B}^3_g) \wedge {\bf P}({\cal A}^1_e)\nonumber\\
&= - eg I({\cal B}^3_g, {\cal A}^1_e) = -egn,
\end{align}
where $n\in \mathds{Z}$ and we used (\ref{c1s2e09}). Here ${\cal A}^1_e$ is the world line traced by the electric charge and ${\cal B}^3_g$ is the surface Poincar\`e -dual to the 3-current $\Lambda_g$ representing the volume spanned by the deformation of the world surface of the Dirac string. The variation of the action is a constant and it is innocuous from the point of view of classical mechanics. However, from the point of view of quantum mechanics the variation (\ref{c1s2e19}) could be observed in interferometry experiments and the theory would be inconsistent, since these transformations only represent a redundancy in the definition of the variables describing the system. In a path integral formulation of the quantum theory, the action appears as a phase and thus, the theory will be invariant under brane transformations and hence consistent, if:
\begin{align}
\label{c1s2e20}
eg = 2\pi m; \;\;\;\; m \in \mathds{Z},
\end{align}
which is the well known Dirac quantization condition \cite{dirac}. Thus, we conclude that the brane symmetry implies electric charge quantization as integer multiples of the inverse of the magnetic charge.

Notice also that the magnetic brane symmetry is realized in (\ref{c1s2e16}) in a Stuckelberg-like structure \cite{Stueckelberg:1900zz} (for a recent discussion see \cite{Ruegg:2003ps}). In the next section we shall see that this structure is at the origin of the rank jump phenomenon associated to the mass generation in the system as a consequence of the monopole condensation. A consequence of this process is the spontaneous breaking of the brane symmetry and the consequent confinement of electric charges in a new condensed phase.

Electromagnetism in the presence of sources also admits a reformulation in terms of dual variables. We follow the same steps as before and define a master action physically equivalent to (\ref{c1s2e16}) introducing an auxiliary field $G$,
\begin{align}
\label{c1s2e21}
S_{M}=\int_{{\cal M}_4} G \wedge  {}^\ast (dA + g{}^\ast \Sigma_g) + \frac 12 G \wedge {}^\ast G + eA\wedge {}^\ast J_e.
\end{align}
The equations of motion of $G$ constrain it to be equal to $F = dA + g{}^\ast \Sigma_g$ and substituting it into (\ref{c1s2e21}) we reobtain (\ref{c1s2e16}). On the other hand, integrating by parts the first term we see that $A$ becomes a Lagrange multiplier producing the constraint:
\begin{align}
\label{c1s2e22}
d^{\dagger} G = e J_e \Rightarrow  d {}^\ast G = -e {}^\ast J_e.
\end{align}
It follows that:
\begin{align}
\label{c1s2e23}
d({}^\ast G + e{}^\ast \Sigma_e) = 0 \Rightarrow {}^\ast G = - dB - e{}^\ast \Sigma_e,
\end{align}
where $B$ is a 1-form introduced due to the Poincar\`e's Lemma. The dual action is:
\begin{align}
\label{c1s2e24}
S_{M} &\rightarrow \frac 12 (G ,G) + g (G, {}^\ast \Sigma_g) = -\frac 12 ({}^\ast G ,{}^\ast G) + g ({}^\ast G, \Sigma_g) \nonumber\\
&= -\frac 12 ({}^\ast G ,{}^\ast G) - g(dB, \Sigma_g) - eg ({}^\ast \Sigma_e, \Sigma_g)\nonumber\\
&= \int_{{\cal M}_4} -\frac 12 (dB + e{}^\ast \Sigma_e) \wedge {}^\ast (dB + e{}^\ast \Sigma_e) + gB\wedge {}^\ast J_g,
\end{align}
where we have discarded the last term in the penultimate line since, due to the Dirac quantization condition, this term is an integer multiple of $2\pi n$. Note that in the dual formulation the couplings have their roles exchanged: the magnetic charges appear minimally coupled to the dual gauge potential $B$ while the electric charges appear non-minimally coupled.

The quantum formulation of the theory is formally defined through the partition function, which has the following general form:
\begin{align}
\label{c1s2e25}
{\cal Z} = \sum_{\{{\cal B}_e\}} \sum_{\{{\cal B}_g\}} \int {\cal D}A e^{i\left[S_{EM}(A, \Sigma_e, \Sigma_g) + S_e(J_e) + S_g(J_g) \right]},
\end{align}
where $S_{EM}(A, \Sigma_e, \Sigma_g)$ is given by (\ref{c1s2e16}) or its dual (\ref{c1s2e24}) and $S_e(J_e)$ and $S_g(J_g)$ are effective actions for the electric and magnetic branes, which must depend only on the brane invariant quantities $J_e$ and $J_g$. The formal sums include all the the surfaces Poincar\`e -dual to the currents $\Sigma_e$ and $\Sigma_g$ (${\bf P}({\cal B}_e) = {}^\ast \Sigma_e$ and ${\bf P}({\cal B}_g) = {}^\ast \Sigma_g$) and thus, just like the functional integral over $A$, they need some fixing to eliminate redundancies: the physically relevant configurations correspond to the gauge and brane invariants.

An important gauge invariant nonlocal operator is the Wilson loop \cite{Wilson:1974sk}:
\begin{align}
\label{c1s2e29}
W(C) = e^{i\lambda\int_C A},
\end{align}
where $\lambda$ is a parameter and $C$ is a loop (topologically equivalent to the circle $S^1$). The surface $C$ is Poincar\`e -dual to a conserved 1-current and thus,
\begin{align}
\label{c1s2e30}
W(C) = W(J) = e^{i\int_{{\cal M}_4} A \wedge {}^\ast J}.
\end{align}
Hence, the insertion of this operator corresponds to the introduction of a minimal coupling in the action and we know that this kind of coupling manifests itself as a non-minimal coupling in the dual formulation. The non-minimal coupling is a manifestation of the presence of defects in spacetime as seem by the dual field to the gauge potential $A$. Thus, we have the curious result that the VEV of a Wilson loop is represented in the dual picture by the insertion of defects (with associated Dirac branes) into the theory (for a discussion on this subject see Witten's notes in \cite{Deligne:1999qp} and also \cite{Kapustin:2005py, Kapustin:2006pk}). We shall see an example of this result in the next section.

\section{Application I - Abelian Higgs Mechanism as a condensation phenomenon}
\label{sec:1}

We begin this section with a brief review of some characteristic features of superconductivity, a phenomenon that plays a fundamental role in the phenomenological understanding of confinement \cite{conf}.

As it happens with any thermodynamical system, the superconductivity phenomenon can be effectively characterized by a certain number of macroscopical variables. Beyond the most common variables like temperature, the superconductor is also characterized by a complex scalar field. Just like temperature has its origin in the microscopical properties of matter, the scalar field has also a microscopical origin in the well established BCS theory proposed by Bardeen, Cooper and Schrieffer \cite{Cooper:1956zz} and corresponds to an effective representation of the Cooper pairs.

The superconducting state is established when the following conditions become energetically favorable:
\begin{align}
\label{c2s1e1a}
 |\phi|&=v,\\
 \label{c2s1e1b}
 d_A \phi &\equiv d\phi -ieA \phi =0,
\end{align}
where $\phi$ is the complex scalar field describing the Cooper pair condensate, $v$ is a positive real constant determined by the properties of the superconducting material and its temperature, $e$ is the Cooper pair charge (twice the electron charge) and $A$ is the electromagnetic gauge potential. Equation (\ref{c2s1e1a}) means that the superconducting state is a coherent state of Cooper pairs, an electric condensate.

The most characteristic phenomenon associated with a superconductivity is the \textit{Meissner effect}. This effect follows as an immediate consequence of the fact that in a superconductor, $\phi \neq 0$ and $d_A \phi =0$ and, hence,
\begin{align}
\label{c2s1e2}
  d^2_A \phi = -ie\phi dA - ie d\phi \wedge A -ieA \wedge d\phi = -ie\phi F = 0,
\end{align}
where $F=dA$. Thus, $F$ is zero since $\phi\neq 0$. In a perfect conductor only the electric field is zero while in a superconductor the total electromagnetic field vanishes. This effect is a consequence of an energy balance and, if the electromagnetic field is intense enough, it can be violated in regions where vortices are formed \cite{Abrikosov:1956sx} (for a review, see chapter 5 of \cite{mvf}) endowing the theory with non-trivial fluxes.
\begin{align}
\label{c2s1e6}
\int_{S^1} A = \frac{2\pi}{e}n; \;\;\;\; n\in \mathds{Z}.
\end{align}
In the infrared (IR) limit these classical vortex solutions appear as objects without structure corresponding to singularities in spacetime and are represented by the $p$-currents we discussed in the previous section.

We are going to write down now an effective theory for the IR limit of the relativistic generalization of the phenomenological model of superconductivity proposed by Ginzburg and Landau \cite{Ginzburg:1950sr}. This relativistic generalization corresponds to the Abelian Higgs Model, whose action defined in $\mathrm{I\!R}^{1,3}$ is given by:
\begin{align}
\label{c2s1e11}
  S_{Higgs}= \int_{{\cal M}_4} \left( -\frac 12 dA \wedge {}^\ast dA - \frac 12 d_A \phi \wedge {}^\ast \overline{d_A \phi} - \frac{\lambda}{4} (\phi\overline{\phi} - v^2)\wedge {}^\ast (\phi\overline{\phi} - v^2) \right).
\end{align}

We want to analyze the low energy limit of this model. This limit corresponds to consider an energy scale where the fluctuations of the modulus of the scalar field $\phi$ are frozen, \emph{i.e.}, there are only fluctuations in the phase of the field. This corresponds to study the system in the so-called \textit{London limit} defined by the condition $\lambda \rightarrow \infty$. Hence, $|\phi| = v$ is a constant and the covariant derivative reads:
\begin{align}
\label{c2s1e12}
  d_A \phi = ive^{i\theta} (d\eta - eA + 2\pi n {}^\ast \Lambda),
\end{align}
where $n \in \mathds{Z}$, $\eta$ is the regular part of the angle $\theta$, while the 3-current $\Lambda$ codifies the fact that $\theta$, being the inverse of a periodic function, is multivalued \cite{mvf}. ${}^\ast \Lambda$ represents the vortex contribution in the system and the non-trivial fluxes are given by
\begin{align}
\label{c2s1e13}
  \int_{S^1} {}^\ast \Lambda = 1.
\end{align}
It is important to notice that ${}^\ast \Lambda$ is only associated to closed flux lines; indeed, the magnetic flux described by ${}^\ast \Lambda$ is proportional to $d{}^\ast \Lambda$ and hence the flux lines are closed, since $d^2=0$.

The IR effective action has the form:
\begin{align}
\label{c2s1e14}
S_{eff}= \int_{{\cal M}_4} \left( -\frac 12 dA \wedge {}^\ast dA - \frac{v^2}{2} (d \eta - eA + 2\pi n {}^\ast \Lambda) \wedge {}^\ast (d \eta - eA + 2\pi n {}^\ast \Lambda) \right).
\end{align}
If there are many vortices in the system we must consider a sum over them. The partition function consists of a sum over all configurations of the system and hence it is given by:
\begin{align}
\label{c2s1e15}
 {\cal Z} = \sum_{\{{\cal C}_3\}} \int {\cal D} A \int {\cal D} \eta e^{iS_{eff}},
\end{align}
where ${\bf P}({\cal C}_3) = {}^\ast \Lambda$. The sum over ${\cal C}_3$ is a way of parameterizing the electric condensation: if the number of vortices vanish we have a perfect superconducting state while if the vortices proliferate the superconducting state is destroyed. This information about the vortex configurations is left open and will be codified in the definition of such a sum. This point will become clearer as we proceed (see the discussion after (\ref{c2s1e23})).

Notice that the gauge dependent fields $A$ and $\eta$ can be combined in the definition of a new massive gauge invariant field $B = A - \frac 1e d\eta$. In this case, the gauge symmetry is hidden into the gauge invariant $B$ and it is this hidden realization of the gauge symmetry that is called the spontaneous breaking of the gauge symmetry. This is a standard but unfortunately misleading nomenclature since the gauge symmetry is not really broken but only hidden. In fact, since the gauge symmetry is a local symmetry, it can not be broken according to Elitzur's theorem \cite{elitzur} (for a recent discussion on the subject, see \cite{Chernodub:2008rz}), what happens here is that the original gauge dependent variables $A$ and $\eta$ are no longer convenient to describe the system in this energy scale, and it is more convenient to introduce the gauge invariant combination of $A$ and $\eta$ given by $B$: under the gauge transformation $A \rightarrow A + \frac 1e d\chi$ and $\eta \rightarrow \eta + \chi$ we have $B \rightarrow B$. Hence, we write:
\begin{align}
\label{c2s1e16}
 {\cal Z} = \sum_{\{{\cal C}_3\}} \int {\cal D} B e^{i\int_{{\cal M}_4} \left( -\frac 12 dB \wedge {}^\ast dB - \frac{e^2 v^2}{2} (B - \frac{2\pi n}{e} {}^\ast \Lambda) \wedge {}^\ast (B - \frac{2\pi n}{e} {}^\ast \Lambda) \right)},
\end{align}
which describes an ensemble of magnetic flux loops in a superconductor. The misleading nomenclature of ``spontaneous breaking of the gauge symmetry" can be loosely related to the fact that (\ref{c2s1e16}) does not possess gauge invariance in terms of the field $B$. But as we have seen in the above construction, the original gauge invariance in terms of the fields $A$ and $\eta$ is present although hidden in $B$. Since this is a standard nomenclature we will also use it throughout this work but keeping its right meaning in mind. Notice from (\ref{c2s1e11}) and (\ref{c2s1e16}) that the superconductor features two characteristic mass scales: the mass $M_\rho=\sqrt{2\lambda}v$ of the excitations associated with the fluctuations $\rho=|\phi| - v$ of the modulus  of the scalar field around the vacuum $|\phi| = v$ and the mass $M_B=ev$ of the excitations of the gauge invariant field $B$, so that the London limit amounts to $M_\rho \gg M_B$.

We are going to insert now an external monopole-antimonopole pair (prescribed magnetic charges) in the system and evaluate the static energy associated with this configuration. The presence of monopoles disturbs the system and to evaluate the energy variation associated with this perturbation we can use the following reasoning: the transition amplitude from the vacuum state before the monopole insertion, which happens at the instant $-\frac T2$, to the vacuum state after the annihilation of the monopoles, at the instant $\frac T2$, can be expressed as a coherent sum of the probabilities of the monopole pair state to be found with energy $E_n$ relatively to the state without monopoles. Explicitly, if $\Phi^\dagger$ denotes the creation operator of the monopole-antimonopole pair, we have:
\begin{align}
\label{c2s1e17}
 \langle 0|\Phi\left(\frac T2\right)\Phi^\dagger\left(-\frac T2\right)|0\rangle &= \langle 0|\Phi(0)e^{-i\hat{H}T}\Phi^\dagger(0)|0\rangle\nonumber\\
 &=  \sum_n |\langle 0|\Phi|n\rangle|^2 e^{-iE_n T},
\end{align}
where $\hat{H}$ is the Hamiltonian operator of the system in the presence of the monopole-antimonopole pair. In the second line we have used the completeness relation of the Hamiltonian eigenstates $|n\rangle$. $E_n$ represents the difference between the energy of the $n$-th excited state of the system in the presence of monopoles and the vacuum energy without monopoles which is included in a normalization factor omitted in the above expression. For the state of the system with monopoles to be considered a stationary state (the new vacuum of the system) we must take the limit $T \rightarrow \infty$. In this limit all the excited states can be ignored in the above expression and the only relevant contribution is given by the ground state $|0\rangle$. $E_0$ is then the difference between the vacuum energy of the system with monopoles and the vacuum energy of the system without monopoles (to be more precise, this reasoning involves formulating the theory in the Euclidean spacetime, taking the indicated limit and then to Wick rotating back Minkowski spacetime):
\begin{align}
\label{c2s1e18}
 \langle 0|\Phi\left(\frac T2\right)\Phi^\dagger\left(-\frac T2\right)|0\rangle  \sim  e^{-iE_0 T}; \;\;\;\; T \rightarrow \infty.
\end{align}
The energy $E_0$ contains the information about the interaction energy between the static monopole-antimonopole pair and it is our aim to determine this energy in the sequel.

The left hand side of (\ref{c2s1e18}) represents the partition function with monopole insertions. As discussed in the previous section, monopoles are seen as defects by the electromagnetic potential and thus, they are inserted in this representation via their associated Dirac strings through a non-minimal coupling as in (\ref{c1s2e16}). So, in the limit $T \rightarrow \infty$, we have:
\begin{align}
\label{c2s1e19}
 e^{-iE_0 T} &= \sum_{\{{\cal C}_3\}} \int {\cal D} B e^{i\int_{{\cal M}_4} \left( -\frac 12 (dB + g{}^\ast \Sigma_g) \wedge {}^\ast (dB + g{}^\ast \Sigma_g) - \frac{e^2 v^2}{2} (B - \frac{2\pi n}{e} {}^\ast \Lambda) \wedge {}^\ast (B - \frac{2\pi n}{e} {}^\ast \Lambda) \right)}\\
 &={\cal Z} (J_g),
\end{align}
where ${}^\ast J_g = d{}^\ast \Sigma_g$ is the monopole current. Notice that due to Elitzur's theorem \cite{elitzur} the brane symmetry,
\begin{align}
\label{c2s1e20}
{}^\ast \Sigma_g  &\rightarrow {}^\ast \Sigma_g  + d{}^\ast \Lambda_g;\\
\label{c2s1e21}
B &\rightarrow  B - g{}^\ast \Lambda_g,
\end{align}
must be maintained, since it is a local symmetry. This constitutes a physical constraint in our system. The sum defining the ensemble of flux loops must be such that all configurations of the Dirac string ${}^\ast \Sigma_g$ that can be reached by the transformations (\ref{c2s1e20}), (\ref{c2s1e21}) are contained in the ensemble. This is a consequence of the external perturbation introduced by the monopoles. The brane symmetry is preserved due to the ensemble of magnetic flux loops if $g = \frac{2\pi n}{e}$. This is why we used the notation ${\cal Z} (J_g)$ calling attention to the fact that the partition function depends only on $J_g$ and not on $\Sigma_g$. As before, we see that the Dirac quantization condition (\ref{c1s2e20}) is a consequence of the brane symmetry.

\begin{figure}[h]  % label dentro da caption para assegurar referencia correta no corpo do texto
       \centering
       \includegraphics[scale=0.6]{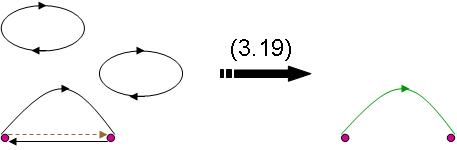}
       \caption{Full black lines represent closed vortex lines, dashed brown lines represent Dirac strings connecting monopoles and antimonopoles and full green lines represent the brane invariants.
\label{fig:0}}
\end{figure}

In order to reveal the spontaneous breaking of the brane symmetry, we shift the field $B$ by:
\begin{align}
\label{c2s1e22}
B \rightarrow  B + \frac{2\pi n}{e} {}^\ast \Lambda,
\end{align}
resulting in:
\begin{align}
\label{c2s1e23}
{\cal Z} (J_g) = \sum_{\{{\cal C}_3\}} \int {\cal D} B e^{i\int_{{\cal M}_4} \left( -\frac 12 (dB + g({}^\ast \Sigma_g + d{}^\ast \Lambda)) \wedge {}^\ast (dB + g({}^\ast \Sigma_g + d{}^\ast \Lambda)) - \frac{e^2 v^2}{2} B \wedge {}^\ast B \right)},
\end{align}
with $g = \frac{2\pi n}{e}$. In the presence of external monopoles the electric condensation is at most capable of expelling, through the Meissner effect, the vortex configurations corresponding to closed flux loops disconnected from the Dirac string. Since the remained vortex configurations in the electric condensate correspond to loops connected to the Dirac string, the sum over the 3-branes ${\cal C}_3$, which are Poincar\`e -dual to the 3-currents $\Lambda$, plays the role of summing over all the surfaces Poincar\`e -dual to the currents $\Sigma_g$ constrained by ${}^\ast J_g = d{}^\ast \Sigma_g$ and hence, we can make the following substitution (except for a possible normalization constant):
\begin{align}
\label{c2s1e24}
\sum_{\{{\cal C}_3\}} \rightarrow \sum_{\{{\cal B}_2\}} {}',
\end{align}
where ${\bf P}({\cal B}_2) = {}^\ast \tilde{\Sigma}_g \equiv {}^\ast \Sigma_g + d{}^\ast \Lambda$ and the notation $({}')$ indicates that the sum is constrained over open surfaces which have as boundary the world lines of the monopole and the antimonopole. This is illustrated in FIG. \ref{fig:0}. Hence, we have the equivalent form:
\begin{align}
\label{c2s1e25}
{\cal Z} (J_g) = \sum_{\{{\cal B}_2\}} {}' \int {\cal D} B e^{i\int_{{\cal M}_4} \left( -\frac 12 (dB + g{}^\ast \tilde{\Sigma}_g) \wedge {}^\ast (dB + g{}^\ast \tilde{\Sigma}_g) - \frac{e^2 v^2}{2} B \wedge {}^\ast B \right)}.
\end{align}

The physical scenario is clear \cite{jt-cho}: it is impossible to have a complete electric condensation when we include external monopoles in the system since the magnetic fields generated by them, although expelled from almost all the space by the Meissner effect, can not simply vanish: as we are going to see in a moment, they are confined into magnetic flux tubes connecting monopoles of opposite sign immersed in the electric condensate. In (\ref{c2s1e25}) $\tilde{\Sigma}_g$ is a brane invariant (notice that this is a brane invariant at the level of the partition function and not at the level of the action, the same happening to $B$ after the shift (\ref{c2s1e22})). In this way, the brane symmetry is preserved but its realization is hidden, just like it happens to the gauge symmetry. Thus, we say that the \emph{brane symmetry was spontaneously broken}.

In order to integrate out the quadratic gauge invariant field $B$ in the partition function (\ref{c2s1e25}), we write:
\begin{align}
S&=-\frac{1}{2}(dB,dB)-g(dB,{}^\ast\tilde{\Sigma}_g)-\frac{g^2}{2}({}^\ast\tilde{\Sigma}_g,{}^\ast\tilde{\Sigma}_g)-\frac{e^2v^2}{2}(B,B)
\nonumber\\
&=-\left(B,\frac{\Delta+e^2v^2}{2}B+gd^\dagger {}^\ast\tilde{\Sigma}_g\right)-\frac{g^2}{2}({}^\ast\tilde{\Sigma}_g,{}^\ast\tilde{\Sigma}_g),
\end{align}
where we made use of the Proca's constraint, $d^\dagger B=0$, writing: $(dB,dB)=(B,\Delta B)$, where $\Delta= d^\dagger d + dd^\dagger =  - \square = - g^{\mu\nu}\partial_{\mu}\partial_{\nu}$ is the Laplacian operator.

The integration of a Gaussian structure can be achieved by substituting the extremum path,
\begin{align}
\delta_B S=0 \Rightarrow (\Delta + e^2v^2)B=-gd^\dagger {}^\ast\tilde{\Sigma}_g\Rightarrow B = \left(-\frac{g}{\Delta + e^2v^2}\right)d^\dagger {}^\ast\tilde{\Sigma}_g,
\end{align}
into the action $S$, from which we obtain the following effective action:
\begin{align}
S_{eff}&=-\left(-\frac{g}{\Delta +e^2v^2}d^\dagger {}^\ast\tilde{\Sigma}_g,\frac{\Delta +e^2v^2}{2}\left(\frac{-g}{\Delta + e^2v^2}\right)d^\dagger {}^\ast\tilde{\Sigma}_g+gd^\dagger {}^\ast\tilde{\Sigma}_g\right)-\frac{g^2}{2}({}^\ast\tilde{\Sigma}_g,{}^\ast\tilde{\Sigma}_g)\nonumber\\
&=\frac{g^2}{2}\left(d^\dagger {}^\ast\tilde{\Sigma}_g,\frac{1}{\Delta + e^2v^2}d^\dagger {}^\ast\tilde{\Sigma}_g\right)-\frac{g^2}{2}({}^\ast\tilde{\Sigma}_g,{}^\ast\tilde{\Sigma}_g)\nonumber\\
&=\frac{g^2}{2}\left(\tilde{\Sigma}_g,\frac{{}^\ast d{}^\ast d{}^\ast{}^\ast}{\Delta+e^2v^2}\tilde{\Sigma}_g-{}^\ast{}^\ast\tilde{\Sigma}_g\right)\nonumber\\
&=\frac{g^2}{2}\left(\tilde{\Sigma}_g,\frac{-d^\dagger d}{\Delta+e^2v^2}\tilde{\Sigma}_g+\tilde{\Sigma}_g\right)\nonumber\\
&=\frac{g^2}{2}\left(\tilde{\Sigma}_g,\frac{(dd^\dagger+e^2v^2)}{\Delta+e^2v^2}\tilde{\Sigma}_g\right)\nonumber\\
&=\frac{1}{2}\left(d^\dagger\tilde{\Sigma}_g,\frac{g^2}{\Delta+e^2v^2}d^\dagger\tilde{\Sigma}_g\right)+
\frac{1}{2}\left(\tilde{\Sigma}_g,\frac{g^2e^2v^2}{\Delta+e^2v^2}\tilde{\Sigma}_g\right)\nonumber\\
&=\int_{\mathcal{M}_4}\left[-\frac{1}{2}{}^\ast J_g\wedge\left(\frac{g^2}{\Delta+e^2v^2}\right)J_g+
\frac{1}{2}{}^\ast\tilde{\Sigma}_g\wedge\left(\frac{g^2e^2v^2}{\Delta+e^2v^2}\right)\tilde{\Sigma}_g\right],
\end{align}
where $J_g=d^\dagger\tilde{\Sigma}_g$. Hence, the integration of the field $B$ in the partition function (\ref{c2s1e25}) gives (except for a normalization constant):
\begin{align}
\label{c2s1e27}
{\cal Z} (J_g) = e^{-\frac i2 \int_{{\cal M}_4} {}^\ast J_g \wedge \left(\frac{g^2}{(\Delta + e^2 v^2)}\right) J_g}\sum_{\{{\cal B}_2\}} {}' e^{\frac i2 \int_{{\cal M}_4} {}^\ast \tilde{\Sigma}_g \wedge \left(\frac{g^2e^2v^2}{(\Delta + e^2 v^2)}\right) \tilde{\Sigma}_g}.
\end{align}
The first term corresponds to an Yukawa-like interaction between the monopoles. The second term corresponds to an Yukawa-like interaction between the brane invariants $\tilde{\Sigma}_g$ containing the Dirac strings: this is the term responsible for the monopole confinement in the electric condensate. \emph{We consider that this is a signature of the confinement phenomenon in Abelian theories involving condensates}. Notice that the sum in the second term shows that the energy of the system in the presence of monopoles involves the contributions of all configurations of the brane invariant $\tilde{\Sigma}_g$ for a given external configuration of the monopole-antimonopole pair. In fact, the brane invariant $\tilde{\Sigma}_g$ corresponds to the magnetic flux tube between the monopole-antimonopole pair \cite{jt-cho}. Notice also that in the limit $v\rightarrow 0$ the electric condensate is destroyed and the interaction reduces to the Coulomb interaction between the monopoles with no confinement, as expected.

\begin{figure}[h]  % label dentro da caption para assegurar referencia correta no corpo do texto
       \centering
       \includegraphics[scale=0.5]{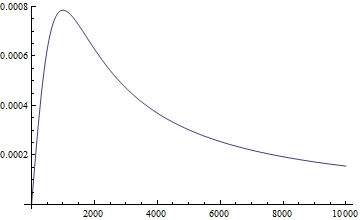}
       \caption{Behavior of the integrand in (\ref{c2s1e36}) as a function of $y$ with the exact form $\cos(y) + y Si(y)$.
\label{fig:1}}
\end{figure}

We now consider a stationary monopole configuration where the monopole-antimonopole pair is separated by a fixed distance $R$:
\begin{align}
\label{c2s1e28}
J_g^{\mu}=\delta^{0\mu}\left[\delta(\vec{x} - \vec{x}_1) - \delta(\vec{x} - \vec{x}_2)\right]; \;\;\;\; R = |\vec{x}_1 - \vec{x}_2|.
\end{align}
The Yukawa interaction between the monopole currents in the first term is easily evaluated:
\begin{align}
\label{c2s1e29}
-\frac i2 ({}^\ast J_g, \left(\frac{g^2}{(\Delta + e^2 v^2)}\right) {}^\ast J_g) &= -\frac i2 \int d^4 x J_{g\; \mu}\frac{g^2}{(\square - e^2 v^2)}J_g^{\mu}\nonumber\\
&= \frac i2 T \int d^3 x J_g^0 \frac{g^2}{(\nabla^2 - e^2 v^2)}J_g^0\nonumber\\
&= (\mbox{self-energy}) + ig^2 T \int \frac{d^3 k}{(2\pi)^3} \frac{e^{-i\vec{k}\cdot\vec{R}}}{k^2 + e^2v^2}\nonumber\\
&= (\mbox{self-energy}) + i T \frac{g^2}{4\pi R} e^{-evR},
\end{align}
where the self-energy does not depend on $R$ and contributes only to the energy renormalization. On the other hand, the terms in the sum in (\ref{c2s1e27}), which represent the brane-brane Yukawa interaction, are very complicated to be calculated in general. Notice, however, that this sum simply reflects the sum over energy states according to (\ref{c2s1e17}). We are looking for the term in this sum that gives the main contribution in the limit $T \rightarrow \infty$. This term corresponds to the flux tube configuration that minimizes the interaction energy between the brane invariants. It is expected that this configuration minimizes the brane invariant surface area. Such a configuration, at a fixed time, corresponds to a straight flux tube connecting the monopole-antimonopole pair. Since the calculation involving arbitrary surface configurations is very difficult, a rigorous proof that the minimal surface is the configuration that minimizes the energy is unknown. However, this is a reasonable hypothesis adopted in the literature \cite{Gubarev:1998ss, ripka} and we shall adopt it here as well. The straight flux tube which has as its boundary the monopole configuration (\ref{c2s1e28}) can be constructed inverting the formula ${}^\ast J_g = d{}^\ast \tilde{\Sigma}_g$, which for a static configuration has the form:
\begin{align}
\label{c2s1e30}
J_g^0 = -\partial_i \tilde{\Sigma}^{i0}.
\end{align}
The vector $\vec{R} = \vec{x}_1 - \vec{x}_2$ that connects the monopoles can be used to invert this formula:
\begin{align}
\label{c2s1e31}
\tilde{\Sigma}^{i0} = - \frac{R^i}{\vec{R} \cdot \nabla} J_g^0.
\end{align}
Notice that this procedure introduces a singularity related to the Fourier modes of the monopole density that are orthogonal to $\vec{R}$. To avoid this singularity, a possibility would be to smooth out the monopole density (\ref{c2s1e28}). However, this will not be necessary here since we are only interested in the energy and we can simply absorb these singular contributions in the energy renormalization. Substituting this expression in the brane-brane interaction we get:
\begin{align}
\label{c2s1e32}
- \frac i2\left( {}^\ast \tilde{\Sigma}_g , \left(\frac{g^2e^2v^2}{(\Delta + e^2 v^2)}\right) {}^\ast \tilde{\Sigma}_g \right) &=  -\frac i4 \int d^4 x \tilde{\Sigma}_{g\; \mu\nu}\frac{g^2e^2v^2}{(\square - e^2 v^2)}\tilde{\Sigma}_g^{\mu\nu}\nonumber\\
&=  \frac i2 \int d^4 x \tilde{\Sigma}_g^{i 0}\frac{g^2e^2v^2}{(\nabla^2 - e^2 v^2)}\tilde{\Sigma}_g^{i 0}\nonumber\\
&= (\mbox{self-energy}) + i T \int \frac{d^3 k}{(2\pi)^3} \frac{(gev)^2 e^{-i\vec{k}\cdot\vec{R}}}{(k^2 + e^2v^2)}\frac{\vec{R}^2}{(\vec{R} \cdot \vec{k})^2}.\nonumber\\
\end{align}
The self-energy term is absorbed in the energy renormalization. The integral in the last line can be rewritten as:
\begin{align}
\label{c2s1e33}
\int \frac{d^3 k}{(2\pi)^3} \frac{e^{-i\vec{k}\cdot\vec{R}}}{(k^2 + e^2v^2)}\frac{\vec{R}^2}{(\vec{R} \cdot \vec{k})^2} = 4\pi \int_0^{\infty} \frac{d k}{(2\pi)^3} \frac{1}{(k^2 + e^2v^2)}\int_0^1 dx \frac{\cos(kRx)}{x^2}.
\end{align}
Noticing that $x$ is the cosine of the angle between $k$ and $R$, the singularity in $x=0$ in the integral over $x$ corresponds to the singularity discussed above; explicitly:
\begin{align}
\label{c2s1e34}
\int_{\epsilon}^1 dx \frac{\cos(kRx)}{x^2} = -\cos(kR) + \frac{\cos(kR\epsilon)}{\epsilon} - kR Si(kR).
\end{align}
The second term above is absorbed in the energy renormalization as anticipated. In the last term we have the Integral Sine function defined by:
\begin{align}
\label{c2s1e35}
Si(t) \equiv \int_0^t dx \frac{\sin(x)}{x}.
\end{align}
At this point we have to pay attention to the scales involved in the problem. Remember we are working in the London limit in which $M_{\rho} \rightarrow \infty$. This means that $M_{\rho}$ defines the ultraviolet (UV) scale of our problem and hence the integral over $k$ in (\ref{c2s1e33}) has an UV cutoff given by $M_{\rho}$. The integral (\ref{c2s1e33}) reads:
\begin{align}
\label{c2s1e36}
-4\pi \int_0^{M_{\rho}} \frac{d k}{(2\pi)^3} \frac{1}{(k^2 + e^2v^2)}(\cos(kR) + kR Si(kR))\nonumber\\
 = -4\pi R \int_0^{RM_{\rho}} \frac{d y}{(2\pi)^3} \frac{1}{(y^2 + R^2M_B^2)}(\cos(y) + y Si(y)),
\end{align}
where we have made explicit the other relevant mass scale in our problem: $M_B = ev$. The confining potential is dominant in the infrared (IR) limit of large distances where the Yukawa potential (\ref{c2s1e29}) effectively vanishes. Thus, the relevant distance scale to identify the confining potential is such that $R \gg \frac{1}{M_B}$. This together with the London limit condition $M_{\rho} \gg M_B$ gives us an idea of the the behavior of the integrand in (\ref{c2s1e36}). Exemplifying these scale conditions by $RM_B = 1000$ and $RM_{\rho} =10000$, we see that the integrand behaves as depicted in FIG. \ref{fig:1}.

\begin{figure}[h]  % label dentro da caption para assegurar referencia correta no corpo do texto
       \centering
       \includegraphics[scale=0.5]{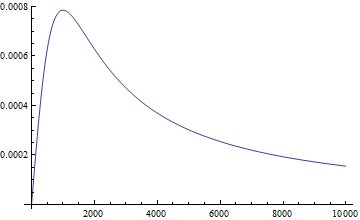}
       \caption{Behavior of the integrand in (\ref{c2s1e36}) as a function of $y$ with the asymptotic value $\cos(y) + y Si(y) \xrightarrow[y \rightarrow \infty]{} y\frac{\pi}{2}$. \label{fig:2}}
\end{figure}

On the other hand, notice that substituting the expression $\cos(y) + y Si(y)$ by its asymptotic value $y\frac{\pi}{2}$, the integrand behaves as depicted in FIG. \ref{fig:2}.

Hence, we conclude that the behavior of the integrands is approximately the same in both cases within the region of integration considered. Thus, we substitute this asymptotic value into the integral and solve it easily:
\begin{align}
\label{c2s1e37}
-4\pi R \int_0^{RM_{\rho}} \frac{d y}{(2\pi)^3} \frac{1}{(y^2 + R^2M_B^2)}y\frac{\pi}{2} = -R\frac{1}{8\pi} \ln\left(\frac{M_{\rho}^2 + M_B^2}{M_B^2}\right).
\end{align}
In this way we have obtained the expression of the linear confining potential. Gathering the results obtained above we write the final expression for the energy $E_0$:
\begin{align}
\label{c2s1e37A}
E_0 = - \frac{g^2}{4\pi R} e^{-M_B R} + \sigma R,
\end{align}
where $\sigma = \frac{g^2 M_B^2}{8\pi} \ln\left(\frac{M_{\rho}^2 + M_B^2}{M_B^2}\right)$ is the so-called string tension of the linear confining flux tube. The asymptotic character of this energy is what defines the confining regime of the theory.

Remember that the above procedure used to identify the energy $E_0$ evaluating the value of $e^{-iE_0 T}$ corresponds to the calculation of the VEV of the operator that creates an open magnetic flux tube connecting a monopole-antimonopole pair at the time $t \rightarrow -\infty$ and destroy it at $t \rightarrow \infty$ (\ref{c2s1e18}). We have seen in section \ref{sec:dual} that this corresponds to the insertion of defects in the partition function and that this procedure is the dual of the calculation of the VEV of the Wilson loop. In this sence what we have obtained here is simply the area law for the Wilson loop which defines the confining regime for the monopoles, $e^{-iE_0 T} \rightarrow e^{-i\sigma R T}$ in the IR limit. The Wilson loop we have calculated corresponds to the holonomy associated to the dual vector potential $\tilde{A}$ (which couples minimally to monopoles) whose expression in terms of the original vector potential $A$ (which couples minimally to electric charges) would be non-local. In this context, the Wilson loop receives another name, to differentiate it from the original Wilson loop, and is called \textit{'t Hooft loop}. Hence, we see that while the Wilson loop is associated to electric current lines, the 't Hooft loop is associated to magnetic flux lines.\\

The interpretation of the system in the low energy limit as a gas of magnetic flux loops was essential in to obtain the results above. We are going to reinterpret this result now under the perspective of the relation between electric currents and magnetic fluxes discussed above. The partition function of the magnetic flux loops gas was defined in (\ref{c2s1e16}). As we have discussed, ${}^\ast \Lambda$ in (\ref{c2s1e16}) represents the closed lines of magnetic flux. Introducing an auxiliary 1-form $H$ we can rewrite (\ref{c2s1e16}) in the equivalent form:
\begin{align}
\label{c2s1e39}
 {\cal Z} &= \sum_{\{{\cal C}_3\}} \int {\cal D} B {\cal D} H e^{i\int_{{\cal M}_4} \left( -\frac 12 dB \wedge {}^\ast dB - e(B - \frac{2\pi n}{e} {}^\ast \Lambda) \wedge {}^\ast H  +  \frac{1}{2v^2} H \wedge {}^\ast H\right)}\nonumber\\
  &=  \int {\cal D} B {\cal D} H e^{i\int_{{\cal M}_4} \left( -\frac 12 dB \wedge {}^\ast dB - e B \wedge {}^\ast H  +  \frac{1}{2v^2} H \wedge {}^\ast H\right)}\sum_{\{{\cal C}_3\}} e^{2\pi i n\int_{{\cal M}_4}  {}^\ast \Lambda \wedge {}^\ast H }.
\end{align}
Notice that the term containing the sum over surfaces can be rewritten using the GPI (\ref{c1s2eB}) and we get:
\begin{align}
\label{c2s1e40}
 {\cal Z} =  \int {\cal D} B {\cal D} H e^{i\int_{{\cal M}_4} \left( -\frac 12 dB \wedge {}^\ast dB - e B \wedge {}^\ast H  +  \frac{1}{2v^2} H \wedge {}^\ast H\right)}\sum_{\{{\cal A}_1\}} \delta ({}^\ast J - {}^\ast H),
\end{align}
where ${}^\ast J = {\bf P}({\cal A}_{1})$. Integrating over $H$ we obtain:
\begin{align}
\label{c2s1e41}
 {\cal Z} = \sum_{\{{\cal A}_1\}} \int {\cal D} B e^{i\int_{{\cal M}_4} \left( -\frac 12 dB \wedge {}^\ast dB - e B \wedge {}^\ast J  +  \frac{1}{2v^2} J \wedge {}^\ast J\right)}.
\end{align}
This theory represents an ensemble of electric currents and is an alternative representation to the ensemble of magnetic flux loops \cite{dafdc}. Apparently this theory does not have a gauge ambiguity, since the sum over the surfaces ${\cal A}_{1}$ is unrestricted and includes configurations of open currents. However, remember that $B$ is a gauge invariant combination of the gauge dependent fields $A$ and $\eta$ and hence, the gauge symmetry is present, although hidden also in the above formulation. We can reveal it by noticing that the sum over open surfaces constitutes a redundancy in the theory. Indeed, since $B$ appears in a gaussian structure, all the physical information of the theory is contained in the stationary point of the action with respect to variations of $B$ and it is trivially integrated out in (\ref{c2s1e41}). Its equation of motion,
\begin{align}
\label{c2s1e43}
 d^{\dagger} d B = - e J,
\end{align}
implies the current conservation,
\begin{align}
\label{c2s1e44}
 d {}^\ast J = 0,
\end{align}
and thus, only the sum over closed surfaces contributes physically: the current conservation is a consequence of the gauge symmetry.

The connection between flux loops and currents becomes manifest by noticing that the flux loops dilution, which establishes the complete transition to the condensed phase, is represented in the formulation (\ref{c2s1e41}) by the electric condensation. Indeed, the dilution of the flux loops in (\ref{c2s1e16}) is represented by the decoupling of the sum over surfaces, which is then absorbed in the normalization of the partition function, resulting in the establishment of the Proca theory as the effective theory, that is, the magnetic fluctuations represented by the vortices are frozen. On the other hand, in the formulation (\ref{c2s1e41}), this effect manifests itself through the condensation of the current lines: formally, the sum over the 1-branes ${\cal A}_{1}$ is substituted by an integral measure over $J$, which in this context becomes a field describing the electric condensate. The result of the integration over $J$ is again the Proca theory. This procedure corresponds to one of the faces of the JTA, which we shall discuss soon in subsection \ref{JTA}. We have in this way a kind of \emph{order-disorder connection} between fluxes and currents \cite{dafdc}. A particular version of this order-disorder map was used by Banks, Kogut and Myerson in \cite{Banks:1977cc} to study phase transitions in relativistic lattice Abelian gauge theories and also by Kleinert in \cite{ksc, Kiometzisprl} to investigate the phase structure of the superconductor and its critical exponents (see also chapter 5 of \cite{mvf} for a review).

The electric condensed state realizes the monopole confinement as we have discussed previously. If monopoles, taken as external magnetic charges, are immersed in the superconductor, the dilution of the magnetic flux loops has as consequence the selection of a fixed configuration for the magnetic flux tube connecting them. The effective theory describing this scenario is the Proca theory non-minimally coupled to the brane invariant as in equation (\ref{c2s1e25}), where the brane symmetry is hidden, a fact we called the spontaneous breaking of the brane symmetry. This means that the monopoles are no longer the adequate degrees of freedom for the description of the system in the confining phase: the magnetic flux tube connecting them acquires the status of relevant degree of freedom. On the other hand, if we introduce external electric charges in the system they will minimally couple to the Proca field and consequently interact through a Yukawa-like potential. At asymptotically large distances the potential vanishes except for a possible constant: this is the screening phenomenon defining the Higgs phase of the system. Physically, the electric condensate can be seen as a reservoir of electric charges and the introduction of external charges disturbs the system inducing the creation of pairs which screen the charge of the external sources at large distances.

We conclude that while the electric charges see the superconducting state as a Higgs phase, the monopoles see it as a confining phase. This is the Higgs-confinement connection which generalizes for the condensed phase the Coulomb-Coulomb connection discussed in section \ref{sec:dual} in the context of the diluted phase. We have then, the following phase structure defined by the asymptotic behavior of external charges \cite{Intriligator:1995au}:
\begin{align}
\begin{array}{c|c}
\label{c2s1e42}
  \text{monopoles} & \text{electric charges}  \\
  &\\
  \text{Coulomb   }  V(R) \sim \frac 1R  & \text{Coulomb   }  V(R) \sim \frac 1R \\
  &\\
  \text{Confinement   } V(R) \sim \sigma R & \text{Higgs   } V(R) \sim const.\\
\end{array}
\end{align}

As we have seen in section \ref{sec:dual} the electromagnetic duality maps minimal coupling into non-minimal coupling and vice-versa. This means that charges are mapped into defects and vice-versa, but in the context of that section both were in the Coulomb phase. However, in the superconductor there is a qualitative difference between the behavior of external charges: electric charges are condensed and monopoles are confined. On the other hand, we could have studied the dual superconductor and everything would be mathematically identical except for the fact that we would be talking about a monopole condensate and electric charge confinement. The phase structure of the dual superconductor is then:
\begin{align}
\begin{array}{c|c}
\label{c2s1e43A}
  \text{monopoles} & \text{electric charges}  \\
  &\\
  \text{Coulomb   }  V(R) \sim \frac 1R  & \text{Coulomb   }  V(R) \sim \frac 1R \\
  &\\
  \text{Higgs   } V(R) \sim const. & \text{Confinement   } V(R) \sim \sigma R\\
\end{array}
\end{align}

\subsection{The Julia-Toulouse Approach}
\label{JTA}

The Proca theory, which describes the electric condensate, has its origin in the Higgs Mechanism. This mechanism connects the Maxwell theory to the Proca theory through an electric condensation process. The mechanism is implemented through the introduction of a scalar field with a $\phi^4$ potential that introduces an energy scale $v$, which must be experimentally determined, and a dimensionless parameter $\lambda$ that is the relative weight of the energy contributions of the terms present in the energy functional. If we are not interested in the details of the condensation process we can ask ourselves if, having the knowledge of the model that describes the system before the condensation, we are able to determine the effective model describing the system in the condensed phase. This is the idea of the Julia-Toulouse prescription \cite{jt}. The condensation of topological defects establishes a new medium in which the defects constitute a continuous distribution in space. The low energy excitations of this medium represent the new degrees of freedom of the condensed phase. Julia and Toulouse specified a prescription to identify these new degrees of freedom, knowing the model that describes the diluted phase. This prescription does not deal with the dynamical reasons responsible for the condensation process: this is considered a separate issue, beyond the scope of the prescription. This prescription concerns only the properties of the new degrees of freedom once the condensation of topological defects has taken place. In the example discussed above, this means that we do not look for the value of the parameter $\lambda$: it will not be present in the effective theory. Only the scale $v$ shall be present since it determines the characteristic scale of the condensate. In fact, as we have seen previously, $\lambda$ defines the mass scale $M_{\rho} = \sqrt{2\lambda}v$ which is seen as an UV cutoff for the effective theory signaling the scale where the low energy excitations of the condensate ceases to be a good description of the system.

The work of Julia and Toulouse has taken place in the context of ordered systems in condensed matter and due to the possible non-linearity of the topological currents, the absence of relativistic symmetry and the need for the introduction of dissipative external defects in this scenario, the construction of effective actions can be very complicated. However, Quevedo and Trugenberger \cite{qt} showed that in theories involving $p$-forms, which are very common in effective descriptions of string theories (for a recent review see \cite{Becker:2007zj}), these difficulties do not show up. They showed that in this context the prescription can be defined into a more precise form, which leads to the determination of the effective action describing the system in the condensed phase. They have also showed that this leads naturally to a dual interpretation of the Higgs Mechanism.

Here we shall present a new and more complete form of this prescription which will allow us to consistently approach the condensation process and its consequences in a great variety of physical systems. Our generalized formulation of the JTA makes it possible to:
\begin{enumerate}
\item deal with general sorts of condensates, including condensates which may break space-time symmetries;
\item consistently define the effective action for the condensed phase, preserving the brane symmetry (being its realization explicit or hidden), in consonance with Elitzur's theorem \cite{elitzur} (see \cite{jt-cho} for a discussion about the explicit breaking of the brane symmetry present in the original formulation and its consequent inconsistencies);
\item deal with the condensation of sources minimally coupled to gauge fields and not only with condensation of defects.
\end{enumerate}

The key concept for this generalization is the recognition of the importance of the brane symmetry. This is the main guiding principle in the construction of the effective action describing the condensed phase. The brane symmetry is a local symmetry and hence it constitutes an ambiguity in the definition of the variables describing the system. Similarly to what happens to the gauge symmetry, there is no physical process capable of eliminating such ambiguity. However, the brane symmetry can be realized in a hidden fashion, akin to the gauge symmetry in the so-called spontaneous breaking of the gauge symmetry. We have seen an example of such hidden realization in (\ref{c2s1e25}). In the same way that the spontaneous breaking of the gauge symmetry gives us an observable massive gauge invariant field, the spontaneous breaking of the brane symmetry gives us an observable brane invariant carrying energy content, which constitutes a signature of the confinement phenomenon.

To illustrate the JTA in a first example, we are going to write the dual formulation of the system described by (\ref{c2s1e25}). The action defining the partition function is the Proca-like action, given by:
\begin{align}
\label{c2s2e02}
S_{Proca} = \int_{{\cal M}_4} \left( -\frac 12 (dB + g{}^\ast \tilde{\Sigma}_g) \wedge {}^\ast (dB + g{}^\ast \tilde{\Sigma}_g) - \frac{e^2 v^2}{2} B \wedge {}^\ast B \right).
\end{align}
Following the procedure given in section \ref{sec:dual}, we introduce an auxiliary 2-form $G$:
\begin{align}
\label{c2s2e03}
{\cal Z} (J_g) = \sum_{\{{\cal B}_g\}} {}' \int {\cal D} B \int {\cal D} G e^{i\left(\left(G, dB + g{}^\ast \tilde{\Sigma}_g \right) + \frac 12 \left(G, G \right) - \frac{e^2 v^2}{2} \left(B, B \right)\right)}.
\end{align}
We can turn $B$ into an auxiliary field integrating by parts in the first term, and then $B$ is trivially integrated out:
\begin{align}
\label{c2s2e04}
{\cal Z} (J_g) = \sum_{\{{\cal B}_g\}} {}' \int {\cal D} G e^{i\left(\frac{1}{2e^2v^2} \left(d^{\dagger} G, d^{\dagger} G \right) + \frac12 \left(G, G \right) + g\left(G, {}^\ast \tilde{\Sigma}_g \right)\right)},
\end{align}
or renaming ${}^\ast G = H$,
\begin{align}
\label{c2s2e05}
{\cal Z} (J_g) = \sum_{\{{\cal B}_g\}} {}' \int {\cal D} H e^{i\int_{{\cal M}_4} \left(- \frac{1}{2e^2v^2} d H \wedge {}^\ast d H - \frac12 H \wedge {}^\ast H + g H \wedge {}^\ast \tilde{\Sigma}_g \right)}.
\end{align}
This formulation is the dual of (\ref{c2s1e25}). The action defining this partition function is the dual of the Proca action non-minimally coupled to the brane invariant in $(3+1)D$ (\ref{c2s2e02}), that is, it is the massive Kalb-Ramond action minimally coupled to the brane invariant \cite{jt-cho}.

We are going to show now that it is possible to directly obtain the massive Kalb-Ramond theory (\ref{c2s2e05}) from the Dual Maxwell theory through the JTA. The Dual Maxwell theory in the presence of electric defects and external magnetic currents has the general form:
\begin{align}
\label{c2s2e06}
{\cal Z}_{DualMax} (J_g) = \sum_{\{{\cal B}_{e}\}} \int {\cal D} \tilde{A}   e^{i\int_{{\cal M}_4} \left( -\frac 12 (d\tilde{A} + e{}^\ast \Sigma_e) \wedge {}^\ast (d\tilde{A} + e{}^\ast \Sigma_e) + g \tilde{A} \wedge {}^\ast J_g \right) +  iS(J_e)},
\end{align}
where ${\cal B}_{e}$ are 2-branes Poincar\`e -dual to the 2-currents ${}^\ast \Sigma_e$. The action $S(J_e)$ is a function only of the brane invariants ${}^\ast J_e = d{}^\ast \Sigma_e$. The 2-currents ${}^\ast \Sigma_e$ behave as defects as seen by the dual electromagnetic gauge potential $\tilde{A}$, meaning that the field $\tilde{A}$ is singular where ${}^\ast \Sigma_e$ is nonzero, such that the combination $H = d\tilde{A} + e{}^\ast \Sigma_e$ is regular in the whole spacetime (indeed, this is the observable electromagnetic field). The field $H$ is an electric brane invariant. Consider now the case in which the electric defects condense. This means that the field $\tilde{A}$ becomes more and more singular until it is not defined anywhere. The JTA consists in considering that the new excitations of the electric condensate shall be described by the regular field $H$, the only one that remains physically well defined after the condensation process has taken place, and we must therefore construct an effective action for the system in terms of this new fundamental field. Notice that the minimal coupling with the monopole current $J_g$ must be rewritten in terms of brane invariant fields. This can be done as follows:
\begin{align}
\label{c2s2e07}
g (\tilde{A}, J_g) &= g (\tilde{A}, {}^\ast  d {}^\ast \Sigma_g) = g (\tilde{A}, d^{\dagger} \Sigma_g)\nonumber\\
&=  g (d\tilde{A}, \Sigma_g + d^{\dagger} \Lambda_g)\nonumber\\
&= g (d\tilde{A} + e{}^\ast \Sigma_e, \Sigma_g + d^{\dagger} \Lambda_g),
\end{align}
where in the passage to the second line we have taken care of evidencing the ambiguity in the definition of the magnetic brane $\Sigma_g$ through the 3-current $\Lambda_g$. In order to obtain the third line in the equation above, we added the electric brane term which does not contribute due to the Dirac quantization condition. Indeed, to obtain the dual formulation of the Maxwell action in the form (\ref{c2s2e06}) we discarded exactly this term for the same reason (see (\ref{c1s2e24})), and here we reintroduced it. Due to the ambiguity in the definition of the brane $\Sigma_g$ we have in fact a family of physically equivalent actions parameterized by the different configurations of $\Lambda_g$. Hence, we sum over the 3-branes ${\cal C}_{g}$ Poincar\`e -dual to $\Lambda_g$ in the partition function:
\begin{align}
\label{c2s2e08}
{\cal Z}_{DualMax} (J_g) &= \sum_{\{{\cal C}_{g}\}} \sum_{\{{\cal B}_{e}\}} \int {\cal D} \tilde{A}   e^{i\int_{{\cal M}_4} \left( -\frac 12 (d\tilde{A} + e{}^\ast \Sigma_e) \wedge {}^\ast (d\tilde{A} + e{}^\ast \Sigma_e) + g (d\tilde{A} + e{}^\ast \Sigma_e)  \wedge {}^\ast (\Sigma_g + d^{\dagger} \Lambda_g) \right) +  iS(J_e)}.\nonumber\\
\end{align}
The JTA in this context consists in promoting the electric 2-current ${}^\ast \Sigma_e$ to the field category representing the establishment of the electric condensate. Formally, we have:
\begin{align}
\label{c2s2e09}
\sum_{\{{\cal B}_{e}\}} \rightarrow \int {\cal D} \Sigma_e.
\end{align}
In this way the electric brane symmetry is realized in a Stuckelberg-like structure. The integrals over $\Sigma_e$ and $\tilde{A}$ are substituted by an integral over the electric brane invariant field $H = d\tilde{A} + e{}^\ast \Sigma_e$ which realizes the electric brane symmetry in a hidden fashion. This is a characteristic of the JTA when there is mass generation in the system which we call the \emph{rank jump}: the field describing the condensed phase is the 2-form $H$, while the one describing the diluted phase is the 1-form $A$. The partition function acquires the following form after the electric condensation:
\begin{align}
\label{c2s2e10}
{\cal Z} (J_g) = \sum_{\{{\cal C}_{g}\}} \int {\cal D} H   e^{i\int_{{\cal M}_4} \left( -\frac 12 H \wedge {}^\ast H + g H  \wedge {}^\ast (\Sigma_g + d^{\dagger} \Lambda_g) \right) +  iS(dH)}.
\end{align}
This mass generation mechanism shown here has as one of its consequences the spontaneous breaking of the magnetic brane symmetry: the sum over ${\cal C}_{g}$ is equivalent to span all the surfaces $\Sigma_g$ which have as boundary the monopole current $J_g$, thus, as we have done previously, we can rewrite it as a sum over the 2-branes ${\cal B}_{g}$ Poincar\`e -dual to the magnetic brane invariants $\tilde{\Sigma}_g \equiv \Sigma_g + d^{\dagger} \Lambda_g$ constrained by the condition ${}^\ast J_g = d{}^\ast \tilde{\Sigma}_g$:
\begin{align}
\label{c2s2e11}
{\cal Z} (J_g) = \sum_{\{{\cal B}_{g}\}} {}' \int {\cal D} H   e^{i\int_{{\cal M}_4} \left( -\frac 12 H \wedge {}^\ast H + g H  \wedge {}^\ast \tilde{\Sigma}_g \right) +  iS(dH)}.
\end{align}
The last step in the prescription consists in noticing that we are looking for an effective description of the low energy modes of the condensate. Therefore, it is a good approximation to consider the action $S(dH)$ as a power series in the derivatives of the condensate field $H$ and keep only the first non-trivial term consistent with the expected symmetries of the condensed phase. In this case, we have:
\begin{align}
\label{c2s2e12}
 S(dH) = -\frac{1}{2M^2} dH \wedge {}^\ast dH,
\end{align}
where $M$ is a phenomenological characteristic scale of the condensate. We can identify it with the phenomenological parameter appearing in (\ref{c2s2e05}), $M \equiv ev$, reobtaining in this way the theory (\ref{c2s2e05}).

This procedure reveals that the condensation process described by the JTA is the dual of the Higgs Mechanism in the present situation, in the sense that it acts in the dual picture. This is the conclusion achieved by Quevedo and Trugenberger \cite{qt}.

The approach we followed here can also be used to deal with the condensation of $p$-currents minimally coupled. We can obtain the theory (\ref{c2s1e25}) starting from the Maxwell system minimally coupled to electric charges and non-minimally coupled to external monopoles:
\begin{align}
\label{c2s2e13}
{\cal Z}_{Max} (J_g) = \sum_{\{{\cal A}_{e}\}} {}' \int {\cal D} A   e^{i\int_{{\cal M}_4} \left( -\frac 12 (dA + g{}^\ast \Sigma_g) \wedge {}^\ast (dA + g{}^\ast \Sigma_g) + e A \wedge {}^\ast J_e  \right) +  iS(J_e)},
\end{align}
where ${\cal A}_{e}$ are 1-branes Poincar\`e -dual to the electric currents $J_e$ and the primed sum indicates that it is constrained to be taken only over closed (conserved) currents (and hence $A$ has a gauge ambiguity). In the action present in (\ref{c2s2e13}) the magnetic brane symmetry is realized in the form:
\begin{align}
\label{c2s2e14}
{}^\ast \Sigma_g  &\rightarrow {}^\ast \Sigma_g  + d{}^\ast \Lambda_g;\\
A &\rightarrow  A - g{}^\ast \Lambda_g.
\end{align}
The term featuring the minimal coupling with the electric current is a brane invariant due to the Dirac quantization condition. We use this fact and the electric current conservation to rewrite the minimal coupling term as:
\begin{align}
\label{c2s2e15a}
 e A \wedge {}^\ast J_e  \rightarrow e\left(A + d\phi + g{}^\ast \Omega_g \right)\wedge {}^\ast J_e,
\end{align}
including in the partition function a sum over all the branes ${\cal E}_{g}$ Poincar\`e -dual to the currents $\Omega_g$ and connected to the Dirac strings $\Sigma_g$, as well as a functional integral over the auxiliary field $\phi$. In introducing the sum and integral in the partition function we are essentially fixing the gauge and brane symmetries in the Stuckelberg form. Physically, this sum represents the embedding of the system in a ``virtual gas" of closed magnetic fluxes $d {}^\ast \Omega_g$ (the expression ``virtual" here accounts for the fact that these objects are unphysical and can be reabsorbed in a redefinition of the fields, just like it happens with the pure gauge $\phi$). The reason for the introduction of these objects is to reveal the physical variables (gauge and brane invariants) that shall define the system after the condensation process has taken place. The partition function acquires now the following form:
\begin{align}
\label{c2s2e15b}
{\cal Z}_{Max} (J_g) = \sum_{\{{\cal A}_{e}\}} \sum_{\{{\cal E}_{g}\}} \int {\cal D} A \int {\cal D} \phi  e^{i\int_{{\cal M}_4} \left( -\frac 12 (dA + g{}^\ast \Sigma_g) \wedge {}^\ast (dA + g{}^\ast \Sigma_g) + e\left(A + d\phi + g{}^\ast \Omega_g\right) \wedge {}^\ast J_e  \right) +  iS(J_e)}.\nonumber\\
\end{align}
We can reveal the physical variables defining them as follows:
\begin{align}
\label{c2s2e15c}
B &\equiv A + d\phi + g{}^\ast \Omega_g, \\
{}^\ast\tilde{\Sigma}_g &\equiv {}^\ast\Sigma_g - d{}^\ast \Omega_g,
\end{align}
such that $B$ and $\tilde{\Sigma}_g$ are gauge and brane invariants. As we know, the sum over ${\cal E}_{g}$ can be rewritten as a sum over the 2-branes ${\cal B}_{g}$ Poincar\`e -dual to the magnetic brane invariants $\tilde{\Sigma}_g$ constrained by ${}^\ast J_g = d{}^\ast \tilde{\Sigma}_g$, while the integrals over $A$ and $\phi$ are translated into an integral over $B$. The partition function reads:
\begin{align}
\label{c2s2e16}
{\cal Z}_{Max} (J_g) = \sum_{\{{\cal A}_{e}\}} \sum_{\{{\cal B}_{g}\}} {}' \int {\cal D} B   e^{i\int_{{\cal M}_4} \left( -\frac 12 (dB + g{}^\ast \tilde{\Sigma}_g) \wedge {}^\ast (dB + g{}^\ast \tilde{\Sigma}_g) + e B \wedge {}^\ast J_e  \right) +  iS(J_e)}.
\end{align}
The sum over ${\cal A}_{e}$ is no longer constrained, since the gauge symmetry was fixed and the fixing of the brane symmetry had the effect of introducing the sum over all the surfaces which have as boundary $J_g$. To proceed with the JTA we must specify a form for the action $S(J_e)$. As before, our aim is to describe an electric condensate at low energies. A term of the form $(J_e, J_e)$ in the action effectively represents the activation energy of electric excitations. This is a kind of ``chemical potential" term such that the expression (\ref{c2s2e16}) can be regarded as a grand canonical partition function. Due to dimensional reasons, this term requires the introduction of a mass scale $M$, which at this stage is related to the activation energy of the electric excitations. Higher powers in $J_e$ represent corrections to this energy and shall be suppressed by higher powers of $M$. For the establishment of the electric condensate and its description at the lowest energies it suffices to consider only the quadratic term \cite{dafdc, jt-cho}. We have then:
\begin{align}
\label{c2s2e17}
{\cal Z}_{Max} (J_g) = \sum_{\{{\cal A}_{e}\}} \sum_{\{{\cal B}_{g}\}} {}' \int {\cal D} B   e^{i\int_{{\cal M}_4} \left( -\frac 12 (dB + g{}^\ast \tilde{\Sigma}_g) \wedge {}^\ast (dB + g{}^\ast \tilde{\Sigma}_g) + e B \wedge {}^\ast J_e  + \frac{1}{2M^2} J_e \wedge {}^\ast J_e \right)}.
\end{align}
We can formally condense $J_e$ imposing:
\begin{align}
\label{c2s2e18}
\sum_{\{{\cal A}_{e}\}} \rightarrow \int {\cal D} J_e.
\end{align}
Integrating over $J_e$ we get (\ref{c2s1e25}). We call this new procedure dealing with the condensation of currents minimally coupled as the \textit{Dual Julia-Toulouse Approach (DJTA)}.

The JTA can be seen as a ``generator of effective theories". This prescription suggests the answer to the question about which theory describes the system after the condensation of $p$-currents. This is a phenomenological approach whose validity depends on \emph{a posteriori} verification. Still, due to the fact that it is a systematic procedure, it can be very useful in guiding us to obtain non-trivial results. This situation is not different than the process of quantization of a classical theory. There is also in this case a large arbitrariness involved in the process: in promoting classical dynamical variables to the operator category, ambiguities can emerge in the ordering of these operators which can only be fixed through comparison with experimental results.

We summarize our results in this section in the following schematic picture:
\begin{align}
\label{c2s2e20}
     \xymatrix{*+[F-:<10pt>]{\rm{Maxwell}}+\ar[rrr]_-{\rm{Dual\; JTA}}^-{\rm{Higgs\; Mechanism}}\ar[dd] & & &
*+[F-:<10pt>]{\rm{Proca}} \ar[dd]\\
 &  & \\
*+[F-:<10pt>]{\rm{Dual\; Maxwell}} \ar[uu]^-{\rm{Duality}} \ar[rrr]_-{\rm{JTA}}& & &
*+[F-:<10pt>]
{\rm{Massive\; Kalb-Ramond}}  \ar[uu]_-{\rm{Duality}}\\
         }
\end{align}

As a final remark for this section, we mention that recently \cite{jt-cho} we have applied the (dual of the) above picture to study the monopole condensation in the so-called $SU(2)$ \emph{restricted gauge theory} defined by means of the \emph{Cho decomposition} \cite{cho} of the non-Abelian connection, confirming it as the subsector of the complete gauge theory responsible for the confinement physics at large distances.

\section{Application II - The Polyakov Model}
\label{sec:2}

In this section we shall discuss a very important result obtained by Polyakov \cite{Polyakov:1975rs}. The Polyakov Model describes a system that exhibits confinement as a result of the collective behavior of defects, which in this case are instantons. An important fact not emphasized in the literature is that this model features the rank jump phenomenon which, as we have seen in the previous section, is a signature of the mass generation within the JTA. We are going to see that the JTA can also be applied in this case and represents another interpretation of a more recent result obtained by Polyakov \cite{Polyakov:1996nc}.

In this and the next three sections we shall work in the Euclidean spacetime ${\cal M}_3 = \mathrm{I\!R}^{3}$. The Polyakov Model is proposed to describe the IR limit of the $SO(3)$ Georgi-Glashow theory defined by the action:
\begin{align}
\label{c2s3e01}
S =\int_{{\cal M}_3} \frac{1}{2e^2} Tr G \wedge {}^\ast G  + \frac 12 d_W \phi^a \wedge {}^\ast d_W \phi^a + \frac{\lambda}{4} \left( |\vec{\phi}|^2 - v^2\right) \wedge {}^\ast \left( |\vec{\phi}|^2 - v^2\right),
\end{align}
where the trace is taken over the components of $G^a = dW^a + \frac 12 \varepsilon^{abc} W^b \wedge W^c$, $a, b, c = 1, 2, 3$; the real field $\vec{\phi}$ transforms in the \textbf{3} of $SO(3)$; and the coupling between $\phi^a$ and $W^a$ is given by the covariant derivative $d_W \phi^a = d\phi^a + \varepsilon^{abc} W^b \phi^c$.

We want to study the effective theory describing the system in the low energy limit, which means that we are considering an energy scale much lower than energies of the order of $\sqrt{\lambda} v$, which are associated with fluctuations of the field $\vec{\phi}$. In this low energy scale the field $\vec{\phi}$ is in a configuration such that $|\vec{\phi}| = v$. Hence, we are going to work in a region where the $SO(3)$ gauge symmetry is broken to $SO(2) \simeq U(1)$ (which corresponds to the rotation symmetry in the plane orthogonal to the arbitrary internal axis defined by the nonzero VEV of $|\vec{\phi}|$) and thus, we are looking for an $U(1)$ effective gauge theory.

Notice that there are defects in the system: in $3D$ the asymptotic space is homeomorphic to $S^2$ and hence we have, as a consequence of the gauge symmetry breaking in the IR, the non-trivial maps $\Pi_2(SO(3)/SO(2)) = \mathds{Z}$, which labels the non-trivial fluxes of the theory. This means that the usual Maxwell theory is not a good candidate for the effective theory in this case, since it is not capable of describing these fluxes. In the original complete theory these defects are classical solutions of the non-linear differential equations of motion. These topological solutions correspond to the \textit{'t Hooft-Polyakov monopoles} \cite{'tHooft:1974qc} in a stationary configuration in $4D$; in this context, they appear as localized objects in the $3D$ spacetime and are called \textit{instantons}. In the IR limit they appear as point singularities in the fields of the effective theory.

The instanton contribution to the partition function is of the form $e^{-\frac{I}{e^2}}$, where $I$ is a constant independent of the coupling constant $e$. This structure is obtained substituting the classical solution in the action (\ref{c2s3e01}). $e^{-\frac{I}{e^2}}$ is proportional to the probability density for the creation of an instanton. More precisely, the average instanton number in a volume $V$ is of the order of $V\mu e^{-\frac{I}{e^2}}$, where $\mu$ is a scale of inverse volume determined by the normalization of the probability density. If the coupling is weak the instanton number is only relevant for large volumes, \emph{i.e.}, the instanton contribution is fundamental in the IR limit. However, the instanton contribution is not given by the addition of local functionals of the Maxwell field to the Maxwell theory. The terms related to the instanton contribution should be relevant in the renormalization group sense, since the relevant terms have their contribution amplified in the IR limit. Also, the $U(1)$ gauge symmetry must be maintained. The only term with these characteristics is the Chern-Simons term, but it breaks the $P$ and $T$ symmetries and there is no physical reason to expect this to happen in this case. These observations indicate that a 1-form does not constitute a good effective description of the system in the low energy limit, although it is present in the original $SO(3)$ action. In fact, these considerations show that the instanton contribution is of a non-perturbative nature.

More precisely, these arguments mean that it is natural to define the effective theory as formally described by the following partition function:
\begin{align}
\label{c2s3e02}
{\cal Z}_A =  \sum_{\{{\cal B}_{g}\}} \int {\cal D} A   e^{-\int_{{\cal M}_3} \left(\frac 12 (dA + g{}^\ast \Sigma_g) \wedge {}^\ast (dA + g{}^\ast \Sigma_g) \right) - S_g(J_g)},
\end{align}
where $g \sim \frac 1e$ and the sum is taken over the surfaces Poincar\`e -dual to $\Sigma_g$. Notice that this sum is unrestricted reflecting the fact that these are objects internal to the system (we have not introduced external charges in the system at this stage). The closed surfaces, $d{}^\ast\Sigma_g = 0$, do not contribute and can be decoupled being absorbed in $A$. In this way, this sum is equivalent to a restricted sum over the configurations of $\Sigma_g$ with the constraint ${}^\ast J_g = d{}^\ast\Sigma_g$ followed by a sum over all configurations of $J_g$. The reasoning we followed in the previous paragraph imply that an effective theory, written in terms of $A$ only, which would result from the sum over instanton configurations in (\ref{c2s3e02}), does not exist.

Since the formulation in terms of $A$ has a non-perturbative character regarding the instanton contribution, it is a good idea to look for a dual formulation. This was the path followed by Polyakov. The dual of the theory (\ref{c2s3e02}) is easily obtained with the methods of section \ref{sec:dual}:
\begin{align}
\label{c2s3e05}
{\cal Z}_{\eta} =  \sum_{\{{\cal A}_{g}\}} \int {\cal D} \eta   e^{-\int_{{\cal M}_3} \left(\frac 12 d\eta \wedge {}^\ast d\eta - ig \eta \wedge {}^\ast J_g\right) - S_g(J_g)},
\end{align}
where $\eta$ is a 0-form and since the dual action appearing in the partition function depends only on $J_g$, the sum over surfaces in (\ref{c2s3e02}) can be rewritten here as a sum over the surfaces ${\cal A}_{g}$ Poincar\`e -dual to $J_g$. This sum has a formal character and it is only properly defined on a lattice \cite{Banks:1977cc,mvf,dafdc}. However, in the present case, notice that $J_g$ is a 0-current. More precisely, by the definition (\ref{c1s2e05}), an instanton localized in the event $x_0$ in spacetime is simply represented by:
\begin{align}
\label{c2s3e06}
 J_g(x) = \delta(x-x_0).
\end{align}
It follows that we can construct an explicit realization of the sum describing an arbitrary number of instantons in spacetime, which constitutes an adequate description of the system in the IR limit. For only one instanton, the contribution to the partition function (\ref{c2s3e05}) would be:
\begin{align}
\label{c2s3e07}
1:\;\; \sum_{\{{\cal A}_{g}\}} e^{ ig (\eta, J_g) - S_g(J_g)}\rightarrow \int d^3 x_0 \mu e^{-\frac{I}{e^2}} e^{ig\eta(x_0)},
\end{align}
that is, a sum over all the possible configurations with weight given by the probability density of existence of an instanton in spacetime, $\mu e^{-\frac{I}{e^2}}$. This information should be formally codified in $S_g(J_g)$. For two instantons, we would have:
\begin{align}
\label{c2s3e08}
2:\;\; \sum_{\{{\cal A}_{g}\}} e^{ ig (\eta, J_g) - S_g(J_g)} \rightarrow \int d^3 x_0 \int d^3 x'_0 \frac{\mu^2 e^{-\frac{2I}{e^2}}}{2!} e^{ig\eta(x_0)}e^{ig\eta(x'_0)},
\end{align}
where the factor $2!$ designates the fact that the two instantons are indistinguishable. We could have also antinstantons with charge $-g$. Therefore, in a situation in which the system has $n_{+}$ instantons and $n_{-}$ antinstantons, the contribution shall be of the form:
\begin{align}
\label{c2s3e09}
\frac{\left(\mu e^{-\frac{I}{e^2}}\right)^{n_{+} + n_{-}}}{n_{+}! n_{-}!}\prod_{i=1}^{n_{+}}\prod_{j=1}^{n_{-}}\int d^3 x^{i}_0 \int d^3 x^{j}_0  e^{ig\eta(x^{i}_0)}e^{-ig\eta(x^{j}_0)}.
\end{align}
Hence, for an arbitrary number of instantons the sum over all configurations can be explicitly written:
\begin{align}
\label{c2s3e10}
\sum^{\infty}_{n_{+},n_{-} = 0}\frac{\left(\mu e^{-\frac{I}{e^2}}\right)^{n_{+} + n_{-}}}{n_{+}! n_{-}!}\prod_{i=1}^{n_{+}}\prod_{j=1}^{n_{-}}\int d^3 x^{i}_0 \int d^3 x^{j}_0  e^{ig\eta(x^{i}_0)}e^{-ig\eta(x^{j}_0)}.
\end{align}
Notice that we are only considering instantons (antinstantons) with charge $g$ ($-g$), and to really take into account all the configurations, the sum would span all the charges with integer multiples of $g$. However, for a weak coupling $e$, the contributions of higher orders in the charge will be exponentially suppressed. This means that it is energetically favorable to create $n$ instantons with charge $g$ spread over a large volume than to create only one instanton with charge $ng$.

The most interesting fact here is that, in this case, the sum can be explicitly evaluated. Using that the instantons and antinstantons contributions are independent. Noticing that,
\begin{align}
\label{c2s3e11}
\sum^{\infty}_{n = 0}\frac{\left(\mu e^{-\frac{I}{e^2}}\right)^{n}}{n!}\prod_{i=1}^{n}\int d^3 x^{i}_0 e^{ig\phi(x^{i}_0)} = e^{\mu e^{-\frac{I}{e^2}} \int d^3 x e^{ig\eta(x)}},
\end{align}
we get the total contribution of instantons and antinstantons:
\begin{align}
\label{c2s3e12}
\sum^{\infty}_{n_{+},n_{-} = 0}\frac{\left(\mu e^{-\frac{I}{e^2}}\right)^{n_{+} + n_{-}}}{n_{+}! n_{-}!}\prod_{i=1}^{n_{+}}\prod_{j=1}^{n_{-}}\int d^3 x^{i}_0 \int d^3 x^{j}_0  e^{ig\eta(x^{i}_0)}e^{-ig\eta(x^{j}_0)} = e^{2\mu e^{-\frac{I}{e^2}} \int d^3 x \cos{g\eta(x)}}.
\end{align}
The effective IR theory is, therefore:
\begin{align}
\label{c2s3e13}
{\cal Z}_{\eta} =  \int {\cal D} \eta   e^{-\int_{{\cal M}_3} \left(\frac 12 d\eta \wedge {}^\ast d\eta \right) + 2\mu e^{-\frac{I}{e^2}} \int d^3 x \cos{g\eta(x)}}.
\end{align}

A very natural question regards the fate of the field $A$. The instanton gas discussed above gives a mass to the scalar field with the value $2g^2\mu e^{-\frac{I}{e^2}}$, and also produces self-interaction terms. We know that in $3D$ a massive scalar field is dual to a massive 2-form (remember from section \ref{sec:dual} that for massive theories the duality relation is $p+q+1=D$, where $p$ and $q$ are the ranks of the dual forms and $D$ is the spacetime dimension). Hence, we have here the rank jump phenomenon.

We are going to show now that this system can also be analyzed via the JTA. This is a new result that supports the generality of the JTA. It is possible to represent the collective behavior of the instantons as a condensate (described by a continuous field) even in the dual picture. This would correspond to the DJTA discussed in the previous section. Indeed, let us go back to (\ref{c2s3e05}). The DJTA consists in taking $J_g$ as a field and to formally consider the sum over surfaces as an integral over $J_g$. The question at this point is: what is the form of the functional $S_g(J_g)$ such that the integration over $J_g$ gives us the effective theory (\ref{c2s3e13})? $J_g$ is an auxiliary field since it does not have derivatives in the action and, hence, an integration over $J_g$ means solving the equation:
\begin{align}
\label{c2s3e14}
\frac{\delta S_g}{\delta J_g} = ig\eta.
\end{align}
We want to determine $J_g$ as a function of $\eta$ such that substituting it into the action we get the cosine term. Hence, we must solve (\ref{c2s3e14}) in such a way that:
\begin{align}
\label{c2s3e15}
-ig\int d^3 xJ_g(x)\eta(x) + S_g(J_g) = -\epsilon \int d^3 x \cos{g\eta(x)},
\end{align}
where we have defined $\epsilon \equiv 2\mu e^{-\frac{I}{e^2}}$. Differentiating (\ref{c2s3e15}) with respect to $\eta$ and using (\ref{c2s3e14}) we obtain:
\begin{align}
\label{c2s3e16}
J_g(x) = i\epsilon \sin{g\eta(x)}.
\end{align}
Substituting in (\ref{c2s3e15}) we get the complete form of the action $S_g(J_g)$:
\begin{align}
\label{c2s3e17}
S(J_g) =\int d^3 x \left( J_g\; \textrm{arcsinh}\left(\frac{J_g}{\epsilon}\right) - \epsilon \sqrt{1 + \frac{J_g^2}{\epsilon^2}} \right).
\end{align}

Now it is easy to understand what happened in the original model (\ref{c2s3e02}) in terms of $A$. The JTA is implemented in its original form, \emph{i.e.}, $\Sigma_g$ becomes a field and the new fundamental field describing the condensed phase is the brane and gauge invariant combination $B \equiv dA + g{}^\ast \Sigma_g$ such that $g {}^\ast J_g \equiv H = dB$: in this way, the brane and gauge symmetries are realized in a hidden fashion in $B$. The sum over branes becomes an integral over $\Sigma_g$, which together with the integral over $A$ becomes effectively an integration over the brane and gauge invariant 2-form $B$. The effective theory is then:
\begin{align}
\label{c2s3e18}
{\cal Z}_B =  \int {\cal D} B  e^{-\int_{{\cal M}_3} \left(\frac 12 B \wedge {}^\ast B  \right) - \int d^3 x  \left(\frac{{}^\ast H}{g}\; \textrm{arcsinh}\left(\frac{{}^\ast H}{g\epsilon}\right) - \epsilon \sqrt{1 + \frac{{}^\ast H^2}{g^2\epsilon^2}}\right)}.
\end{align}
This is the result recently obtained by Polyakov \cite{Polyakov:1996nc}. It describes the dynamics of the relevant degrees of freedom in the IR limit of the theory (\ref{c2s3e01}). This theory describes a massive Kalb-Ramond field with mass $g^2\epsilon$, the same mass of the scalar field, since they are dual.

Notice that the action (\ref{c2s3e17}) involves a multivalued function (the inverse of a periodic function). To go to the corresponding single-valued representation \cite{mvf}, we must reveal the different branches of this function. For any complex number $z$:
\begin{align}
\label{c2s3e19}
\textrm{arcsinh}\left(z\right) = \ln\left(z + \sqrt{z^2 +1}\right) + 2i\pi n; \;\;\; n \in \mathds{Z},
\end{align}
where $n$ parametrizes the different branches. We can promote this function to the functional category and in this case $n$ becomes a distribution such that its integral in a region involving the singularity defined by the distribution is an integer (imagine a lattice where $z$ assumes complex values on the sites while $n$ assumes integer values and then extrapolate to the continuum). This is the definition of a 0-current. Hence, we have:
\begin{align}
\label{c2s3e20}
\frac{{}^\ast H}{g}\; \textrm{arcsinh}\left(\frac{{}^\ast H}{g\epsilon}\right) = \frac{{}^\ast H}{g}\; \ln\left(\frac{{}^\ast H}{g\epsilon} + \sqrt{\left(\frac{{}^\ast H}{g\epsilon}\right)^2 + 1}\right) + \frac{2i\pi n}{g}{}^\ast H \Lambda ; \;\;\; n \in \mathds{Z}.
\end{align}
A sum over the branches must be included in the definition of the partition function when we write it in its single-valued representation. This sum becomes a sum over the 0-surfaces Poincar\`e -dual to the 0-currents $\Lambda$ which define the branches. These are internal defects of the system and have their origin in the periodicity of the field $\eta$ (which is an angle, since it is the argument of a cosine); they are analogous to the flux loops we have encountered in the superconductor in section \ref{sec:1}. Remember that in that context, in the calculation of the confining potential, the loops were incorporated into a sum over branes whose boundary were the monopole world-lines. The same phenomenon will happen here in a dual version: when evaluating the Wilson loop associated to an electric charge, we shall need to express this charge in terms of its electric Dirac brane $\Sigma_e$ (${}^\ast J_e = d {}^\ast \Sigma_e$). In this case, the sum over the different branches will be translated into a sum over surfaces whose boundary are the electric currents $J_e$. This observation was originally made by Polyakov in \cite{Polyakov:1996nc}. Notice that also here the consistency of the formulation of the theory requires the quantization of the external electric charges as can be seen from the last term of (\ref{c2s3e20}): $e = \frac{2i\pi n}{g}$.

Explicitly, inserting the Wilson loop $W(C) = e^{ie(A, J_e)}$ in (\ref{c2s3e02}) and summing over the instanton gas we can obtain the VEV of the Wilson loop in this system. It will be expressed in terms of the brane $\Sigma_e$ due to the rank jump (the same procedure used in (\ref{c2s2e07})). The expression for the VEV of the Wilson loop in terms of $B$ is:
\begin{align}
\label{c2s3e21}
\langle W(C) \rangle =  \sum_{\{{\cal B}_{e}\}} {}' \int {\cal D} B  e^{-\frac 12\left( B, B  \right) - \int d^3 x  \left(\frac{{}^\ast H}{g}\; \ln\left(\frac{{}^\ast H}{g\epsilon} + \sqrt{\left(\frac{{}^\ast H}{g\epsilon}\right)^2 + 1}\right) - \epsilon \sqrt{1 + \frac{{}^\ast H^2}{g^2\epsilon^2}}\right) + ie(B, \tilde{\Sigma}_e)},
\end{align}
where the sum is taken over the surfaces ${\cal B}_{e}$ Poincar\`e -dual to $\tilde{\Sigma}_e$ constrained by ${}^\ast J_e = d {}^\ast \tilde{\Sigma}_e$, which is an external charge here, and we have defined the Stuckelberg-like electric brane invariant $\tilde{\Sigma}_e = \Sigma_e + d^{\dagger} \Lambda$. This brane invariant is constructed with the help of the branches and the sum over the branches has become the sum over $\tilde{\Sigma}_e$. Notice that, due to the minimal coupling with the massive Kalb-Ramond field $B$, the electric brane invariant carries energy, being an observable. As in the superconductor, there is also here a spontaneous breaking of the brane symmetry. This model, as it is well known, exhibits confinement in the sense that the VEV of the Wilson loop obeys an area law \cite{Polyakov:1975rs}. The important point we emphasize here is that, like in the superconductor, a signature of the confinement phenomenon is the spontaneous breaking of the brane symmetry. Indeed, we can consider (\ref{c2s3e20}) as an expansion in derivatives of $B$. It is a good approximation for the IR region to consider only the first order in this expansion if $B$ varies slowly. We assume that this is the case. Hence,
\begin{align}
\label{c2s3e22}
\int d^3 x  \left(\frac{{}^\ast H}{g}\; \ln\left(\frac{{}^\ast H}{g\epsilon} + \sqrt{\left(\frac{{}^\ast H}{g\epsilon}\right)^2 + 1}\right) - \epsilon \sqrt{1 + \frac{{}^\ast H^2}{g^2\epsilon^2}}\right) \approx \int d^3 x \frac{{}^\ast H^2}{2g^2\epsilon},
\end{align}
and, except for an overall constant, we get the massive Kalb-Ramond model with minimal coupling:
\begin{align}
\label{c2s3e23A}
\langle W(C) \rangle \approx  \sum_{\{{\cal B}_{e}\}} {}' \int {\cal D} B  e^{-\int_{{\cal M}_3} \frac 12 \left( B \wedge {}^\ast B  + \frac{1}{g^2\epsilon} H \wedge {}^\ast H \right) + i(B, \tilde{\Sigma}_e)}.
\end{align}
As discussed by Polyakov \cite{Polyakov:1996nc} this approximation suffices to identify the confinement phenomenon. This shows that the determining factor for the confinement of external electric charges embedded in the instanton gas is the mass term for the Kalb-Ramond field $B$ and this, in turn, is responsible for the spontaneous breaking of the electric brane symmetry.

\section{Application III - Radiative corrections in QED$_3$ as a condensation phenomenon}
\label{sec:3}

The effective theory describing QED$_3$ at low energies is the Maxwell-Chern-Simons theory (MCS) \cite{mcs}, which includes the lowest order terms from an expansion in the inverse fermion mass. This is a good approximation provided the fermions are very massive (notice, however, that even for massless fermions the Chern-Simons (CS) term \cite{Chern:1974ft} is induced due to the parity anomaly, in this case, by the Pauli-Villars regulator fermions \cite{Redlich:1983kn} - see also the earlier discussion in \cite{Niemi:1983rq}). In saying that some excitations are very massive relatively to other possible excitations of the system, we are effectively considering that they are dynamically inert. However, their presence affects the states of the system through quantum fluctuations and this disturbs the propagation of lighter particles. In the case of QED$_3$ the quantum fermionic fluctuations induce the CS term which, in turn, introduces inertia in the electromagnetic propagations and the photons acquire mass. This is similar to the Higgs Mechanism in the superconductor and hence it is natural to suppose that an effective description of the system can be made in terms of a condensate.

The duality between the MCS theory and the self-dual model (SD) \cite{sd} has as a particular case the duality between the Maxwell theory and the massless scalar field theory, which we have encountered in the previous section in the analysis of the Polyakov Model. In that case, we saw that a system described by a massless scalar field minimally coupled to the instanton gas is effectively described by a condensate whose excitations are spinless massive bosons and we say that the scalar field acquired mass due to the instanton gas. Here we have something similar: the Maxwell theory minimally coupled to massive fermions is effectively described by a condensate whose excitations are massive vector particles described by the MCS theory. In the previous section, we also saw that the dual theory is expressed in terms of a massive Kalb-Ramond field, realizing the rank jump phenomenon that characterizes the JTA when there is mass generation. It is therefore natural to conjecture that the same happens here, that is, there must be a dual formulation of the radiative corrections and, by duality, this formulation should connect the massless scalar theory to the SD theory realizing the rank jump phenomenon as depicted in the following schematic picture:
\begin{align}
\label{c2s3e23}
 \xymatrix{*+[F-:<10pt>]{\textrm{Maxwell}}+\ar[rrr]^-{\textrm{Quantum fluctuations}}\ar[dd] & & &
*+[F-:<10pt>]{\textrm{MCS}} \ar[dd]\\
 &  & \\
*+[F-:<10pt>]{\textrm{Scalar}} \ar[uu]^-{\textrm{Duality}} \ar[rrr]_-{\textrm{JTA}}& & &
*+[F-:<10pt>]
{\textrm{SD}}  \ar[uu]_-{\textrm{Duality}}\\}
\end{align}

We are going to study now the Scalar-SD connection indicated in the bottom of the above diagram. Since fermions are minimally coupled to the Maxwell gauge potential, the expected structure of the dual scalar theory features a non-minimal coupling:
\begin{align}
\label{c2s3e24}
{\cal Z}_{\phi} =  \sum_{\{{\cal B}_{e}\}} \int {\cal D} \phi   e^{-\int_{{\cal M}_3} \left(\frac 12 (d\phi + e{}^\ast \Sigma_e) \wedge {}^\ast (d\phi + e{}^\ast \Sigma_e) \right) - S_e(J_e)},
\end{align}
where $\Sigma_e$ are 2-currents such that ${}^\ast J_e = d{}^\ast \Sigma_e$ is the electric current and ${\bf P}({\cal B}_e) = {}^\ast \Sigma_e$. We can use the flux-current relation to rewrite this expression in terms of other variables. Inserting an identity in the form of an integral of a delta, we can write the partition function as:
\begin{align}
\label{c2s3e25}
{\cal Z}_{\phi} =  \sum_{\{{\cal B}_{e}\}} \int {\cal D} \phi \int {\cal D} H \delta({}^\ast H -{}^\ast \Sigma_e)  e^{-\int_{{\cal M}_3} \left(\frac 12 (d\phi + e{}^\ast H) \wedge {}^\ast (d\phi + e{}^\ast H) \right) - S_e(d{}^\ast H)},
\end{align}
and using now the GPI (\ref{c1s2eB}),
\begin{align}
\label{c2s3e26}
\sum_{\{{\cal B}_{e}\}} \delta({}^\ast H -{}^\ast \Sigma_e) = \sum_{\{{\cal F}_{e}\}} e^{-2\pi i \int_{{\cal M}_3} {}^\ast \omega \wedge {}^\ast H},
\end{align}
where ${\cal F}_{e}$ are 1-branes Poincar\`e -dual to the 1-current $\omega$, the resulting action reads:
\begin{align}
\label{c2s3e27}
S = \frac 12 \left( d\phi + e{}^\ast H, d\phi + e{}^\ast H \right) + 2\pi i \left( {}^\ast \omega,  H \right) + S_e(d{}^\ast H).
\end{align}
Redefining:
\begin{align}
\label{c2s3e28}
{}^\ast H \rightarrow f - \frac 1e d\phi,
\end{align}
we get:
\begin{align}
\label{c2s3e29}
S = \frac{e^2}{2} \left( f, f \right) + 2\pi i \left( \omega,  f \right) - \frac{2\pi i}{e} \left( d^{\dagger}\omega,  \phi \right) + S_e(d{}^\ast f).
\end{align}
Notice that now $\phi$ is a Lagrange multiplier imposing the constraint:
\begin{align}
\label{c2s3e30}
d {}^\ast \omega = 0,
\end{align}
that is, the fluxes are closed. This constrains the sum over ${\cal F}_{e}$ to be taken only over closed surfaces. Hence, we can introduce surfaces ${\cal G}_{e}$ which have the surfaces ${\cal F}_{e}$ as boundaries, through the introduction of a 2-current $\Lambda$ Poincar\`e -dual to ${\cal G}_{e}$ such that ${}^\ast \omega = d{}^\ast \Lambda$. The partition function now reads:
\begin{align}
\label{c2s3e31}
{\cal Z}_{f} =  \sum_{\{{\cal G}_{e}\}} \int {\cal D} f e^{-\int_{{\cal M}_3} \left(\frac{e^2}{2} f \wedge {}^\ast f  + 2\pi i f\wedge d{}^\ast \Lambda \right) - S_e(d{}^\ast f)}.
\end{align}
The condensate described by the MCS theory breaks $P$ and $T$ and hence, if we want to represent the same effect in the dual formulation, this information must be present. Being a property of the condensate, the breaking of these symmetries should be contained in the action $S_e(d{}^\ast f)$, which is seen as a derivative expansion since we are looking for an effective description of the low energy excitations of the system. Each term must be a brane invariant which, in the present representation, means that the terms in $S_e$ are invariant under $f \rightarrow f + d\chi$ where $\chi$ is a 0-form. The first term in the derivative expansion satisfying all these requirements is the CS term. Therefore, $S_e$ has the form:
\begin{align}
\label{c2s3e32}
S_e = i\theta \int_{{\cal M}_3} f \wedge df,
\end{align}
where $\theta$ is a phenomenological parameter (notice also that the CS term does not involve the dual Hodge star operator in its definition and hence \emph{it does not depend on the metric}, since it is naturally a 3-form, being therefore classified as a \emph{topological term}). We have then:
\begin{align}
\label{c2s3e33}
{\cal Z}_{f} =  \sum_{\{{\cal G}_{e}\}}  \int {\cal D} f e^{-\int_{{\cal M}_3} \left(\frac{e^2}{2} f \wedge {}^\ast f  + i\theta f \wedge df + 2\pi i f\wedge d{}^\ast \Lambda \right) }.
\end{align}
Notice that again the non-minimally coupled field has experienced a rank jump. Notice also that the flux-like defects are present in the form of a minimal coupling. These objects parametrize the phases of the system: if they proliferate and become a continuous distribution, the sum over ${\cal G}_{e}$ becomes an integral over $\Lambda$ and the last term imposes the constraint $df=0$, which implies that $f = d\phi$ and we recover the scalar field theory. On the other hand, if we formally set to zero the contributions of the internal defects $\omega$ we obtain the SD theory:
\begin{align}
\label{c2s3e34}
{\cal Z}_{SD} =  \int {\cal D} f e^{-\int_{{\cal M}_3} \left(\frac{e^2}{2} f \wedge {}^\ast f  + i\theta f \wedge df  \right) }.
\end{align}
As we know, a similar reasoning can be applied to the original theory written in terms of current variables instead of the Poisson-dual flux ones. There the parameter controlling the condensation is the electric brane $\Sigma_e$ (Poisson-dual to $\omega$). As we have discussed before, the Poisson-dual currents $\Sigma_e$ and $\omega$ have a kind of order-disorder relation. Indeed, we see from (\ref{c2s3e24}) that if we dilute $\Sigma_e$ (or, equivalently, if we proliferate $\omega$), we obtain the scalar field theory. On the other hand, if $\Sigma_e$ proliferates becoming a continuous distribution, then the sum over ${\cal B}_{e}$ becomes an integral over $\Sigma_e$ and this, together with the integral over $\phi$, becomes effectively an integral over the invariant $ef \equiv d\phi + e{}^\ast \Sigma_e$ and we reobtain the SD theory (\ref{c2s3e34}). This is the dual of the MCS theory as one can easily verify using the methods of section \ref{sec:dual}. The field $f$ has mass $\frac{e^2}{2\theta}$ ($e^2$ has dimension of mass). The MCS theory is obtained in lowest order in the inverse fermion mass integrating the fermions in the partition function of QED$_3$. What we have seen here is that the vacuum with the quantum fluctuations can be expressed in terms of a condensate that breaks $P$ and $T$. This condensate has the effect of generating a mass to the photon through the CS term. The similarity with the Higgs Mechanism is clear. Notice also that the complete transition to the SD theory (or to its dual, MCS theory) is only established in the limit where a complete condensation of the electric defects (or a complete dilution of the fluxes, depending on which picture we are working) occurs.

We are going to analyze now the Maxwell-MCS connection indicated in the top of the diagram (\ref{c2s3e23}). In the analysis of the Polyakov Model in the previous section, we have seen that it was possible to interpret the modification in the scalar theory due to the instanton contribution through a condensate, represented by a minimal coupling, and an action for the currents that was nothing more than the Legendre transform of the potential (see (\ref{c2s3e15})). In the present case, we have the Maxwell theory minimally coupled:
\begin{align}
\label{c2s3e35}
{\cal Z}_{A} =  \sum_{\{{\cal B}_{e}\}}  \int {\cal D} A e^{-\int_{{\cal M}_3} \left(\frac 12 dA \wedge {}^\ast dA  - ie A \wedge {}^\ast J_e \right) - S_e(J_e)},
\end{align}
that is the dual formulation of (\ref{c2s3e24}). If we want the coupling with the current $J_e$, \emph{after its condensation}, to represent the effect of the fermions, then the form of $S_e$ must be:
\begin{align}
\label{c2s3e36}
-ie \int_{{\cal M}_3}  A \wedge {}^\ast J_e  + S_e(J_e) =  \ln \det (\Dslash + M),
\end{align}
such that determining $J_e$ as a function of $A$ and inverting, it is possible, in principle, to determine the form of the action $S_e$. In (\ref{c2s3e36}), $M$ is the mass of the fermions and $\Dslash = \gamma^\mu(\partial_\mu + ieA_\mu)$ is Dirac's covariant derivative. We are only interested in low energy excitations and hence we are only going to consider the first order term in the derivative expansion of the right hand side of the above equation. This is the CS term:
\begin{align}
\label{c2s3e37}
\ln \det (\Dslash + M) \approx i \frac{e^2}{8\pi} \frac{M}{|M|} \int_{{\cal M}_3} A \wedge dA.
\end{align}
In this approximation, we immediately determine $J_e$:
\begin{align}
\label{c2s3e38}
{}^\ast J_e = d{}^\ast \Sigma_e = -\frac{e}{4\pi} \frac{M}{|M|} dA.
\end{align}
Substituting in (\ref{c2s3e36}) we determine $S_e$:
\begin{align}
\label{c2s3e39}
S_e(J_e) = i2\pi \frac{M}{|M|} \int_{{\cal M}_3} {}^\ast \Sigma_e \wedge d{}^\ast \Sigma_e.
\end{align}
Notice that this is a formal analysis and it only makes sense if $J_ e$ and $\Sigma_e$ are (continuous) fields. $S_e$ is a brane invariant that depends only on $J_e$. If we compare the result above with (\ref{c2s3e32}), it seems that we can make the following identification, $\theta = 2\pi \frac{M}{|M|}$. However, since there is a large arbitrariness in the passage to the continuum via the JTA, nothing guarantees that this identification is exact. It turns out that with the introduction of magnetic defects in the system, brane symmetry consistency seems to fix the value of the $\theta$ parameter (see \ref{sec:5}). Note, however, that this is a different system than the one considered here since the presence of instantons will affect the evaluation of the fermionic determinant, whose expression is not known in this case. The important result at this point is the identification of the CS structure of $S_e$, adequate to simulate the fermionic lowest energy effective contribution as seen as a condensate breaking the $P$ and $T$ symmetries. This is essentially the result that some of us reported in \cite{Gamboa:2008ne}. In section \ref{sec:5}, we shall see how these concepts allow us to approach the issue of defining the MCS theory in the presence of magnetic defects.

\section{Application IV - Condensation of the condensate: the hierarchy structure in the Fractional Quantum Hall Effect}
\label{sec:4}

The analysis presented in the previous section finds perfect consonance in the theory of the Fractional Quantum Hall Effect (FQHE). Indeed, we are going to show now that it is possible to obtain the hierarchy structure proposed by Haldane \cite{Haldane:1983xm} from effective theories derived via the JTA. While what we are going to discuss is essentially contained in the review paper written by Wen \cite{Wen:1995qn}, it is important to notice that here this procedure is seen as a particular application of the formalism we have developed.

The Quantum Hall Effect occurs in electronic systems at low temperatures with effective planar dynamics subjected to an intense magnetic field orthogonal to the plane. The characteristic feature of the Quantum Hall Effect is the quantization of the Hall conductivity, which represents the electronic response of the system to the applied electromagnetic field. Explicitly, for an electric current $J_e$ in the plane in the presence of an electromagnetic field $F = dA$, we have:
\begin{align}
\label{c2s3e39A}
e{}^\ast J_e = \sigma F,
\end{align}
where:
\begin{align}
\label{c2s3e39B}
\sigma = \frac{\nu e^2}{2\pi},
\end{align}
is the Hall conductivity (in units of $\hbar = 1$), $\nu$ is the so-called filling fraction parameter and the energetically favorable states of the system correspond to incompressible states where this parameter assumes integer or fractional values. The integer case, called Integer Quantum Hall Effect, can be understood with a model of independent electrons subjected to an external magnetic field, the Landau theory. However, to explain the FQHE the interactions between the electrons cannot be neglected. Indeed, this is a system of strongly correlated electrons and it is not productive to insist in regarding the electrons as the relevant degrees of freedom in this case. The system constitutes a \emph{topological fluid} and its excitations (quasiparticles) are the new degrees of freedom. These excitations are characterized by fractional quantum numbers with respect to the electrons and this is directly related to the fractional values of the conductivity. Laughlin \cite{Laughlin:1983fy} proposed wave functions to define these states. For the ground state with filling fraction $\nu = \frac{1}{m}; \; m \in \mathds{Z}$, for example, the wave function has the form:
\begin{align}
\label{c2s3e39C}
\Psi = \prod_{i<j} (z_i - z_j)^m e^{-\frac 14 \sum_i |z_i|^2},
\end{align}
where $z_i = x_i + i y_i$ denotes the position of the $i$-th electron in the plane and $m$ should be an odd number in order for the wave function to be antisymmetric under particle exchange (see \cite{johnson} for a brief didactic exposition). We can construct an effective action for this state. Notice that the equation (\ref{c2s3e39A}) has exactly the same structure of (\ref{c2s3e38}) and hence, a way of describing the electronic linear response to an external electromagnetic field is given by the action:
\begin{align}
\label{c2s3e39D}
S = -ie \int_{{\cal M}_3} A \wedge {}^\ast J_e + i m\pi \int_{{\cal M}_3} {}^\ast \Sigma_e \wedge d{}^\ast \Sigma_e,
\end{align}
where ${}^\ast J_e = d{}^\ast \Sigma_e$. In this action we must interpret $\Sigma_e$ as a field representing the electronic condensate that constitutes the Hall fluid. Extremizing this action with respect to $\Sigma_e$ we obtain (\ref{c2s3e39A}) with $\nu = \frac 1m$. The dynamical field in (\ref{c2s3e39D}) is $\Sigma_e$, being $A$ an external field, therefore the partition function of the system has the form:
\begin{align}
\label{c2s3e39E}
{\cal Z} = \int {\cal D} \Sigma_e e^{-S(\Sigma_e, A)}.
\end{align}
We want to understand how the quasiparticles emerge in this system. The quasiparticles represent small excitations around the fluid ground state represented by (\ref{c2s3e39C}). The introduction of quasiparticles disturbs the fluid and can be represented by the insertion of 1-currents coupled to the fluid field $\Sigma_e$. Hence, the action is modified to:
\begin{align}
\label{c2s3e39F}
S \rightarrow S' = S - i2\pi q_1 \int_{{\cal M}_3} {}^\ast \Sigma_e \wedge {}^\ast J_{1},
\end{align}
where $q_1$ is the charge of the current $J_1$ with respect to the condensate field $\Sigma_e$. It can be shown that $q_1$ must be quantized in integer values, but for simplicity we shall take $q_1 = 1$ from now on. The proposal of Haldane \cite{Haldane:1983xm} is that the very presence of these excitations is the prelude of the transition to other Hall states characterized by another filling fractions. The excitations disturb the energetic balance and due to the emergence of more and more excitations, the system eventually becomes incompressible again, reaching a new energetically favorable configuration which establishes a new condensed state for the system. This can be easily understood via the DJTA. We introduce into the action a term characterizing the quasiparticles:
\begin{align}
\label{c2s3e39G}
S' \rightarrow  S^{(1)}= S - i2\pi \int_{{\cal M}_3} {}^\ast \Sigma_e \wedge {}^\ast J_{1} + i m_1 \pi \int_{{\cal M}_3} {}^\ast \Sigma_{1} \wedge d{}^\ast \Sigma_{1}.
\end{align}
The partition function acquires now the form:
\begin{align}
\label{c2s3e39H}
{\cal Z}^{(1)} = \sum_{\{{\cal B}_{1}\}} \int {\cal D} \Sigma_e e^{-S^{(1)}(\Sigma_e, \Sigma_{1}, A)},
\end{align}
where ${\cal B}_{1}$ are the surfaces Poincar\`e -dual to $\Sigma_{1}$. The last term in (\ref{c2s3e39G}), being a term of self-intersection between $p$-currents at this stage of the construction, is directly related to the statistics of the quasiparticles which results to be fractional \cite{Wilczek:1990ik}. Following the DJTA we consider that these excitations proliferate forming a condensate. This is a ``condensate over another condensate''. In this case the sum over ${\cal B}_{1}$ becomes an integral over $\Sigma_{1}$ such that we have now two dynamical fields in the theory, $\Sigma_e$ and $\Sigma_{1}$. Extremizing the action with respect to these fields we get the following equations of motion:
\begin{align}
\label{c2s3e39I}
e{}^\ast J_e &= \sigma dA + \frac{e}{m} {}^\ast J_{1},\\
\label{c2s3e39Ia}
{}^\ast J_{1} &= \frac{1}{m_1} {}^\ast J_e.
\end{align}
Substituting (\ref{c2s3e39Ia}) in (\ref{c2s3e39I}), we obtain:
\begin{align}
\label{c2s3e39J}
e{}^\ast J_e = \sigma^{(1)} dA,
\end{align}
where:
\begin{align}
\label{c2s3e39K}
\sigma^{(1)} = \frac{e^2}{2\pi} \left(\frac{1}{m-\frac{1}{m_1}}\right),
\end{align}
is the new Hall conductivity. This is the first level in the Hall hierarchy. It is evident that the procedure can be continued. We can introduce excitations $J_2$ in the condensate minimally coupled to the field $\Sigma_{1}$. These excitations eventually condense producing the next level in the Hall hierarchy with:
\begin{align}
\label{c2s3e39L}
\sigma^{(2)} = \frac{e^2}{2\pi} \left(\frac{1}{m-\frac{1}{m_1-\frac{1}{m_2}}}\right).
\end{align}
Generalizations of this model can be made in order to consider more complicated situations. The example discussed above regards the FQHE in only one material layer. For a more general case of multiple layers the action of the system has the form:
\begin{align}
\label{c2s3e39M}
S = -ie  t_I \int_{{\cal M}_3} A \wedge {}^\ast J^I + i \pi K_{IJ} \int_{{\cal M}_3} {}^\ast \Sigma^I \wedge d{}^\ast \Sigma^J,
\end{align}
where the vector $t_I$ codifies the information about the current carriers in the other layers of the sampling (for only one layer we have $t_I = \delta_{I1}$) and is called charge vector. The matrix $K_{IJ}$ gives information about the filling fraction through the matrix expression $\nu = {\bf t}^{T} K^{-1} {\bf t}$. The Hall fluid is a system featuring topological order. This kind of system does not have a description in terms of order parameters in the usual sense. Indeed, the states of this topological quantum fluid are characterized by the topological action (\ref{c2s3e39M}) defined by the quantities $t_I$ and $K_{IJ}$. This is an almost complete description of the Hall fluid (for systems defined over curved surfaces it may be necessary the introduction of a new topological quantity called spin vector, see \cite{Wen:1995qn} for details).

Notice the crucial role of the JTA in this construction. Although the ``condensate over another condensate'' idea was already present in the literature, it is important to notice that the JTA allows the embedding of this concept into a more general formalism. Notice also that the system does not exhibit brane symmetry breaking nor confinement.

\section{Application V - Instantons in the Maxwell-Chern-Simons theory}
\label{sec:5}

The formalism we have developed here also allows us to approach a controversial issue in the literature: the definition of the MCS theory in the presence of magnetic defects. It seems that there is no consensus regarding the definition of the MCS theory with non-minimal coupling: the problem is to try to directly incorporate an external instanton non-minimally coupled into the theory. In doing so we explicitly break the magnetic brane symmetry due to the presence of the CS term. Furthermore, a non-conserved electric current seems to emerge in the system. This current is localized in the magnetic Dirac brane and hence, the latter would become observable. Therefore, by naively formulating the MCS theory with non-minimal coupling we get an inconsistent result.

In the literature this problem was firstly approached in \cite{Henneaux:1986tt}, where the conservation of the electric current (and the gauge invariance) was reobtained by the \textit{ad hoc} introduction of an external current. The problem was also approached in \cite{Pisarski:1986gr}, where it was first recognized that the Dirac brane would become observable in the presence of the CS term and that the resulting action of the system would be proportional to the distance between the instantons connected by this brane. In \cite{Affleck:1989qf} another important effect was observed: the CS term destroys the electric charge confinement. We have seen in section \ref{sec:2} that the Maxwell theory in $3D$ in the presence of magnetic defects confines electric charges minimally coupled. What was noticed in \cite{Affleck:1989qf} is that by adding a CS term in this theory (obtaining therefore the MCS theory), electric confinement is destroyed. We have seen that the origin of electric confinement lies in the magnetic condensate, hence a CS term suppresses the condensation of magnetic defects. The formulation of this problem on the lattice, using duality and working with the SD representation, was worked out in \cite{Diamantini:1993iu} and the confining behavior for the instantons, suggested by Pisarski \cite{Pisarski:1986gr}, was found in this context. This explains why these objects do not condense: they are confined. However, the same consistency problems persist: in the SD representation the SD field couples minimally to the Dirac brane and thus, the latter would be observable. Furthermore, the questions regarding the gauge symmetry are not answered since the SD theory does not have gauge invariance.

The approach we have developed here allows us to clarify these questions as some of us reported in \cite{Grigorio:2008gd}. We want to formulate the MCS theory in the presence of magnetic defects and the MCS theory, in turn, as we have seen in section \ref{sec:3}, can be regarded as describing an electric condensate. Starting from a phase where both the magnetic defects and the electric charges are discretely distributed, we shall be able to construct an adequate formulation of the system.

Thus, we want to study the Maxwell theory (\ref{c2s3e35}) in the presence of external magnetic defects. Consider then the following partition function:
\begin{align}
\label{c2s3e40}
{\cal Z}^g_{A}(J_g) =  \sum_{\{{\cal B}_{e}\}} \int {\cal D} A e^{-\int_{{\cal M}_3} \left(\frac 12 (dA + g{}^\ast\Sigma_g)\wedge {}^\ast (dA + g{}^\ast\Sigma_g)  - ie A \wedge {}^\ast J_e  + i\theta {}^\ast \Sigma_e \wedge d{}^\ast \Sigma_e\right)}.
\end{align}
where $\Sigma_e$ and $J_e$ are $p$-currents and $\theta$ is seen as a phenomenological parameter. The magnetic defects $\Sigma_g$ define the instantons which are simply the boundary of the surface Poincar\`e -dual to the 1-current $\Sigma_g$, that is, the instanton density is given by the 0-current $J_g$ such that ${}^\ast J_g = d {}^\ast \Sigma_g$. We are going to use the GPI to rewrite this theory. Inserting an identity in the form of a delta we have:
\begin{align}
\label{c2s3e41}
{\cal Z}^g_{A}(J_g) =  \sum_{\{{\cal B}_{e}\}} \int {\cal D} A \int {\cal D} H \delta({}^\ast H -{}^\ast \Sigma_e) e^{-\int_{{\cal M}_3} \left(\frac 12 (dA + g{}^\ast\Sigma_g)\wedge {}^\ast (dA + g{}^\ast\Sigma_g)  - ie A \wedge d{}^\ast H  + i \theta{}^\ast H \wedge d{}^\ast H\right)},\nonumber\\
\end{align}
and using (\ref{c2s3e26}), we obtain:
\begin{align}
\label{c2s3e42}
{\cal Z}^g_{A}(J_g) = \sum_{\{{\cal F}_{e}\}} \int {\cal D} A \int {\cal D} H e^{-\int_{{\cal M}_3} \left(\frac 12 (dA + g{}^\ast\Sigma_g)\wedge {}^\ast (dA + g{}^\ast\Sigma_g)  - ie A \wedge d{}^\ast H  + i\theta {}^\ast H \wedge d{}^\ast H -2\pi i {}^\ast \omega \wedge {}^\ast H\right)}.\nonumber\\
\end{align}
Now, we are going to integrate out $H$. The action that appears in (\ref{c2s3e42}) is:
\begin{align}
\label{c2s3e43}
S &=\int_{{\cal M}_3} \left( \frac 12 (dA + g{}^\ast\Sigma_g)\wedge {}^\ast (dA + g{}^\ast\Sigma_g)  - ie A \wedge d{}^\ast H\right.\nonumber\\
&\;\;\;\;  + i\theta {}^\ast H \wedge d{}^\ast H -2\pi i {}^\ast \omega \wedge {}^\ast H \Big).
\end{align}

The equation of motion of $H$ is:
\begin{align}
\label{c2s3e44}
 - ie dA  + 2i\theta d{}^\ast H - 2\pi i {}^\ast \omega = 0,
\end{align}
which tells us that $d {}^\ast \omega = 0$ and hence we can write ${}^\ast \omega = d{}^\ast \Lambda$ where $\Lambda$ is a 2-form. The solution for $H$ is therefore:
\begin{align}
\label{c2s3e45}
{}^\ast H =  \frac{e}{2\theta} A  + \frac{\pi}{\theta} {}^\ast \Lambda,
\end{align}
except for an exact differential form that does not contribute to the action. Hence, the action reads:
\begin{align}
\label{c2s3e46}
S = \int_{{\cal M}_3} \left(\frac 12 (dA + g{}^\ast\Sigma_g)\wedge {}^\ast (dA + g{}^\ast\Sigma_g)  - i\frac{e^2}{4\theta} \left(A + \frac{2\pi}{e} {}^\ast\Lambda\right) \wedge\left(dA + \frac{2\pi}{e} d{}^\ast\Lambda\right)\right),
\end{align}
and we get:
\begin{align}
\label{c2s3e47}
{\cal Z}^g_{A}(J_g) = \sum_{\{{\cal G}_{e}\}} \int {\cal D} A e^{-\int_{{\cal M}_3} \left(\frac 12 (dA + g{}^\ast\Sigma_g)\wedge {}^\ast (dA + g{}^\ast\Sigma_g)  - i\frac{e^2}{4\theta} \left(A + \frac{2\pi}{e} {}^\ast\Lambda\right) \wedge\left(dA + \frac{2\pi}{e} d{}^\ast\Lambda\right)\right)},
\end{align}
where the sum is taken over the surface configurations ${\cal G}_{e}$ Poincar\`e -dual to $\Lambda$. Defining now the brane invariant field $B$:
\begin{align}
\label{c2s3e48}
B \equiv A + \frac{2\pi}{e} {}^\ast\Lambda,
\end{align}
we have the equivalent form:
\begin{align}
\label{c2s3e49A}
{\cal Z}^g_{A}(J_g) = \sum_{\{{\cal G}_{e}\}} \int {\cal D} B e^{-\int_{{\cal M}_3} \left(\frac 12 (dB + g{}^\ast\Sigma_g - \frac{2\pi}{e} d{}^\ast\Lambda)\wedge {}^\ast (dB + g{}^\ast\Sigma_g - \frac{2\pi}{e} d{}^\ast\Lambda)  - i\frac{e^2}{4\theta} B \wedge dB\right)}.
\end{align}
In this representation all the information about the electric condensate is contained in the magnetic fluxes $d{}^\ast\Lambda$. These can be seen as the parameters controlling the condensation, exactly in the same way as we have seen in the case of the superconductor: diluting $d{}^\ast\Lambda$ we get deeply into the electric condensed phase. As in the case of the superconductor, we see that in the presence of external magnetic defects it is impossible to realize a complete electric condensation: this is forbidden by the brane symmetry \cite{jt-cho}. The best we can do is to dilute all the magnetic flux loops disconnected from the Dirac brane ${}^\ast\Sigma_g$. In this way, as long as the Dirac quantization condition is satisfied, the sum over the surfaces ${\cal G}_{e}$ (now involving only the ones connected to the Dirac brane) are translated into a sum over the different configurations of the surfaces $\Sigma_g$, exactly in the same way as we have seen before. Hence, we have:
\begin{align}
\label{c2s3e49}
{\cal Z}^g_{A}(J_g) = \sum_{\{{\cal B}_{g}\}} {}' \int {\cal D} B e^{-\int_{{\cal M}_3} \left(\frac 12 (dB + \frac{2\pi }{e}{}^\ast\tilde{\Sigma}_g)\wedge {}^\ast (dB + \frac{2\pi }{e}{}^\ast\tilde{\Sigma}_g) - i\frac{e^2}{4\theta} B \wedge dB\right)},
\end{align}
where ${\cal B}_{g}$ is the surface Poincar\`e -dual to the brane invariant $\tilde{\Sigma}_g \equiv {}^\ast\Sigma_g - d{}^\ast\Lambda$ and the sum is constrained such that ${}^\ast J_g = d {}^\ast \tilde{\Sigma}_g$. These are by construction brane invariant surfaces. The field $B$ is also a brane invariant. The theory is hence written only in terms of brane invariants but it has explicit or manifest gauge symmetry. We have then a spontaneous breaking of the brane symmetry in the same way we have seen in the superconductor and in the Polyakov Model: the internal defects $\Lambda$ of the system constitute together with the Dirac branes $\Sigma_g$ a brane invariant $\tilde{\Sigma}_g$ which hides the realization of the brane symmetry in the condensed regime and carries energy, being therefore an observable. Notice, however, that in this case the gauge symmetry was not spontaneously broken.

Notice that, thanks to the GPI, we have been able to map a discrete distribution (a gas) of electric currents $J_e$ into a distribution of closed magnetic fluxes $d{}^\ast\Lambda$. This is an order-disorder map. In starting with the diluted system we got a better control over what is happening in the system: we have started with the instantons $J_g$ as the only physical information regarding the magnetic content of the system (the magnetic Dirac brane $\Sigma_g$ is unphysical, only its boundary $J_g$ is physical). Through the GPI, the electric sector was mapped into a gas of magnetic loops. The components of this gas combine themselves with the Dirac brane $\Sigma_g$ and with the original gauge field $A$ revealing the physically relevant variables of this system, the brane invariants $\tilde{\Sigma}_g$ and $B$. Hence, we propose the theory (\ref{c2s3e49}) as the adequate model to describe the MCS system in the presence of magnetic defects.

To have a better understanding of the relation between this result and the one obtained by Henneaux and Teitelboim \cite{Henneaux:1986tt}, we go back to the action before the redefinition of $A$ in (\ref{c2s3e48}). Henneaux and Teitelboim worked at the level of the action and analyzed subsequently the Hamiltonian structure. The action postulated by them is related to the action (\ref{c2s3e46}). This action is written in terms of brane dependent variables. Making explicit the terms, we have:
\begin{align}
\label{c2s3e50}
S = \int_{{\cal M}_3} \left( \frac 12 (dA + g{}^\ast\Sigma_g)\wedge {}^\ast (dA + g{}^\ast\Sigma_g)  - i\frac{e^2}{4\theta} A \wedge dA - i\frac{e\pi}{\theta} A \wedge d{}^\ast\Lambda -i\frac{\pi^2}{\theta} {}^\ast\Lambda \wedge d{}^\ast\Lambda \right).
\end{align}

\begin{figure}[h]  % label dentro da caption para assegurar referencia correta no corpo do texto
       \centering
       \includegraphics[scale=0.5]{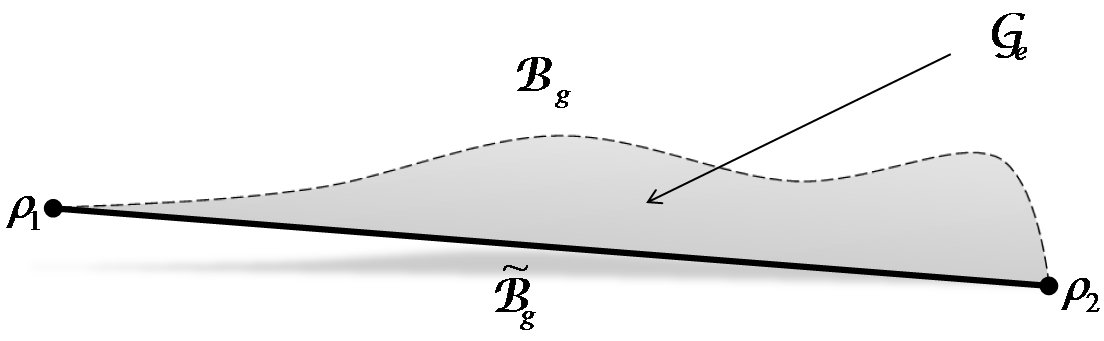}
       \caption{Two instantons $\rho_1$ and $\rho_2$ connected by a confining physical string ${\cal \tilde{B}}_{g}$ Poincar\`e -dual to $\tilde{\Sigma}_g$. Total current conservation is achieved by taking into account the induced electric current $\Sigma_g$, Poincar\`e -dual to the surface ${\cal B}_{g}$. These currents are boundaries of the surface ${\cal G}_{e}$, Poincar\`e -dual to $\Lambda$. \label{fig:3}}
\end{figure}

Only the first three terms correspond to the action postulated by Henneaux e Teitelboim. Notice, however, that the last term is necessary to maintain the brane symmetry, satisfying Elitzur's theorem. The third term corresponds to an interaction between the gauge field $A$ and the conserved current $d {}^\ast\Lambda$. Notice that this current is geometrically the boundary between $\Sigma_g$ and $\tilde{\Sigma}_g$, since it is defined by ${}^\ast\tilde{\Sigma}_g = {}^\ast\Sigma_g - d{}^\ast\Lambda$. In the formulation proposed by Henneaux and Teitelboim, $\Sigma_g$ represents an electric current induced in the Dirac brane. This current is not conserved since it has as boundary the instantons $J_g$. They introduced in an \textit{ad hoc} manner an external non-conserved current $\tilde{\Sigma}_g$, such that the total current taken as the difference between $\Sigma_g$ and $\tilde{\Sigma}_g$ is conserved. Within the interpretation we are presenting here $\Sigma_g$ is an unphysical object, the only physical object being $\tilde{\Sigma}_g$, which indeed represents a non-conserved current emanating from the instantons which constitute its boundary. And this is consistent with the gauge symmetry since it is not broken (explicitly or spontaneously). In the present formulation there is a natural interpretation for these currents as illustrated in FIG. \ref{fig:3}.

It is important to mention that one of the main results obtained by Henneaux and Teitelboim was the quantization of the topological mass, considered by them as an independent parameter. They obtained the relation:
\begin{align}
\label{c2s3e50A}
m = \frac{2\pi n}{g^2}; \;\;\; n\in \mathds{Z},
\end{align}
where $m$ is the topological mass and $g$ is the instanton charge. This result was obtained assuming that the Dirac brane carries an electric charge $mg$ and this must be subjected to the same quantization condition as any other electric charge. It must be pointed out that Pisarski \cite{Pisarski:1986gr} obtained a more restrictive quantization, viz. $m = \frac{4\pi n}{g^2}$ for $n\in \mathds{Z}$, but he considered the system at finite temperature, that is, with a compact time direction.

In the analysis presented here the mass is defined in terms of the charge $e$, which has the minimum value allowed by the dirac charge quantization: $e=\frac{2\pi}{g}$. This is the charge of the $P$ and $T$ symmetry breaking condensate. Thus, if we introduce an external electric charge $q$ in the system, brane symmetry demands this charge to be quantized in multiples $e$, that is, $q=ne$, with $n\in \mathds{Z}$. Now, notice that the system has an external electric current already which is carried by the magnetic brane as discussed after (\ref{c2s3e50}). The electric charge associated with this current is given by the coefficient of the third term of (\ref{c2s3e50}) and reads $q=\frac{e\pi}{\theta}$. The mass can be read from the CS coefficient to be given by $m=\frac{e^2}{2\theta}$. Combining these expressions and taking into account the quantization of $q$, we deduce that
\begin{align}
\label{c2s3e50C}
 m = \frac{qe}{2\pi} = \frac{2\pi n}{g^2}; \;\;\; n\in \mathds{Z},
\end{align}
which reproduces the mass quantization (\ref{c2s3e50A}); note that this quantization is a consequence of the brane symmetry.

A few observations are in order at this point. Notice that this mass quantization fixes the value of the $\theta$ parameter to be given by $\theta=\frac{\pi}{n}$. This suggests that the charges comprising the condensate have an anyonic character in the presence of instantons. Notice also that, with this value of $\theta$, the last term of (\ref{c2s3e50}) has a coefficient of $\pi n$. Since this it is a self-linking number, it informs us that the statistical nature of the instantons can be ferminonic or bosonic for $n$ odd or even, respectively. It is interesting to observe that for bosonic instantons the last term of (\ref{c2s3e50}) turns out to be innocuous, thus justifying its omission in the Henneaux and Teitelboim formulation in this case. Also in this case, the mass quantization becomes equivalent to the Pisarski quantization (since $n = 2m$ with $m\in \mathds{Z}$). Due to the phenomenological approach we have followed here, these conclusions are to be taken with caution and deserve a further investigation. Nevertheless we feel that these are important conclusions that follow from the brane symmetry of the theory.

We are going to discuss now the confining character of the theory (\ref{c2s3e49}) as interpreted by Pisarski. We already know that the spontaneous breaking of the brane symmetry is a defining factor of the confinement of the charges that define the boundary of this brane; in this case, the instantons. Let us reveal the observability of the brane invariant $\tilde{\Sigma}_g$ formulating the theory in the dual picture. The action defining (\ref{c2s3e49}) can be rewritten introducing an auxiliary field:
\begin{align}
\label{c2s3e51}
S =  \int_{{\cal M}_3} \left( i\left(dB + \frac{2\pi }{e}{}^\ast\tilde{\Sigma}_g \right)\wedge {}^\ast\Pi + \frac 12 \Pi\wedge {}^\ast\Pi - i\frac{e^2}{4\theta} B\wedge dB\right).
\end{align}

We can redefine $B$:
\begin{align}
\label{c2s3e52}
{}^\ast B \equiv \frac{2\theta}{e^2}(\Pi - C),
\end{align}
and the action in terms of $C$ and $f \equiv {}^\ast\Pi$ reads:
\begin{align}
\label{c2s3e53}
S = \int_{{\cal M}_3} \left( -i\frac{\theta}{e^2} C \wedge dC + i\frac{\theta}{e^2} f \wedge df + \frac 12 f \wedge {}^\ast f + \frac{2\pi i}{e} f \wedge {}^\ast \tilde{\Sigma}_g \right).
\end{align}

Hence, $C$ decouples from the theory carrying the gauge symmetry with it. This is simply the MCS-SD duality. The brane invariant $\tilde{\Sigma}_g$ couples minimally to the SD field $f$ revealing its physical character. This form of the action reveals the fact that there is a direct interaction between the brane invariants $\tilde{\Sigma}_g$ mediated by the massive SD field $f$. Indeed, integrating $f$ in the partition function:
\begin{align}
\label{c2s3e54}
{\cal Z}^g_f(J_g) = \sum_{\{{\cal B}_{g}\}} {}' \int {\cal D} f e^{-\int_{{\cal M}_3} \left(  i\frac{\theta}{e^2} f \wedge df + \frac 12 f \wedge {}^\ast f + \frac{2\pi i}{e} f \wedge {}^\ast \tilde{\Sigma}_g \right)},
\end{align}
we get the following effective theory for the brane invariants:
\begin{align}
\label{c2s3e55}
{\cal Z}^g(J_g) = \sum_{\{{\cal B}_{g}\}} {}' e^{-\int_{{\cal M}_3} \left( \frac{2\pi^2}{e^2} J_g\wedge \left(\frac{1}{\Delta + \left(\frac{e^2}{2\theta}\right)^2}\right) {}^\ast J_g - i\frac{\pi^2}{\theta} \tilde{\Sigma}_g \wedge \left(\frac{1}{\Delta + \left(\frac{e^2}{2\theta}\right)^2}\right) d\tilde{\Sigma}_g + \frac{e^2\pi^2}{2\theta^2} \tilde{\Sigma}_g \wedge \left(\frac{1}{\Delta + \left(\frac{e^2}{2\theta}\right)^2}\right) {}^\ast \tilde{\Sigma}_g\right)}.\nonumber\\
\end{align}

The effective action in the exponential is the same obtained in \cite{Diamantini:1993iu}, where the authors worked with the system defined on the lattice. The last term of the effective action, as we already know, is responsible for the spontaneous breaking of the brane symmetry and hence by the confining potential. Suppose that $J_g$ describes an instanton-antinstanton pair separated by a spacetime interval $L$. These instantons are points (events) in spacetime. Each term in the sum over branes in (\ref{c2s3e55}) represents the probabilistic weight attributed to the configurations of the brane invariant $\tilde{\Sigma}_g$ connecting these instantons. To explicitly evaluate this contribution we notice that this is exactly the same situation we have found in the calculation of the confining potential in the Abelian Higgs Model. In that case we considered a configuration of static monopoles, that led us to an effective $3D$ calculation. This is exactly the same calculation we need to do here. As in the case of the superconductor, we expect that the term in the sum (\ref{c2s3e55}) that contributes most is associated to the brane invariant with the shortest length connecting the instanton-antinstanton pair. This configuration corresponds to a straight line with length $L$. Hence, for this configuration of the brane invariant we have asymptotically:
\begin{align}
\label{c2s3e56}
e^{-\int_{{\cal M}_3} \left(\frac{e^2\pi^2}{2\theta^2} \tilde{\Sigma}_g \wedge \left(\frac{1}{\Delta + \left(\frac{e^2}{2\theta}\right)^2}\right) {}^\ast \tilde{\Sigma}_g\right)} \xrightarrow[L \rightarrow \infty]{} e^{-\sigma L},
\end{align}
where $\sigma = \frac{g^2 m^2}{8\pi} \ln\left(\frac{M^2 + m^2}{m^2}\right)$ is the string tension of the confining potential, with $m=\frac{e^2}{2\theta}$ being the topological mass and $g$ the instanton charge. Here $M$ is an UV cutoff. Following Pisarki's interpretation \cite{Pisarski:1986gr}, this result tells us that the probability of creation of a instanton-antinstanton pair is strongly suppressed if the separation $L$ between them is large. In particular, the probability of creation of an isolated instanton is zero. Hence, we say that they are confined. We see here that this is a direct consequence of the spontaneous breaking of the brane symmetry.

\section{Application VI - Monopoles in the Carroll-Field-Jackiw model}
\label{sec:6}

In this section we are going to approach an emblematic model that violates Lorentz and discrete spacetime symmetries in $(3+1)D$: the Carroll-Field-Jackiw model (CFJ) \cite{Carroll:1989vb}, defined by the gauge invariant action,
\begin{align}
 \label{c3s3e01}
  S_{CFJ}= \int d^4x\left( -\frac 14 F^{\mu\nu}F_{\mu\nu} + p_{\mu}\varepsilon^{\mu\nu\rho\sigma}A_{\nu}\partial_{\rho}A_{\sigma} \right).
\end{align}
The CS-like term is responsible for the breaking of Lorentz symmetry due to the presence of the constant vector $p_{\mu}$, which selects a preferred direction in each Lorentz reference frame. If $p_{\mu}$ is a vector it also breaks $CPT$ invariance, if it is a pseudo-vector it breaks $P$ and $CP$ invariance (but $CPT$ holds). Due to the asymmetry introduced by the Lorentz invariance violation (LIV), we are going to work in this section with the explicit tensorial index notation.

The majority of the studies about this model focus on considering it as a modification of the usual electromagnetism. An immediate consequence that emerges from the analysis of the dispersion relations of this model is the birefringence phenomenon: the propagation of an electromagnetic wave has different speeds for the two polarizations, even in the vacuum. This effect is parameterized by the Lorentz violating vector $p^{\mu}$ and hence, this effect can be used to define limits on the magnitude of this vector. In \cite{Carroll:1989vb} only a time-like $p^{\mu}$ was considered and it was argued that astronomical observations of polarized light and also geomagnetic data seem to exclude the possibility of a nonzero magnitude for $p^{\mu}$. However, more recently, the authors of \cite{Feng:2006dp} have claimed that they have found signs in the data of WMAP and BOOMERANG favoring a nonzero value. For the space-like case, astronomical observations seemed to support the idea that the universe is not isotropic \cite{Nodland:1997cc}, favoring a nonzero value for $p^{\mu}$, however, this result was contested in \cite{Wardle:1997gu}. For a recent discussion involving the different aspects of the Lorentz violation in the electromagnetic theory regarding the possibility of its detection through astronomical phenomena, see \cite{Kostelecky:2008be}.

Other studies focus on the formal aspects related to the consistency of the model as a quantum field theory in function of the Lorentz character of $p^{\mu}$ (see \cite{Andrianov:1998wj, Adam:2001ma}). It was observed that a time-like $p^{\mu}$ defines a model where microcausality and unitarity cannot be simultaneously satisfied. On the other hand, the field theory seems to be well defined for a space-like $p^{\mu}$.

We could simply postulate this model as being the true electromagnetism as done in \cite{Carroll:1989vb}. Since nowadays we understand the electromagnetism as a low energy effective theory, whose complete theory is the Standard Model (SM), it is more natural to think that the LIV in the electromagnetism is part of a more complete structure that violates the Lorentz symmetry. This is the idea of the Extended Standard Model (ESM) introduced by Colladay and Kostelecky \cite{Colladay:1996iz}. The ESM is a kind of catalogue with all the possible terms we can add to the usual SM such that the resulting action is a Lorentz scalar and hence the Lorentz symmetry is not broken at the level of the observer. Thus, observers related by Lorentz transformations perceive the same physics. One requires also that the gauge symmetry is maintained and that the theory is renormalizable by power counting. This latter requirement is not fundamental since the usual SM is itself an effective theory and hence there is nothing wrong in including non-renormalizable terms (indeed, the neutrino oscillations are probably described by non-renormalizable terms \cite{GonzalezGarcia:2002dz}). The point is that these terms are suppressed at low energies. The ESM is hence seem as an effective theory where the terms that violates the Lorentz symmetry have the general form:
\begin{align}
 \label{c3s3e02}
  T^{\mu\nu\rho...}(fields\; and \; derivatives)_{\mu\nu\rho...},
 \end{align}
where $T^{\mu\nu\rho...}$ is a constant term and consequently it determines preferred directions in each inertial reference frame breaking the Lorentz symmetry. Notice that, being a Lorentz scalar, this term has the same value in all reference frames, then one says that the observer-like Lorentz symmetry is maintained while the particle-like Lorentz symmetry is broken (see \cite{Colladay:1996iz} for details). If we expect this model to be indeed a generalization of the SM, the magnitude of $T^{\mu\nu\rho...}$ has to be very small since we do not observe its effects. There are already many observational limits of the magnitudes of the different tensors that appear in the ESM. For a recent discussion, see \cite{Kostelecky:2008be, Kostelecky:2008ts}. In this context, the CFJ model is identified as the electromagnetic sector of the ESM that violate the Lorentz symmetry and $CPT$ (since $p_{\mu}$ is taken as a vector in ESM framework).

More recently, this model has found its way as an effective description of the physics of the chiral magnetic effect (CME) in hot QCD matter. This is a remarkable effect associated with the interplay of topological properties and axial anomaly in QCD, leading to possible violations of local $P$ and $CP$ in this setting \cite{Kharzeev:2004ey,Kharzeev:2007tn,Kharzeev:2007jp,Fukushima:2008xe,Kharzeev:2009fn}. The CME takes place when a sufficiently strong magnetic field (of order $\sim \Lambda^2_{QCD}$) is applied in a QCD environment with a chiral imbalance (an asymmetry between the numbers of left-handed and right-handed fermions) leading to an observable separation of electric charges. This imbalance is provided by topological fluctuations of the QCD vacuum. A possible source of these fluctuations is the axion field, whose configuration might define regions in spacetime (domain walls) where $P$, $CP$ and Lorentz symmetries are broken. The effective theoretical description of the system stems from the QCD $+$ QED Lagrangian with an axion coupling in the QCD sector \cite{Kharzeev:2009fn}, that is:
\begin{align}
\label{QCDQED}
 {\cal L}_{QCD+QED} &=  - \frac 14 \sum_a F^a_{\mu\nu}F^{a\mu\nu}- \frac{g^2}{32\pi^2} \theta(x) \sum_aF^a_{\mu\nu} {}^{\ast} F^{a\mu\nu}- \frac 14 F_{\mu\nu}F^{\mu\nu}\nonumber\\
 &+\sum_f \bar{\psi}_f\left[i\gamma^{\mu} \left(\partial_{\mu} -ig\sum_a A^{a}_{\mu}t^a-iq_f A_{\mu}-m_f  \right)\right]\psi_f,
\end{align}
where $A^{a}_{\mu}$ are the non-Abelian gluon fields defining the field strength $F^a_{\mu\nu}$ with $a=1...8$ for the $SU(3)$ group and $t^a$ the generators in the fundamental representation. $A_{\mu}$ is the electromagnetic potential defining the electromagnetic field $F_{\mu\nu}$. The fermion field $\psi_f$, with flavor f, has mass $m_f$ and interacts with the gluon fields with strength $g$ and has charge $q_f$ under the electromagnetic field. The axion field $\theta(x)$ is a pseudo-scalar with an external prescribed configuration, that is why it appears without a kinetic term. The fermions provide the bridge connecting the gluonic sector with the photonic sector. The effective photonic theory has thus the general form at lowest order in a derivative expansion
\begin{align}
\label{axionQED}
 {\cal L}_{AxionQED} = - \frac 14 F_{\mu\nu}F^{\mu\nu}- \frac{c}{4} \theta(x) F_{\mu\nu} {}^{\ast} F^{\mu\nu}
 \end{align}
where $c$ is a constant. By considering the axion configuration such that $\partial_{\mu}\theta = p_{\mu}$ is a constant in some region, we see that (after integrating by parts in the last term) (\ref{axionQED}) is the CFJ model (\ref{c3s3e01}). Note however that since $\theta$ is a pseudo-scalar it follows that $p_{\mu}$ is a pseudo-vector and $P$ and $CP$ are broken but not $CPT$. The existence of regions with $\partial_{\mu}\theta \neq 0$ provides the conditions for the chiral imbalance, which in the presence of a magnetic field produce the separation of charges along the direction of the magnetic field, characterizing the chiral magnetic effect. Charge asymmetries have been observed in the STAR and PHENIX collaboration at the RHIC \cite{:2009uh,:2009txa,Ajitanand:2010rc} but the interpretation of the results is still a matter of debate. If the CME turns out to be the correct explanation for this observed charge asymmetry it will represent the first direct observation of the non-trivial topological properties of QCD. For a further discussion on the development of these matters, both theoretically and experimentally, we point the reader to \cite{Kharzeev:2009fn,Kharzeev:2010ym}, on which this discussion was based.

Our interest in the CFJ model regards its non-trivial topological character. We want to investigate the possibility and the consequences of defining magnetic charges in the presence of the CS-like term. This problem was already approached in \cite{Barraz:2007mi} where its similarity with the analogous problem in $(2+1)D$ was exploited via dimensional reduction. Here we are going to obtain new results that suggest that this system features confining properties using the procedures developed in the previous sections.

The dual formulation of the CFJ model was obtained by some of us in \cite{Guimaraes:2006gj}. For the sake of a self-contained presentation, let us review the result here. Following the usual procedures of duality we have studied until now, consider the theory in first order in the derivatives, equivalent to (\ref{c3s3e01}):
\begin{align}
\label{c3s3e32}
 {\cal L}_{MCFJ} = \frac 12 \Pi_{\mu\nu}\varepsilon^{\mu\nu\rho\sigma}\partial_{\rho}A_{\sigma} - \frac 14 \Pi_{\mu\nu}\Pi^{\mu\nu} +
 p_\mu\varepsilon^{\mu\nu\rho\sigma}A_{\nu}\partial_{\rho}A_{\sigma}.
\end{align}
This is the master Lagrangian of the model. $\Pi_{\mu\nu}$ is an auxiliary field that can be integrated out giving us again (\ref{c3s3e01}). On the other hand, we can eliminate $A_{\mu}$ completely as a function of $\Pi_{\mu\nu}$. The Euler-Lagrange equations for $A_{\mu}$ furnish:
\begin{align}
\label{c3s3e48}
 \Lambda^{\mu} \equiv \varepsilon^{\mu\nu\rho\sigma}\partial_{\nu}\Pi_{\rho\sigma} = 4 \varepsilon^{\mu\nu\rho\sigma}p_\nu\partial_{\rho}A_{\sigma}.
\end{align}
Notice that $\Lambda^{\mu}$, by definition, satisfies:
\begin{align}
\label{c3s3e49}
 \partial_{\mu}\Lambda^{\mu} = 0,
\end{align}
and due to the equations of motion for $A^{\mu}$, it also obeys:
\begin{align}
\label{c3s3e50}
 p_{\mu}\Lambda^{\mu} = 0.
\end{align}
Formally, we can substitute (\ref{c3s3e48}) in (\ref{c3s3e32}) and we get:
\begin{align}
\label{c3s3e51}
 {\cal L}_{MCFJ} \rightarrow \frac 14 \Lambda^{\mu}A_{\mu} - \frac 14 \Pi_{\mu\nu}\Pi^{\mu\nu},
\end{align}
where $A^{\mu}\equiv A^{\mu}(\Pi)$ is defined by (\ref{c3s3e48}). To complete the procedure we must rewrite $\Lambda^{\mu}A_{\mu}$ as a function of $\Pi^{\mu\nu}$. It follows from (\ref{c3s3e48}) that the solution of the constraint:
\begin{align}
\label{c3s3e52}
 \varepsilon^{\mu\nu\rho\sigma}\partial_{\nu}(\Pi_{\rho\sigma} + 2 p_{[\rho}A_{\sigma]}) =
 0\nonumber\\
 \Rightarrow \Pi_{\mu\nu} = - 2 p_{[\mu}A_{\nu]} + \partial_{[\mu}B_{\nu]},
\end{align}
introduces a new gauge field $B^{\mu}$, where we used the notation $X_{[\mu} Y_{\nu]}\equiv X_\mu Y_\nu - X_\nu Y_\mu$. Thus,
\begin{align}
\label{c3s3e53}
 p^{\mu}\Lambda^{\nu}\Pi_{\mu\nu} = - 2 p^2\Lambda^{\mu}A_{\mu} + \Lambda^{\mu}(p^{\nu}\partial_{\nu})B_{\mu} - \Lambda^{\mu}\partial_{\mu}(p^{\nu}B_{\nu}),
\end{align}
where (\ref{c3s3e49}) was used. The first term in the right hand side of the above equation contains the structure of the first term in (\ref{c3s3e51}). Substituting it in (\ref{c3s3e51}) we can integrate by parts and the last term of (\ref{c3s3e53}) vanishes due to (\ref{c3s3e49}), and we get:
\begin{align}
\label{c3s3e54}
 {\cal L}_{dualCFJ} &= -\frac{1}{8p^2}(p^{\alpha}\Pi_{\alpha\mu})\varepsilon^{\mu\nu\rho\sigma}\partial_{\nu}\Pi_{\rho\sigma} - \frac 14
 \Pi_{\mu\nu}\Pi^{\mu\nu}\nonumber\\
 &\;\;\;\;+ \frac{1}{8p^2} [(p^{\alpha}\partial_{\alpha})B_{\mu}]\varepsilon^{\mu\nu\rho\sigma}\partial_{\nu}\Pi_{\rho\sigma}.
\end{align}
Notice that the relation (\ref{c3s3e52}) does not contain the component of the field $A_{\mu}$ in the direction of $p^{\mu}$. In fact, this component is a Lagrange multiplier in (\ref{c3s3e32}). However, the constraint imposed by this component establishes a relation between $\Pi_{\mu\nu}$ and $B_{\mu}$ that can be read from (\ref{c3s3e52}) as being:
\begin{align}
\label{c3s3e55}
 \varepsilon^{\mu\nu\rho\sigma}p_{\nu}\Pi_{\rho\sigma} =2
 \varepsilon^{\mu\nu\rho\sigma}p_{\nu}\partial_{\rho}B_{\sigma}.
\end{align}
$\Pi_{\mu\nu}$ is indeed the dual of $A_{\mu}$ but its components must satisfy certain constraints that are more easily addressed with the introduction of the field $B_{\mu}$. Notice that:
\begin{align}
\label{c3s3e56}
(\varepsilon^{\mu\nu\rho\sigma}p_{\nu}\Pi_{\rho\sigma})^2 = 2p^2
\Pi_{\mu\nu}\Pi^{\mu\nu} - 4
(p^{\alpha}\Pi_{\alpha\mu})(p_{\beta}\Pi^{\beta\mu})\nonumber\\
\Rightarrow \Pi_{\mu\nu}\Pi^{\mu\nu} = \frac{2}{p^2}
(\varepsilon^{\mu\nu\rho\sigma}p_{\nu}\partial_{\rho}B_{\sigma})^2 + \frac{2}{p^2}
(p^{\alpha}\Pi_{\alpha\mu})(p_{\beta}\Pi^{\beta\mu}),
\end{align}
where we used (\ref{c3s3e55}). We can rewrite the first term of (\ref{c3s3e54}) as:
\begin{align}
\label{c3s3e57}
-\frac{1}{8p^2}(p^{\alpha}\Pi_{\alpha\mu})\varepsilon^{\mu\nu\rho\sigma}\partial_{\nu}\Pi_{\rho\sigma}
&= -\frac{1}{4p^4}
(p^{\alpha}\Pi_{\alpha\mu})\varepsilon^{\mu\nu\rho\sigma}p_{\nu}\partial_{\rho}[(p^{\beta}\partial_{\beta})B_{\sigma}]\nonumber\\
&\;\;\;\;+ \frac{1}{4p^4}
(p^{\alpha}\Pi_{\alpha\mu})\varepsilon^{\mu\nu\rho\sigma}p_{\nu}\partial_{\rho}(p^{\beta}\Pi_{\beta\sigma}).
\end{align}
In an analogous manner, the last term is rewritten as:
\begin{align}
\label{c3s3e58}
\frac{1}{8p^2}
[(p^{\alpha}\partial_{\alpha})B_{\mu}]\varepsilon^{\mu\nu\rho\sigma}\partial_{\nu}\Pi_{\rho\sigma}
&= \frac{1}{4p^4}
[(p^{\alpha}\partial_{\alpha})B_{\mu}]\varepsilon^{\mu\nu\rho\sigma}p_{\nu}\partial_{\rho}[(p^{\beta}\partial_{\beta})B_{\sigma}]\nonumber\\
&\;\;\;\;- \frac{1}{4p^4}
[(p^{\alpha}\partial_{\alpha})B_{\mu}]\varepsilon^{\mu\nu\rho\sigma}p_{\nu}\partial_{\rho}(p^{\beta}\Pi_{\beta\sigma})\, .
\end{align}
Combining these results, we can rewrite (\ref{c3s3e54}) as:
\begin{align}
\label{c3s3e59}
 {\cal L}_{dualCFJ} &=  \frac{1}{4p^4} (p^{\alpha}\Pi_{\alpha\mu})\varepsilon^{\mu\nu\rho\sigma}p_{\nu}\partial_{\rho}(p^{\beta}\Pi_{\beta\sigma}) - \frac{1}{2p^2} (p^{\alpha}\Pi_{\alpha\mu}) (p_{\beta}\Pi^{\beta\mu})\nonumber\\
  &\;\;\;\;- \frac{1}{2p^2} (\varepsilon^{\mu\nu\rho\sigma}p_{\nu}\partial_{\rho}B_{\sigma})^2 + \frac{1}{2p^4}
  (p^{\alpha}\Pi_{\alpha\mu})\varepsilon^{\mu\nu\rho\sigma}p_{\nu}\partial_{\rho}[(p^{\beta}\partial_{\beta})B_{\sigma}]\nonumber\\
 &\;\;\;\;+ \frac{1}{4p^4}[(p^{\alpha}\partial_{\alpha})B_{\mu}]\varepsilon^{\mu\nu\rho\sigma}p_{\nu}\partial_{\rho}[(p^{\beta}\partial_{\beta})B_{\sigma}].
\end{align}
Redefining $p^{\alpha}\Pi_{\alpha\mu} \equiv f_{\mu}$, we get the form:
\begin{align}
\label{c3s3e44}
 {\cal L}_{dualCFJ} &= \frac{1}{4p^4} f_{\mu}\varepsilon^{\mu\nu\rho\sigma}p_{\nu}\partial_{\rho}f_{\sigma} - \frac{1}{2p^2} f_{\mu} f^{\mu} - \frac{1}{2p^2} (\varepsilon^{\mu\nu\rho\sigma}p_{\nu}\partial_{\rho}B_{\sigma})^2\nonumber\\
  &\;\;\;\;+ \frac{1}{2p^4}
  f_{\mu}\varepsilon^{\mu\nu\rho\sigma}p_{\nu}\partial_{\rho}[(p^{\alpha}\partial_{\alpha})B_{\sigma}]
 + \frac{1}{4p^4}[(p^{\alpha}\partial_{\alpha})B_{\mu}]\varepsilon^{\mu\nu\rho\sigma}p_{\nu}\partial_{\rho}[(p^{\beta}\partial_{\beta})B_{\sigma}].
\end{align}
Notice that the components in the direction defined by $p^{\mu}$ are effectively zero due to the contraction with the tensor $\varepsilon^{\mu\nu\rho\sigma}p_{\nu}$.

The terms involving $\frac{1}{p^2}$ do not allow a direct extension of this result for the case where $p^{\mu}$ is light-like. But this is trivially accomplished working with light-cone coordinates \cite{Guimaraes:2006gj}.

To study this system it is important that we are capable of including external sources. This allows us to study the VEV of operators, like the Wilson loop. The coupling of an electric source to the CFJ system can be defined and its dual representation can also be obtained \cite{Guimaraes:2006gj}. A much more complicated problem that we are going to approach in the sequel regards the definition of the CFJ model in the presence of magnetic monopoles.

The CFJ model is often regarded as an effective field theory originating from quantum fermionic fluctuations. We have already seen an example of this in the discussion of the CME, with $p_{\mu}=\partial_{\mu}\theta$ a pseudo-vector. Also, in the context of the Extended Standard Model, with $p_{\mu}$ a vector, it can be regarded as the effective photonic theory of a Lorentz and $CPT$ violating QED. In section \ref{sec:3} we introduced the idea that fermionic fluctuations could be effectively described via JTA. We saw that this prescription led us to a more precise definition of the MCS theory in the presence of magnetic defects. We want to follow an analogous path here and argue that the CFJ model can be interpreted as the Maxwell theory in $4D$ embedded in an electric condensate that breaks the Lorentz and discrete spacetime symmetries.

The problem we find in trying to define the CFJ model in the presence of magnetic defects is the same we found in the case of the MCS theory: due to the CS-like term, if we try to include a non-minimal coupling directly into the theory, the brane symmetry is explicitly broken and the unphysical magnetic Dirac string would become inconsistently observable. We have seen in the case of the MCS theory that the spontaneous breaking of the brane symmetry is induced by internal defects of the electric condensate represented by closed fluxes. The strategy adopted to obtain this conclusion consists in finding a representation of the system in terms of $p$-currents such that, through the condensation process of these currents implemented via JTA, the desired theory can be reached. This theory, by construction, will be consistent with the brane symmetry. In what follows, we shall firstly discuss the formulation in terms of $p$-currents without introducing monopoles with the aim of understanding the effects of the LIV. Once this is done, the introduction of monopoles is immediate and shall be discussed later in this section.

The first step in the implementation of the JTA in the present case consists in finding a representation of the theory in terms of $p$-currents that is capable of giving us the phenomenology that maps the Maxwell theory into the CFJ model, simulating in this way the role played by the fermions. The very presence of $p$-currents must contain the information about the breaking of Lorentz and discrete spacetime symmetries. Following the same steps of previous sections, we want to find an 1-current $J_e$ and an action $S_e(J_e)$ such that:
\begin{align}
 \label{c3s3e60}
  \int d^4x\left( -\frac 14 F^{\mu\nu}F_{\mu\nu} + e A_{\mu}J_e^{\mu}\right) + S_e(J_e) \rightarrow S_{CFJ},
\end{align}
where $S_{CFJ}$ is defined by (\ref{c3s3e01}) and the arrow indicates the condensation process described via DJTA. Following the discussions of sections \ref{sec:2} and \ref{sec:3}, if we interpret $J_e$ as a (continuous) field, we must have:
\begin{align}
 \label{c3s3e61}
  \int d^4x\; e A_{\mu}J_e^{\mu} + S_e(J_e) = \int d^4x\; p_{\mu}\varepsilon^{\mu\nu\rho\sigma}A_{\nu}\partial_{\rho}A_{\sigma}.
\end{align}
Considering $J_e$ as a function of $A$ and differentiating (\ref{c3s3e61}) with respect to $A$, we get:
\begin{align}
 \label{c3s3e62}
  e J_e^{\mu}  = -2 \varepsilon^{\mu\nu\rho\sigma} p_{\nu}\partial_{\rho}A_{\sigma}.
\end{align}
Notice that the current satisfies:
\begin{align}
 \label{c3s3e63}
  \partial_{\mu}J_e^{\mu}  = 0,\\
  p_{\mu}J_e^{\mu}  = 0.
\end{align}
The first equation denotes the current conservation and tells us that $J_e$ is Poincar\`e -dual to the boundary of a 2-surface $\Sigma_e$:
\begin{align}
 \label{c3s3e64}
  {}^\ast J_e = d{}^\ast \Sigma_e \Rightarrow J_e^{\mu}  = -\partial_{\nu} \Sigma_e^{\nu\mu}.
\end{align}
The second equation fixes the 2-surface Poincar\`e -dual to $\Sigma_e$ as being orthogonal to the direction defined by $p_{\mu}$. This constraint effectively reduces the space where the surface is defined from $4D$ to $3D$. Consequently, we can represent this surface by an 1-current $\Lambda_e$ in the form:
\begin{align}
 \label{c3s3e65}
  \Sigma_e^{\mu\nu} = -\varepsilon^{\mu\nu\rho\sigma} p_{\rho} \Lambda_{e\; \sigma}.
\end{align}
This is the manifestation of the breaking of Lorentz and discrete spacetime symmetries. The current $J_e$ reads:
\begin{align}
 \label{c3s3e66}
  J_e^{\mu}  =  \varepsilon^{\mu\nu\rho\sigma} p_{\nu} \partial_{\rho}\Lambda_{e\; \sigma}.
\end{align}
Substituting in (\ref{c3s3e61}), the action $S_e(J_e)$ reads:
\begin{align}
 \label{c3s3e67}
  S(J_e)  = \int d^4x\; \frac{e^2}{4}\Lambda_{e\; \mu}\varepsilon^{\mu\nu\rho\sigma} p_{\nu} \partial_{\rho}\Lambda_{e\; \sigma}.
\end{align}
With this information we can formally define the partition function of the system:
\begin{align}
 \label{c3s3e68}
 {\cal Z} =  \sum_{\{{\cal A}_{e}\}} \int {\cal D} A e^{i \int d^4x\left( -\frac 14 F^{\mu\nu}F_{\mu\nu} + e A_{\mu}J_e^{\mu}\right) + iS_e(J_e)},
\end{align}
where the sum is taken over the configurations of the 1-surfaces ${\cal A}_{e}$ Poincar\`e -dual to $\Lambda_e$.

Let us study now the system defined by (\ref{c3s3e68}) with respect to its different representations. By construction, the effect of the fermionic fluctuations is simulated here by the condensation of the current $\Lambda_e$, that is, if we consider this current as a field and formally substitute the sum over ${\cal A}_{e}$ by a functional integral over $\Lambda_e$, we reobtain the CFJ model described by the action (\ref{c3s3e01}). On the other hand, the dilution of this current gives us the Maxwell theory in $(3+1)D$. We can go to an alternative representation where we exchange the currents $\Lambda_e$ by magnetic fluxes through the GPI, which in the present case has the form:
\begin{align}
 \label{c3s3e69}
\sum_{\{{\cal A}_{e}\}} \delta(H_{\mu} - \Lambda_{\mu}) = \sum_{\{{\cal C}_{e}\}} e^{2\pi i \int d^4x H_{\mu}\varepsilon^{\mu\nu\rho\sigma} \Omega_{\nu\rho\sigma}},
\end{align}
where $H$ is an 1-form and in the right hand side the sum is taken over the configurations of the 3-surfaces ${\cal C}_{e}$ Poincar\`e -dual to $\Omega$. Inserting the identity $\mathds{1} = \int {\cal D} H \delta(H_{\mu} - \Lambda_{\mu})$ in (\ref{c3s3e68}) and using (\ref{c3s3e69}), we get the equivalent representation:
\begin{align}
 \label{c3s3e70}
 {\cal Z} =  \sum_{\{{\cal C}_{e}\}}\int {\cal D} A \int {\cal D} H  e^{i \int d^4x\left( -\frac 14 F^{\mu\nu}F_{\mu\nu} + e A_{\mu}J_e^{\mu}(H) + 2\pi H_{\mu}\varepsilon^{\mu\nu\rho\sigma} \Omega_{\nu\rho\sigma}\right) + iS_e(J_e(H))},
\end{align}
where the expressions involving the 1-current $\Lambda$ are written now in terms of the 1-form $H$,
\begin{align}
 \label{c3s3e71}
 J_e^{\mu}  &=  \varepsilon^{\mu\nu\rho\sigma} p_{\nu} \partial_{\rho}H_{\sigma}\\
  S(J_e)  &= \int d^4x\; \frac{e^2}{4}H_{\mu}\varepsilon^{\mu\nu\rho\sigma} p_{\nu} \partial_{\rho}H_{\sigma}.
\end{align}
Now we are going to integrate out the field $H$. Since it has a gaussian structure, the integration process is equivalent to solving its equation of motion:
\begin{align}
 \label{c3s3e72}
 e\varepsilon^{\mu\nu\rho\sigma} p_{\nu} \partial_{\rho}A_{\sigma} + 2\pi \varepsilon^{\mu\nu\rho\sigma} \Omega_{\nu\rho\sigma} + \frac{e^2}{2}\varepsilon^{\mu\nu\rho\sigma} p_{\nu} \partial_{\rho}H_{\sigma} = 0.
\end{align}
Notice that this equation tells us that $\Omega$ has a structure such that:
\begin{align}
 \label{c3s3e73}
 \varepsilon^{\mu\nu\rho\sigma} \Omega_{\nu\rho\sigma} = \varepsilon^{\mu\nu\rho\sigma} p_{\nu}\partial_{\rho}\omega_{\sigma},
\end{align}
where $\omega$ is an 1-current. Hence, except for terms that vanish in the action, $H$ is given by:
\begin{align}
 \label{c3s3e74}
 H_{\mu} = -\frac{2}{e} A_{\mu} - \frac{4\pi}{e^2} \omega_{\mu},
\end{align}
and the system is defined in an equivalent form by the partition function:
\begin{align}
 \label{c3s3e75}
 {\cal Z} =  \sum_{\{{\cal D}_{e}\}}\int {\cal D} A e^{i \int d^4x\left[ -\frac 14 F^{\mu\nu}F_{\mu\nu} - \left(A_{\mu} + \frac{2\pi}{e} \omega_{\mu}\right) \varepsilon^{\mu\nu\rho\sigma}p_{\nu}\partial_{\rho} \left( A_{\sigma} + \frac{2\pi}{e} \omega_{\sigma}\right)\right]},
\end{align}
where ${\cal D}_{e}$ are surfaces Poincar\`e -dual to $\omega$. This is the formulation in terms of closed magnetic fluxes defined by $\omega$ (more precisely, the closed fluxes are related to $\varepsilon^{\mu\nu\rho\sigma} \partial_{\rho}\omega_{\sigma}$ which has a vanishing derivative). As we have already mentioned in the previous sections, these two representations of the system, in terms of $\Lambda$ in (\ref{c3s3e68}) and in terms of $\omega$ in (\ref{c3s3e75}), correspond to a order-disorder mapping. Indeed, if the fluxes defined by $\omega$ disappear, then the system is described by the CFJ model. On the other hand, if these fluxes condense, $\omega$ becomes a (continuous) field and the sum over ${\cal D}_{e}$ becomes an integral over $\omega$ and the last term decouples giving us the Maxwell theory. In this sense, these fluxes represent defects over the electric condensate since when they proliferate the condensed phase is destroyed.

This freedom provided by the order parameters is what allows us to adequately define the system in the presence of magnetic monopoles (which correspond to the introduction of open magnetic fluxes). One defines the presence of monopoles with the introduction of a magnetic Dirac brane $\Sigma_g$,
\begin{align}
\label{c3s3e76}
 F_{\mu\nu} \rightarrow G_{\mu\nu} \equiv \partial_{\mu} A_{\nu} - \partial_{\nu} A_{\mu} + g {}^\ast \Sigma_{g\; \mu\nu},
\end{align}
such that Bianchi's identity is violated:
\begin{align}
\label{c3s3e77}
 \partial_{\mu} {}^\ast G^{\mu\nu} = - g\partial_{\mu} \Sigma_g^{\mu\nu} = g J_g^{\nu}.
\end{align}
The magnetic brane symmetry is realized here according to:
\begin{align}
\label{c3s3e78}
 {}^\ast \Sigma_{g\; \mu\nu} &\rightarrow {}^\ast \Sigma_{g\; \mu\nu} + \partial_{\mu}\sigma_{\nu} - \partial_{\nu}\sigma_{\mu}\nonumber\\
 A_{\mu} &\rightarrow A_{\mu} - g\sigma_{\mu},
\end{align}
such that the field $G_{\mu\nu}$, being an observable, remains invariant. The system in the presence of magnetic charges is then described by:
\begin{align}
 \label{c3s3e79}
 {\cal Z}(J_g) =  \sum_{\{{\cal D}_{e}\}}\int {\cal D} A e^{i \int d^4x\left[ -\frac 14 G^{\mu\nu}G_{\mu\nu} - \left(A_{\mu} + \frac{2\pi}{e} \omega_{\mu}\right) \varepsilon^{\mu\nu\rho\sigma}p_{\nu}\partial_{\rho} \left( A_{\sigma} + \frac{2\pi}{e} \omega_{\sigma}\right)\right]}.
\end{align}
The brane symmetry is preserved at the level of the partition function due to the presence of the internal fluxes $\omega$ as long as the Dirac quantization condition, $eg=2\pi$, is satisfied. To obtain the CFJ model we must dilute the fluxes $\omega$. As we know, the brane symmetry prohibits the system to undergo a complete dilution of these defects and the internal fluxes connected to the external Dirac brane remain in the system. Note that the CS-like term does not have components of fields in the direction of $p_{\nu}$. Since this term is responsible for the spontaneous breaking of the brane symmetry, we see that it does not occur in the direction of $p_{\nu}$. This asymmetry introduced by the LIV makes the analysis of the results more complicated.

To have a better understanding of this result it is interesting to look at the dual formulation, where the monopoles appear minimally coupled. We follow here the same ideas discussed around (\ref{c2s2e13}). Consider again (\ref{c3s3e68}), but now in the presence of external monopoles:
\begin{align}
 \label{c3s3e80}
 {\cal Z}(J_g) =  \sum_{\{{\cal A}_{e}\}} \int {\cal D} A e^{i \int d^4x\left( -\frac 14 G^{\mu\nu}G_{\mu\nu} + e A_{\mu}J_e^{\mu}\right) + iS_e(J_e)}.
\end{align}
We shall first express the system in terms of brane invariants. For this, note that since $eg=2\pi$, we can rewrite the electric coupling as:
\begin{align}
 \label{c3s3e81}
  e A_{\mu}J_e^{\mu} \rightarrow e(A_{\mu} + g\Xi_{g\;\mu})J_e^{\mu},
\end{align}
which should be compared with (\ref{c2s2e15a}). $\Xi_g$ is such that, under the brane transformation (\ref{c3s3e78}):
\begin{align}
 \label{c3s3e82}
  \Xi_{g\;\mu} \rightarrow \Xi_{g\;\mu} + \sigma_{\mu}.
\end{align}
Hence, we can construct the following brane invariants:
\begin{align}
 \label{c3s3e83}
 \tilde{A}_{\mu} &\equiv A_{\mu} + g\Xi_{\mu}\nonumber\\
 {}^\ast \tilde{\Sigma}_{g\; \mu\nu} &\equiv {}^\ast \Sigma_{g\; \mu\nu} - \partial_{\mu}\Xi_{\nu} + \partial_{\nu}\Xi_{\mu}.
\end{align}
We introduce in the partition function a sum over the different configurations of the surfaces connected to the Dirac brane  ${}^\ast \Sigma_{g\; \mu\nu}$ and Poincar\`e -dual to $\Xi_g$, represented by ${\cal E}_{g}$, as we did in (\ref{c2s2e15b}). This sum is redundant at this point and constitutes a normalization in the partition function. Thus, the system is described by:
\begin{align}
 \label{c3s3e84}
 {\cal Z}(J_g) =  \sum_{\{{\cal E}_{g}\}} \sum_{\{{\cal A}_{e}\}} \int {\cal D} \tilde{A} e^{i \int d^4x\left( -\frac 14 G^{\mu\nu}(\tilde{A}) G_{\mu\nu}(\tilde{A}) + e \tilde{A}_{\mu}J_e^{\mu}\right) + iS_e(J_e)},
\end{align}
where:
\begin{align}
\label{c3s3e85}
 G_{\mu\nu}(\tilde{A}) \equiv \partial_{\mu} \tilde{A}_{\nu} - \partial_{\nu} \tilde{A}_{\mu} + g {}^\ast \tilde{\Sigma}_{g\; \mu\nu}.
\end{align}
We can rewrite the sum over ${\cal E}_{g}$ as a sum over the surfaces ${\cal B}_{g}$ Poincar\`e -dual to the brane invariant ${}^\ast \tilde{\Sigma}_{g\; \mu\nu}$ constrained by $\partial_{\mu} \tilde{\Sigma}_g^{\mu\nu} =  - J_g^{\nu}$:
\begin{align}
 \label{c3s3e86}
 {\cal Z}(J_g) =  \sum_{\{{\cal B}_{g}\}}{}' \sum_{\{{\cal A}_{e}\}} \int {\cal D} \tilde{A} e^{i \int d^4x\left( -\frac 14 G^{\mu\nu}(\tilde{A}) G_{\mu\nu}(\tilde{A}) + e \tilde{A}_{\mu}J_e^{\mu}\right) + iS_e(J_e)}.
\end{align}
We are going to obtain now the dual formulation of this theory with respect to $\tilde{A}$. The sector of the theory containing $\tilde{A}$ is described by the Lagrangian density:
\begin{align}
 \label{c3s3e87}
 {\cal L} &= -\frac 14 G^{\mu\nu}(\tilde{A}) G_{\mu\nu}(\tilde{A}) + e \tilde{A}_{\mu}J_e^{\mu}\nonumber\\
 &\rightarrow -\frac 12 \Pi^{\mu\nu} G_{\mu\nu}(\tilde{A}) + \frac 14 \Pi^{\mu\nu} \Pi_{\mu\nu} + e \tilde{A}_{\mu}J_e^{\mu},
\end{align}
where we have introduced the auxiliary field $\Pi$. Integrating $\tilde{A}$ we get the following constraint over $\Pi$:
\begin{align}
 \label{c3s3e88}
 \partial_{\mu} \Pi^{\mu\nu} + e\varepsilon^{\nu\mu\rho\sigma} p_{\mu} \partial_{\rho}\Lambda_{e\; \sigma} = 0,
\end{align}
which can be solved with the introduction of a gauge field $B$, dual to the field $\tilde{A}$,
\begin{align}
 \label{c3s3e89}
  \Pi^{\mu\nu} =\varepsilon^{\mu\nu\rho\sigma}\partial_{\rho}B_{\sigma} - \varepsilon^{\mu\nu\rho\sigma} p_{\rho}\Lambda_{e\;\sigma}.
\end{align}
The theory in the dual representation reads then:
\begin{align}
 \label{c3s3e90}
 {\cal Z}(J_g) =  \sum_{\{{\cal B}_{g}\}}{}' \sum_{\{{\cal A}_{e}\}} \int {\cal D} B e^{i \int d^4x\left( -\frac 14 H^{\mu\nu} H_{\mu\nu} + \frac g2 H^{\mu\nu} \tilde{\Sigma}_{g\; \mu\nu}\right) + iS_e(J_e)},
\end{align}
where:
\begin{align}
\label{c3s3e91}
H_{\mu\nu} \equiv \partial_{\mu} B_{\nu} - \partial_{\nu} B_{\mu} - e (p_{\mu} \Lambda_{e\;\nu} - p_{\nu} \Lambda_{e\;\mu}).
\end{align}
Notice that the non-minimal coupling exhibits the consequences of the LIV. The electric brane symmetry follows from the usual definition of the brane transformation for the electric brane $\Sigma_e$ taking into account the definition (\ref{c3s3e65}):
\begin{align}
 \label{c3s3e92}
 p_{\mu} \Lambda_{e\;\nu} - p_{\nu} \Lambda_{e\;\mu} &\rightarrow p_{\mu} \Lambda_{e\;\nu} - p_{\nu} \Lambda_{e\;\mu} + \partial_{\mu} \lambda_{\nu} - \partial_{\nu} \lambda_{\mu}\nonumber\\
 B_{\mu} &\rightarrow B_{\mu} + e\lambda_{\mu},
\end{align}
which clearly leaves $H_{\mu\nu}$ invariant.

The CFJ model must be obtained condensing the electric brane $\Lambda_e$. With the purpose of simplifying the discussion, let us choose $p^{\mu} = (0,0,0,m)$. The action defining the system has then the following form:
\begin{align}
 \label{c3s3e93}
  S &= \int d^4x \left[ -\frac 12 \left(\partial_a B_{3} - \partial_3 B_{a} +em \Lambda_{e\;a} \right)^2 -\frac 14 \left(\partial_a B_{b} - \partial_b B_{a}\right)^2\right.\nonumber\\
   &\left. + g H^{a3} \tilde{\Sigma}_{g\; a3} + \frac g2 H^{ab} \tilde{\Sigma}_{g\; ab} - \frac{me^2}{4}\Lambda_{e\; a}\varepsilon^{abc}\partial_{b}\Lambda_{e\;c} \right],
\end{align}
where the Latin indices denote the components in the hyperplane orthogonal to the direction defined by $p^\mu$. Following the JTA, the electric brane $\Lambda_e$ is promoted to the field category and we formally consider the sum over ${\cal A}_{e}$ as a functional integral over $\Lambda_e$:
\begin{align}
 \label{c3s3e94}
 {\cal Z}(J_g) \rightarrow  \sum_{\{{\cal B}_{g}\}}{}' \int {\cal D} \Lambda_e \int {\cal D} B e^{i S(\Lambda_e, B, \tilde{\Sigma}_g)}.
\end{align}
Notice that all the information about the brane is contained in $H^{a3}$, which is a brane invariant. This means that the integral over $\Lambda_e$ is in fact an integral over $H^{a3}$ once we fix the brane redundancy. The next natural step would be to combine the integrals over $B$ and $\Lambda_e$ into an integral over $H$. However, the LIV does not allow a trivial complete elimination of $B$ and the best we can do is to absorve the component $B_3$. More precisely, renaming $H^{a3} \equiv f^a$ ($\Rightarrow H_{a3} = -f_a$) the action reads:
\begin{align}
 \label{c3s3e95}
  S(f_a, B_a, \tilde{\Sigma}_g) &= \int d^4x \left[\frac 12 f_af^a -\frac 14 \left(\partial_a B_{b} - \partial_b B_{a}\right)^2\right.\nonumber\\
   &\left. + g f^a \tilde{\Sigma}_{g\; a3} - g  B_{b} \partial_a \tilde{\Sigma}_g^{ab} - \frac{1}{4m}\left(f_a - \partial_3 B_{a}\right)\varepsilon^{abc}\partial_{b}\left(f_c - \partial_3 B_{c}\right) \right],
\end{align}
and the partition function of the system is given by:
\begin{align}
 \label{c3s3e96}
 {\cal Z}(J_g) = \sum_{\{{\cal B}_{g}\}}{}' \int {\cal D} f_a \int {\cal D} B_a e^{i S(f_a, B_a, \tilde{\Sigma}_g)}.
\end{align}
We believe that this is the adequate formulation of the CFJ model in the presence of monopoles. Notice that the action (\ref{c3s3e95}) with $g=0$ reduces to the action (\ref{c3s3e44}) with $p^\mu=(0,0,0,m)$. It is interesting to notice how the rank jump characterizing the mass generation in the JTA is realized here. The initial system defined by (\ref{c3s3e90}) has, besides the currents represented by the sums, a massless 1-form gauge field $B$. After the condensation of $\Lambda_e$ a new 1-form is added to the system, although this form is not independent. According to what we have discussed above, the only non-vanishing components of $\Lambda_e$ are orthogonal to $p^{\mu}$ and through the condensation process they absorb the third component of the field $B$. This is the rank jump: $B_3$ is effectively substituted by $f_a$. We also know that $f_a$ and $B_a$ must be understood as components of the 2-form $H$, which have certain constraints that are better expressed through the introduction of the field $B_a$.

Notice that only the components $\tilde{\Sigma}_{g\; a3}$ of the magnetic brane invariant are minimally coupled to $f_a$, which indicates spontaneous breaking of the brane symmetry only in the hyperplane orthogonal to the direction defined by the Lorentz violating vector $p^\mu$. Although a more detailed analysis is necessary, this ``partial spontaneous breaking of the brane symmetry" constitutes a strong evidence that this system exhibits confining properties in the hyperplane orthogonal to $p^\mu$.

\section{Conclusion}
\label{sec:conc}

Throughout this paper we have studied the manifestation of the confinement phenomenon in Abelian theories in diverse examples. In all of them we have identified an universal criterium characterizing the confining regime in formulations involving condensates, namely the spontaneous breaking of the brane symmetry. The concept of brane symmetry is most commonly understood from the Dirac string ambiguity in the description of monopoles in interaction with the electromagnetic field \cite{dirac}. Its breaking is properly understood as a particular realization in which the symmetry is hidden into observable brane invariants composed of Dirac branes and internal defects of the condensate describing regions of the space where the condensate has not been established. These brane invariants carry energy content and are the physical flux tubes present in the condensate connecting the confined charges of opposite sign in their boundaries.

The formalism we developed in the present work generalizes the ideas of Julia and Toulouse \cite{jt} and Quevedo and Trugenberger \cite{qt} to deal with the outcome of condensation of defects taking into proper account Elitzur's theorem \cite{elitzur}. This generalization also provides a more precise implementation of the Julia-Toulouse condensation process by working with ensembles of defects at the level of the partition function. This is built on important contributions made by Banks, Kogut and Myerson \cite{Banks:1977cc} and Kleinert \cite{mvf,Kiometzisprl} on the ensemble formulation and in the use of the order-disorder map codified in the Poisson's Identity and the contribution of Kleinert in the interpretation of brane symmetry \cite{mvf,Kleinert:1992eb}. With the generalization of Poisson's Identity \cite{dafdc} to deal with arbitrary $p$-currents, we were able to construct a parametrization of the condensation-dilution process defining in a very precise way the Julia-Toulouse Approach.

The Abelian Higgs model, which describes a relativistic supercondutor, was studied and it was shown that its confining property can be understood as a natural manifestation of the brane symmetry breaking. We provided a careful computation of the confining potential to illustrate this symmetry breaking. This phenomenon seems to display an universal character and it appears in other systems exhibiting confining properties as well. The Polyakov model, the Maxwell-Chern-Simons theory with external instantons and the Carroll-Field-Jackiw Model with external monopoles, all seem to display confining properties related to the brane symmetry breaking and the rank jump. The maintenance of brane symmetry (in its hidden or explicit realization) is what ensures the charge quantization of external charges in any phase we consider in each one of these systems. The proper treatment of the brane symmetry also led us to obtain a new formulation of the Maxwell-Chern-Simons theory in the presence of magnetic defects and also to recover the topological mass quantization. Also, the presence of instantons seems to endow the condensate with anyonic properties. Another interesting aspect of the last application is that it constitutes an example where the brane symmetry is realized in a hidden fashion while the gauge symmetry is explicitly realized, reinforcing the independence between these two local symmetries.

Another novel result we have presented here using the JTA was the definition of the Carroll-Field-Jackiw Model in the presence of monopoles, where we have also pointed out that a ``partial spontaneous breaking of the brane symmetry" occurs only in the hyperplane orthogonal to the direction defined by the Lorentz violating vector, constituting in this way a strong theoretical evidence that this system exhibits asymmetric confining properties. It is interesting to speculate if this may be relevant to QCD, since the Carroll-Field-Jackiw Model seems to play an important role as the effective description of the chiral magnetic effect in QCD.

Due to the wide scope of this generalized Julia-Toulouse Approach, we believe that many other applications, beyond those we have exploited in this work, can be developed. We also notice that this formalism unifies under common principles diverse results found in the literature, giving us at least a better intuition regarding the nature of the studied phenomena. The use we have made of the JTA to describe the known hierarchy structure of the Fractional Quantum Hall states is just an example of this.

\section{Acknowledgements}

We thank Jorge Noronha for a critical reading of this manuscript and for pointing out reference \cite{Kharzeev:2009fn}. We thank Conselho Nacional de Desenvolvimento Cient\'ifico e Tecnol\'ogico (CNPq) and Funda\c{c}\~ao de Amparo \`a Pesquisa do Estado do Rio de Janeiro (FAPERJ) for financial support.


\begin{thebibliography}{99}

\bibitem{dirac} P. A. M. Dirac, \emph{Quantised Singularities in the Electromagnetic Field}, Proc. Roy. Soc. A \textbf{133}, 60 (1931); P. A. M. Dirac, \emph{The Theory of Magnetic Poles}, Phys. Rev. \textbf{74}, 817 (1948).

\bibitem{elitzur} S. Elitzur, \emph{Impossibility of spontaneously breaking local symmetries}, Phys.
Rev. D \textbf{12}, 3978 (1975).

\bibitem{jt} B. Julia and G. Toulouse, \emph{The Many-Defect Problem: Gauge-Like Variables for
Ordered Media Containing Defects}, J. Physique Lett. \textbf{40}, 395 (1979).

\bibitem{qt} F. Quevedo and C. A. Trugenberger, \emph{Phases of Antisymmetric Tensor Fields Theories}, Nucl. Phys. B \textbf{501}, 143 (1997), [arXiv:hep-th/9604196].

\bibitem{Banks:1977cc} T. Banks, R. Myerson and J. B. Kogut, \emph{Phase Transitions in Abelian
Lattice Gauge Theories}, Nucl. Phys. B \textbf{129}, 493 (1977).

\bibitem{mvf} H. Kleinert, \emph{Multivalued Fields in Condensed Matter, Electromagnetism
and Gravitation} (World Scientific Publishing Company, 2007); H. Kleinert, \emph{Gauge Fields in Condensed Matter} (World Scientific Publishing Company, 1989).

\bibitem{conf} H. B. Nielsen and P. Olesen, \emph{Vortex-Line Models for Dual Strings}, Nucl. Phys. B \textbf{61}, 45 (1973); Y. Nambu, \emph{Strings, Monopoles, and Gauge Fields}, Phys. Rev. D \textbf{10}, 4262 (1974); Y. Nambu, \emph{Magnetic And Electric Confinement Of Quarks}, Phys. Rept. {\bf 23}, 250 (1976); M. Creutz, \emph{Higgs mechanism and quark confinement}, Phys. Rev. D \textbf{10}, 2696 (1974); G. 't Hooft, \emph{High Energy Physics} (Editrice Compositori, Bologna, 1976);  G. 't Hooft, \emph{On The Phase Transition Towards Permanent Quark Confinement}, Nucl. Phys. B {\bf 138}, 1 (1978); G. 't Hooft, \emph{A Property Of Electric And Magnetic Flux In Non-Abelian Gauge Theories}, Nucl. Phys. B {\bf 153}, 141 (1979); G. Parisi, \emph{Quark imprisonment and vacuum repulsion}, Phys. Rev. D \textbf{11}, 970 (1975); A. Jevicki and P. Senjanovic, \emph{String-Like Solution of the Higgs Model with Magnetic Monoples}, Phys. Rev. D \textbf{11}, 860 (1975); S. Mandelstam, \emph{Vortices and quark confinement in non-Abelian gauge theories}, Phys. Rep. C \textbf{23}, 245 (1976); S. Mandelstam, \emph{Charge-Monopole Duality And The Phases Of Non-Abelian Gauge Theories}, Phys. Rev. D {\bf 19}, 2391 (1979).

\bibitem{Gamboa:2008ne} J. Gamboa, L. S. Grigorio, M. S. Guimaraes, F. Mendez and C. Wotzasek, \emph{Radiative processes as a condensation phenomenon and the physical meaning of deformed canonical structures}, Phys. Lett. B {\bf 668}, 447 (2008), [arXiv:0805.0626 [hep-th]].

\bibitem{Grigorio:2008gd} L. S. Grigorio, M. S. Guimaraes and C. Wotzasek, \emph{Monopoles in the presence of the Chern-Simons term via the Julia-Toulouse approach}, Phys. Lett. B {\bf 674}, 213 (2009), [arXiv:0808.3698 [hep-th]].

\bibitem{dafdc}  L. S. Grigorio, M. S. Guimaraes, R. Rougemont and C. Wotzasek, \emph{Dual
approaches for defects condensation}, 	Phys. Lett. B \textbf{690}, 316 (2010) [arXiv:0908.0370v2
[hep-th]].

\bibitem{jt-cho} L. S. Grigorio, M. S. Guimaraes, W. Oliveira, R. Rougemont and C. Wotzasek, \emph{$U(1)$ effective confinement theory from $SU(2)$ restricted gauge theory via the Julia-Toulouse Approach}, Phys. Lett. B \textbf{697}, 392 (2011), [arXiv:1010.6215v1 [hep-th]].

\bibitem{Kleinert:1992eb} H. Kleinert, \emph{Double-Gauge Invariance and Local Quantum Field Theory of Charges and Dirac Magnetic Monopoles}, Phys. Lett. B {\bf 246}, 127 (1990); H. Kleinert, \emph{The Extra Gauge Symmetry of String Deformations in Electromagnetism With Charges And Dirac Monopoles}, Int. J. Mod. Phys. A {\bf 7}, 4693 (1992); H. Kleinert, \emph{Abelian Double-Gauge Invariant Continuous Quantum Field Theory of Electric Charge Confinement}, Phys. Lett. B {\bf 293}, 168 (1992).

\bibitem{oxman} L. E. Oxman, \emph{Large Dual Transformations and the Petrov-Diakonov Representation of the Wilson Loop}, Phys. Rev. D \textbf{82}, 105020 (2010), [arXiv:0909.5171v4 [hep-th]].

\bibitem{ripka} G. Ripka, \emph{Dual Superconductor Models of Color Confinement} (Springer-Verlag, 2005), [arXiv:hep-ph/0310102].

\bibitem{efftheory} C. P. Burgess, \emph{Introduction to effective field theory}, Ann. Rev. Nucl. Part. Sci. {\bf 57}, 329 (2007), [arXiv:hep-th/0701053]; A. V. Manohar, \emph{Effective field theories}, [arXiv:hep-ph/9606222]; D. B. Kaplan, \emph{Effective field theories}, [arXiv:nucl-th/9506035]; I. Z. Rothstein, \emph{TASI lectures on effective field theories}, [arXiv:hep-ph/0308266]; J. Polchinski, \emph{Effective Field Theory And The Fermi Surface}, [arXiv:hep-th/9210046].

\bibitem{cho} Y. M. Cho, \emph{Restricted Gauge Theory}, Phys. Rev. D \textbf{21}, 1080
(1980); Y. M. Cho, \emph{Extended Gauge Theory and Its Mass Spectrum}, Phys. Rev. D \textbf{23},
2415 (1981); W. S. Bae, Y. M. Cho and S. W. Kimm, \emph{Extended QCD Versus Skyrme-Faddeev Theory},
Phys. Rev. D \textbf{65}, 025005 (2001).

\bibitem{Wen:1995qn} X. G. Wen, \emph{Topological orders and edge excitations in FQH states}, Advances in Physics \textbf{44}, 405 (1995), [arXiv:cond-mat/9506066v2].

\bibitem{Kramers:1941kn}
  H.~A.~Kramers and G.~H.~Wannier,
  \emph{Statistics of the two-dimensional ferromagnet. Part 1},
  Phys.\ Rev.\  {\bf 60}, 252 (1941).

\bibitem{Montonen:1977sn}
  C.~Montonen and D.~I.~Olive,
  \emph{Magnetic Monopoles As Gauge Particles?},
  Phys.\ Lett.\  B {\bf 72}, 117 (1977).

\bibitem{Osborn:1979tq}
  H.~Osborn,
  \emph{Topological Charges For N=4 Supersymmetric Gauge Theories And Monopoles Of
  Spin 1},
  Phys.\ Lett.\  B {\bf 83}, 321 (1979).

\bibitem{Vafa:1994tf}
  C.~Vafa and E.~Witten,
  \emph{A Strong coupling test of S duality},
  Nucl.\ Phys.\  B {\bf 431}, 3 (1994),
  [arXiv:hep-th/9408074].

\bibitem{Seiberg:1994rs}
  N.~Seiberg and E.~Witten,
  \emph{Monopole Condensation, And Confinement In N=2 Supersymmetric Yang-Mills Theory},
  Nucl.\ Phys.\  B {\bf 426}, 19 (1994),
  [Erratum-ibid.\  B {\bf 430}, 485 (1994)],
  [arXiv:hep-th/9407087].

\bibitem{Harvey:1996ur}
  J.~A.~Harvey,
  \emph{Magnetic monopoles, duality, and supersymmetry},
  [arXiv:hep-th/9603086].

\bibitem{AlvarezGaume:1996mv}
  L.~Alvarez-Gaume and S.~F.~Hassan,
  \emph{Introduction to S-duality in N = 2 supersymmetric gauge theories: A pedagogical review of the work of Seiberg and Witten},
  Fortsch.\ Phys.\  {\bf 45}, 159 (1997),
  [arXiv:hep-th/9701069].

\bibitem{DiVecchia:1998ky}
  P.~Di Vecchia,
  \emph{Duality in N = 2,4 supersymmetric gauge theories},
  [arXiv:hep-th/9803026].

\bibitem{Kapustin:2006pk} A. Kapustin and E. Witten, \emph{Electric-magnetic duality and the geometric Langlands program}, [arXiv:hep-th/0604151].

\bibitem{Alvarez:1994dn} E. Alvarez, L. Alvarez-Gaume and Y. Lozano, \emph{An introduction to T duality in string theory}, Nucl. Phys. Proc. Suppl. {\bf 41}, 1 (1995), [arXiv:hep-th/9410237].

%\cite{Deser:1997mz}
\bibitem{Deser:1997mz}
  S.~Deser, A.~Gomberoff, M.~Henneaux and C.~Teitelboim,
  \emph{Duality, self-duality, sources and charge quantization in Abelian N-form theories},
  Phys.\ Lett.\ B {\bf 400}, 80 (1997),
  [arXiv:hep-th/9702184].
  %%CITATION = HEP-TH 9702184;%%

%\cite{Wotzasek:1998rj}
\bibitem{Wotzasek:1998rj}
  C.~Wotzasek,
  \emph{On the dimensional dependence of the electromagnetic duality groups},
  Phys.\ Rev.\ D {\bf 58}, 125026 (1998),
  [arXiv:hep-th/9809136].
  %%CITATION = HEP-TH 9809136;%%

%\cite{Noronha:2003vp}
\bibitem{Noronha:2003vp}
  J.~L.~Noronha, D.~Rocha, M.~S.~Guimaraes and C.~Wotzasek,
  \emph{On the dimensional dependence of duality groups for massive p-forms},
  Phys.\ Lett.\ B {\bf 564}, 163 (2003),
  [arXiv:hep-th/0305102].
  %%CITATION = HEP-TH 0305102;%%

%\cite{Wu:1975es}
\bibitem{Wu:1975es}
  T.~T.~Wu and C.~N.~Yang,
  \emph{Concept of nonintegrable phase factors and global formulation of gauge fields},
  Phys.\ Rev.\  D {\bf 12}, 3845 (1975);
  T.~T.~Wu and C.~N.~Yang,
  \emph{Dirac Monopole Without Strings: Monopole Harmonics},
  Nucl.\ Phys.\  B {\bf 107}, 365 (1976).
  %%CITATION = NUPHA,B107,365;%%

%\cite{Bekaert:2002cz}
\bibitem{Bekaert:2002cz}
  X.~Bekaert,
  \emph{Issues in electric-magnetic duality},
  [arXiv:hep-th/0209169];
  K.~Lechner and P.~A.~Marchetti,
  \emph{Duality-invariant quantum field theories of charges and monopoles},
  Nucl.\ Phys.\  B {\bf 569}, 529 (2000),
  [arXiv:hep-th/9906079];
  K.~Lechner and P.~A.~Marchetti,
  \emph{Interacting branes, dual branes, and dyonic branes: A unifying lagrangian approach in D dimensions},
  JHEP {\bf 0101}, 003 (2001),
  [arXiv:hep-th/0007076].
  K.~Lechner and P.~Marchetti,
  \emph{Chern kernels and anomaly cancellation in M-theory},
  Nucl.\ Phys.\  B {\bf 672}, 264 (2003),
  [arXiv:hep-th/0302108].

%\cite{Stueckelberg:1900zz}
\bibitem{Stueckelberg:1900zz}
  E.~C.~G.~Stueckelberg,
  \emph{Interaction energy in electrodynamics and in the field theory of nuclear forces},
  Helv.\ Phys.\ Acta {\bf 11}, 225 (1938).
  %%CITATION = HPACA,11,225;%%

%\cite{Ruegg:2003ps}
\bibitem{Ruegg:2003ps}
  H.~Ruegg and M.~Ruiz-Altaba,
  \emph{The Stueckelberg field},
  Int.\ J.\ Mod.\ Phys.\  A {\bf 19}, 3265 (2004),
  [arXiv:hep-th/0304245].
  %%CITATION = IMPAE,A19,3265;%%

%\cite{Wilson:1974sk}
\bibitem{Wilson:1974sk}
  K.~G.~Wilson,
  \emph{Confinement Of Quarks},
  Phys.\ Rev.\  D {\bf 10}, 2445 (1974).
  %%CITATION = PHRVA,D10,2445;%%

%\cite{Deligne:1999qp}
\bibitem{Deligne:1999qp}
  P.~Deligne {\it et al.},
  \emph{Quantum fields and strings: A course for mathematicians.  Vols. 1, 2},
  (AMS, 1999).

%\cite{Kapustin:2005py}
\bibitem{Kapustin:2005py}
  A.~Kapustin,
  \emph{Wilson-'t Hooft operators in four-dimensional gauge theories and S-duality},
  Phys.\ Rev.\  D {\bf 74}, 025005 (2006),
  [arXiv:hep-th/0501015].

%\cite{Cooper:1956zz}
\bibitem{Cooper:1956zz}
  L.~N.~Cooper,
  \emph{Bound electron pairs in a degenerate Fermi gas},
  Phys.\ Rev.\  {\bf 104}, 1189 (1956);
  J.~Bardeen, L.~N.~Cooper and J.~R.~Schrieffer,
  \emph{Microscopic Theory Of Superconductivity},
  Phys.\ Rev.\  {\bf 106}, 162 (1957);
  J.~Bardeen, L.~N.~Cooper and J.~R.~Schrieffer,
  \emph{Theory Of Superconductivity},
  Phys.\ Rev.\  {\bf 108}, 1175 (1957).

%\cite{Ginzburg:1950sr}
\bibitem{Ginzburg:1950sr}
  V.~L.~Ginzburg and L.~D.~Landau,
  \emph{On the Theory of superconductivity},
  Zh.\ Eksp.\ Teor.\ Fiz.\  {\bf 20}, 1064 (1950);
  V.~L.~Ginzburg,
  \emph{Nobel Lecture: On superconductivity and superfluidity (what I have and have not managed to do) as well as on the 'physical minimum' at the beginning of the XXI century},
  Rev.\ Mod.\ Phys.\  {\bf 76}, 981 (2004).

%\cite{Abrikosov:1956sx}
\bibitem{Abrikosov:1956sx}
  A.~A.~Abrikosov,
  \emph{On the Magnetic properties of superconductors of the second group},
  Sov.\ Phys.\ JETP {\bf 5}, 1174 (1957),
  [Zh.\ Eksp.\ Teor.\ Fiz.\  {\bf 32}, 1442 (1957)];
  A.~A.~Abrikosov,
  \emph{Nobel Lecture: Type-II superconductors and the vortex lattice},
  Rev.\ Mod.\ Phys.\  {\bf 76}, 975 (2004).

%\cite{Chernodub:2008rz}
\bibitem{Chernodub:2008rz}
  M.~N.~Chernodub, L.~Faddeev and A.~J.~Niemi,
  \emph{Non-Abelian Supercurrents and Electroweak Theory},
  JHEP {\bf 0812}, 014 (2008),
  [arXiv:0804.1544 [hep-th]].
  %%CITATION = JHEPA,0812,014;%%

%\cite{Gubarev:1998ss}
\bibitem{Gubarev:1998ss}
  F.~V.~Gubarev, M.~I.~Polikarpov and V.~I.~Zakharov,
  \emph{Monopole-antimonopole interaction in Abelian Higgs model},
  Phys.\ Lett.\  B {\bf 438}, 147 (1998),
  [arXiv:hep-th/9805175].
  %%CITATION = PHLTA,B438,147;%%

%\cite{ksc}
\bibitem{ksc} H. Kleinert, \emph{Disorder Version of the Abelian Higgs Model and the Order of the Superconductive Phase Transition}, Lett. Nuovo Cimento \textbf{35}, 405 (1982)

%\cite{Kiometzisprl}
\bibitem{Kiometzisprl}
  M.~Kiometzis, H.~Kleinert and A.~M.~J.~Schakel,
  \emph{Critical Exponents of the Superconducting Phase Transition},
  Phys.\ Rev.\ Lett.\  {\bf 73}, 1975 (1994);
  M.~Kiometzis, H.~Kleinert and A.~M.~J.~Schakel,
  \emph{Dual description of the superconducting phase transition},
  Fortsch.\ Phys.\  {\bf 43}, 697 (1995).

%\cite{Intriligator:1995au}
\bibitem{Intriligator:1995au}
  K.~A.~Intriligator and N.~Seiberg,
  ``Lectures on supersymmetric gauge theories and electric-magnetic duality,''
  Nucl.\ Phys.\ Proc.\ Suppl.\  {\bf 45BC}, 1 (1996),
  [arXiv:hep-th/9509066].
  %%CITATION = NUPHZ,45BC,1;%%

%\cite{Becker:2007zj}
\bibitem{Becker:2007zj}
  K.~Becker, M.~Becker and J.~H.~Schwarz,
  \emph{String theory and M-theory: A modern introduction}
  (Cambridge University Press, 2007).

%\cite{Polyakov:1975rs}
\bibitem{Polyakov:1975rs}
  A.~M.~Polyakov,
  \emph{Compact gauge fields and the infrared catastrophe},
  Phys.\ Lett.\  B {\bf 59}, 82 (1975);
  A.~M.~Polyakov,
  \emph{Quark Confinement And Topology Of Gauge Groups},
  Nucl.\ Phys.\  B {\bf 120}, 429 (1977);
  A.~M.~Polyakov,
  \emph{Gauge Fields And Strings},
  (CRC Press, 1987).

%\cite{Polyakov:1996nc}
\bibitem{Polyakov:1996nc}
  A.~M.~Polyakov,
  \emph{Confining strings},
  Nucl.\ Phys.\  B {\bf 486}, 23 (1997),
  [arXiv:hep-th/9607049].
  %%CITATION = NUPHA,B486,23;%%

%\cite{'tHooft:1974qc}
\bibitem{'tHooft:1974qc}
  G.~'t Hooft,
  \emph{Magnetic Monopoles In Unified Gauge Theories},
  Nucl.\ Phys.\  B {\bf 79}, 276 (1974);
  A.~M.~Polyakov,
  \emph{Particle spectrum in quantum field theory},
  JETP Lett.\  {\bf 20}, 194 (1974),
  [Pisma Zh.\ Eksp.\ Teor.\ Fiz.\  {\bf 20}, 430 (1974)].

\bibitem{mcs}
  S.~Deser, R.~Jackiw and S.~Templeton,
  \emph{Topologically massive gauge theories},
  Annals Phys.\  {\bf 140}, 372 (1982),
  [Erratum-ibid.\  {\bf 185}, 406.1988\ APNYA,281,409 (1988\ APNYA,281,409-449.2000)];
  S.~Deser, R.~Jackiw and S.~Templeton,
  \emph{Three-Dimensional Massive Gauge Theories},
  Phys.\ Rev.\ Lett.\  {\bf 48}, 975 (1982).

\bibitem{sd}
  S.~Deser and R.~Jackiw,
  \emph{`Selfduality' Of Topologically Massive Gauge Theories},
  Phys.\ Lett.\  B {\bf 139}, 371 (1984).

\bibitem{nakahara} M. Nakahara, \emph{Geometry, topology and physics} (IOP Publishing, 1990).

\bibitem{nash} C. Nash and S. Sen, \emph{Topology And Geometry For Physicists} (Academic Press, 1983).

%\cite{Chern:1974ft}
\bibitem{Chern:1974ft}
  S.~S.~Chern and J.~Simons,
  \emph{Characteristic Forms And Geometric Invariants},
  Annals Math.\  {\bf 99}, 48 (1974).
  %%CITATION = ANMAA,99,48;%%

%\cite{Redlich:1983kn}
\bibitem{Redlich:1983kn}
  A.~N.~Redlich,
  \emph{Gauge Noninvariance And Parity Nonconservation Of Three-Dimensional Fermions},
  Phys.\ Rev.\ Lett.\  {\bf 52}, 18 (1984);
  A.~N.~Redlich,
  \emph{Parity Violation And Gauge Noninvariance Of The Effective Gauge Field Action In Three-Dimensions},
  Phys.\ Rev.\  D {\bf 29}, 2366 (1984).

%\cite{Niemi:1983rq}
\bibitem{Niemi:1983rq}
  A.~J.~Niemi and G.~W.~Semenoff,
  \emph{Axial Anomaly Induced Fermion Fractionization And Effective Gauge Theory Actions In Odd Dimensional Space-Times},
  Phys.\ Rev.\ Lett.\  {\bf 51}, 2077 (1983).
  %%CITATION = PRLTA,51,2077;%%

%\cite{Wilczek:1990ik}
\bibitem{Wilczek:1990ik}
  F.~Wilczek,
  \emph{Fractional statistics and anyon superconductivity}
  (World Scientific Publishing Company, 1990).

%\cite{Haldane:1983xm}
\bibitem{Haldane:1983xm}
  F.~D.~M.~Haldane,
  \emph{Fractional Quantization Of The Hall Effect: A Hierarchy Of Incompressible Quantum Fluid States},
  Phys.\ Rev.\ Lett.\  {\bf 51}, 605 (1983).
  %%CITATION = PRLTA,51,605;%%

%\cite{Laughlin:1983fy}
\bibitem{Laughlin:1983fy}
  R.~B.~Laughlin,
  \emph{Anomalous quantum Hall effect: An incompressible quantum fluid with fractionally charged excitations},
  Phys.\ Rev.\ Lett.\  {\bf 50}, 1395 (1983).
  %%CITATION = PRLTA,50,1395;%%

\bibitem{johnson} B. L. Johnson, \emph{Understanding the Laughlin wave function for the fractional quantum Hall effect}, Am. J. Phys. \textbf{70}, 401 (2002).

%\cite{Henneaux:1986tt}
\bibitem{Henneaux:1986tt}
  M.~Henneaux and C.~Teitelboim,
  \emph{Quantization Of Topological Mass In the Presence Of A Magnetic Pole},
  Phys.\ Rev.\ Lett.\  {\bf 56}, 689 (1986).
  %%CITATION = PRLTA,56,689;%%

%\cite{Pisarski:1986gr}
\bibitem{Pisarski:1986gr}
  R.~D.~Pisarski,
  \emph{Magnetic Monopoles in Topologically Massive Gauge Theories},
  Phys.\ Rev.\  D {\bf 34}, 3851 (1986).
  %%CITATION = PHRVA,D34,3851;%%

%\cite{Affleck:1989qf}
\bibitem{Affleck:1989qf}
  I.~Affleck, J.~A.~Harvey, L.~Palla and G.~W.~Semenoff,
  \emph{The Chern-Simons Term Versus The Monopole},
  Nucl.\ Phys.\  B {\bf 328}, 575 (1989).
  %%CITATION = NUPHA,B328,575;%%

%\cite{Diamantini:1993iu}
\bibitem{Diamantini:1993iu}
  M.~C.~Diamantini, P.~Sodano and C.~A.~Trugenberger,
  \emph{Topological Excitations In Compact Maxwell-Chern-Simons Theory},
  Phys.\ Rev.\ Lett.\  {\bf 71}, 1969 (1993),
  [arXiv:hep-th/9306073].
  %%CITATION = PRLTA,71,1969;%%

%\cite{Carroll:1989vb}
\bibitem{Carroll:1989vb}
  S.~M.~Carroll, G.~B.~Field and R.~Jackiw,
  \emph{Limits on a Lorentz and Parity Violating Modification of Electrodynamics},
  Phys.\ Rev.\  D {\bf 41}, 1231 (1990).
  %%CITATION = PHRVA,D41,1231;%%

%\cite{Colladay:1996iz}
\bibitem{Colladay:1996iz}
  D.~Colladay and V.~A.~Kostelecky,
  \emph{CPT violation and the standard model},
  Phys.\ Rev.\  D {\bf 55}, 6760 (1997),
  [arXiv:hep-ph/9703464];
  D.~Colladay and V.~A.~Kostelecky,
  \emph{Lorentz-violating extension of the standard model},
  Phys.\ Rev.\  D {\bf 58}, 116002 (1998),
  [arXiv:hep-ph/9809521].

%\cite{Feng:2006dp}
\bibitem{Feng:2006dp}
  B.~Feng, M.~Li, J.~Q.~Xia, X.~Chen and X.~Zhang,
  \emph{Searching for CPT violation with WMAP and BOOMERANG},
  Phys.\ Rev.\ Lett.\  {\bf 96}, 221302 (2006),
  [arXiv:astro-ph/0601095].
  %%CITATION = PRLTA,96,221302;%%

%\cite{Nodland:1997cc}
\bibitem{Nodland:1997cc}
  B.~Nodland and J.~P.~Ralston,
  \emph{Indication of Anisotropy in Electromagnetic Propagation over Cosmological Distances},
  Phys.\ Rev.\ Lett.\  {\bf 78}, 3043 (1997),
  [arXiv:astro-ph/9704196].
  %%CITATION = PRLTA,78,3043;%%

%\cite{Wardle:1997gu}
\bibitem{Wardle:1997gu}
  J.~F.~C.~Wardle, R.~A.~Perley and M.~H.~Cohen,
  \emph{Observational Evidence Against Birefringence over Cosmological Distances},
  Phys.\ Rev.\ Lett.\  {\bf 79}, 1801 (1997),
  [arXiv:astro-ph/9705142].
  %%CITATION = PRLTA,79,1801;%%

%\cite{Kostelecky:2008be}
\bibitem{Kostelecky:2008be}
  A.~Kostelecky and M.~Mewes,
  \emph{Astrophysical Tests of Lorentz and CPT Violation with Photons},
  [arXiv:0809.2846 [astro-ph]].
  %%CITATION = ARXIV:0809.2846;%%

%\cite{Andrianov:1998wj}
\bibitem{Andrianov:1998wj}
  A.~A.~Andrianov and R.~Soldati,
  \emph{Patterns of Lorentz symmetry breaking in QED by CPT-odd interaction},
  Phys.\ Lett.\  B {\bf 435}, 449 (1998),
  [arXiv:hep-ph/9804448].
  %%CITATION = PHLTA,B435,449;%%

%\cite{Adam:2001ma}
\bibitem{Adam:2001ma}
  C.~Adam and F.~R.~Klinkhamer,
  \emph{Causality and CPT violation from an Abelian Chern-Simons like term},
  Nucl.\ Phys.\  B {\bf 607}, 247 (2001),
  [arXiv:hep-ph/0101087].
  %%CITATION = NUPHA,B607,247;%%

%\cite{Kharzeev:2004ey}
\bibitem{Kharzeev:2004ey}
  D.~Kharzeev,
  \emph{Parity violation in hot QCD: Why it can happen, and how to look for it},
  Phys.\ Lett.\  B {\bf 633}, 260 (2006)
  [arXiv:hep-ph/0406125].
  %%CITATION = PHLTA,B633,260;%%

%\cite{Kharzeev:2007tn}
\bibitem{Kharzeev:2007tn}
  D.~Kharzeev and A.~Zhitnitsky,
  \emph{Charge separation induced by P-odd bubbles in QCD matter},
  Nucl.\ Phys.\  A {\bf 797}, 67 (2007)
  [arXiv:0706.1026 [hep-ph]].
  %%CITATION = NUPHA,A797,67;%%

%\cite{Kharzeev:2007jp}
\bibitem{Kharzeev:2007jp}
  D.~E.~Kharzeev, L.~D.~McLerran and H.~J.~Warringa,
  \emph{The effects of topological charge change in heavy ion collisions: ``Event by
  event P and CP violation''},
  Nucl.\ Phys.\  A {\bf 803}, 227 (2008)
  [arXiv:0711.0950 [hep-ph]].
  %%CITATION = NUPHA,A803,227;%%

%\cite{Fukushima:2008xe}
\bibitem{Fukushima:2008xe}
  K.~Fukushima, D.~E.~Kharzeev and H.~J.~Warringa,
  \emph{The Chiral Magnetic Effect},
  Phys.\ Rev.\  D {\bf 78}, 074033 (2008)
  [arXiv:0808.3382 [hep-ph]].
  %%CITATION = PHRVA,D78,074033;%%

%\cite{Kharzeev:2009fn}
\bibitem{Kharzeev:2009fn}
  D.~E.~Kharzeev,
  \emph{Topologically induced local P and CP violation in QCD $\times$ QED},
  Annals Phys.\  {\bf 325}, 205 (2010)
  [arXiv:0911.3715 [hep-ph]].
  %%CITATION = APNYA,325,205;%%

%\cite{Kharzeev:2010ym}
\bibitem{Kharzeev:2010ym}
  D.~E.~Kharzeev,
  \emph{Axial anomaly, Dirac sea, and the chiral magnetic effect},
  [arXiv:1010.0943 [hep-ph]].
  %%CITATION = ARXIV:1010.0943;%%

%\cite{:2009uh}
\bibitem{:2009uh}
  B.~I.~Abelev {\it et al.}  [STAR Collaboration],
  \emph{Azimuthal Charged-Particle Correlations and Possible Local Strong Parity
  Violation},
  Phys.\ Rev.\ Lett.\  {\bf 103}, 251601 (2009)
  [arXiv:0909.1739 [nucl-ex]].
  %%CITATION = PRLTA,103,251601;%%

%\cite{:2009txa}
\bibitem{:2009txa}
  B.~I.~Abelev {\it et al.}  [STAR Collaboration],
  \emph{Observation of charge-dependent azimuthal correlations and possible local
  strong parity violation in heavy ion collisions},
  Phys.\ Rev.\  C {\bf 81}, 054908 (2010)
  [arXiv:0909.1717 [nucl-ex]].
  %%CITATION = PHRVA,C81,054908;%%

%\cite{Ajitanand:2010rc}
\bibitem{Ajitanand:2010rc}
  N.~N.~Ajitanand, R.~A.~Lacey, A.~Taranenko and J.~M.~Alexander,
  \emph{A new method for the experimental study of topological effects in the
  quark-gluon plasma},
  [arXiv:1009.5624 [nucl-ex]].
  %%CITATION = ARXIV:1009.5624;%%

%\cite{Barraz:2007mi}
\bibitem{Barraz:2007mi}
  N.~M.~.~Barraz, J.~M.~Fonseca, W.~A.~Moura-Melo and J.~A.~Helayel-Neto,
  \emph{On Dirac-like monopoles in a Lorentz- and CPT-violating electrodynamics},
  Phys.\ Rev.\  D {\bf 76}, 027701 (2007),
  [arXiv:hep-th/0703042].
  %%CITATION = PHRVA,D76,027701;%%

%\cite{Guimaraes:2006gj}
\bibitem{Guimaraes:2006gj}
  M.~S.~Guimaraes, L.~Grigorio and C.~Wotzasek,
  \emph{The dual of the Carroll-Field-Jackiw model},
  [arXiv:hep-th/0609215].
  %%CITATION = HEP-TH/0609215;%%

%\cite{GonzalezGarcia:2002dz}
\bibitem{GonzalezGarcia:2002dz}
  M.~C.~Gonzalez-Garcia and Y.~Nir,
  \emph{Developments in neutrino physics},
  Rev.\ Mod.\ Phys.\  {\bf 75}, 345 (2003),
  [arXiv:hep-ph/0202058].
  %%CITATION = RMPHA,75,345;%%

%\cite{Kostelecky:2008ts}
\bibitem{Kostelecky:2008ts} V. A. Kostelecky and N. Russell, \emph{Data Tables for Lorentz and CPT Violation}, [arXiv:0801.0287v4 [hep-ph]], accepted for publication in Review of Modern Physics.

\end{thebibliography}
\end{document}